\documentclass[11pt]{article}
\pdfoutput=1

\usepackage[utf8]{inputenc}

\usepackage{jheppub}
\usepackage{url,comment}
\usepackage{times}
\usepackage{latexsym}
\usepackage{graphicx, graphics, hyperref, amsmath, amssymb, slashed, xcolor, bbm,bm,amsthm, array}
 \usepackage{subfigure}
\usepackage{listings} 
 \usepackage{trimclip}
 \usepackage{braket}
 \usepackage{mathtools}          

\usepackage{rotating}
\usepackage{afterpage}
 
\newcommand{\nc}{\newcommand}

\nc{\beq}{\begin{equation}}
\nc{\eeq}{\end{equation}}
\nc{\barray}{\begin{eqnarray}}
\nc{\earray}{\end{eqnarray}}
\nc{\barrayn}{\begin{eqnarray*}}
\nc{\earrayn}{\end{eqnarray*}}
\nc{\bcenter}{\begin{center}}
\nc{\ecenter}{\end{center}}
\nc{\mc}{\mathcal}
\nc{\er}[1]{(\ref{eq:#1})}
\nc{\onehalf}{\frac{1}{2}} 
\nc{\partialbar}{\bar{\partial}}
\nc{\psit}{\widetilde{\psi}}
\nc{\Tr}{\mbox{Tr}}
\nc{\hc}{\mbox{H.c.}}
\nc{\ev}{\;\mathrm{eV}}
\nc{\mev}{\;\mathrm{MeV}}
\nc{\gev}{\;\mathrm{GeV}}
\nc{\kev}{\;\mathrm{keV}}
\nc{\tev}{\;\mathrm{TeV}}
\nc{\eval}{\biggr\rvert}

\def\chii0{\chi_i^0}
\def\chij0{\chi_j^0}

\newcommand{\gsim}{\lower.7ex\hbox{$\;\stackrel{\textstyle>}{\sim}\;$}}
\newcommand{\lsim}{\lower.7ex\hbox{$\;\stackrel{\textstyle<}{\sim}\;$}}
\nc{\ttbar}{t\bar t}

\newcommand{\cref}[1]{Chapter~\ref{c.#1}}

\graphicspath{{plots/}}

\title{Twin Higgs Portal Dark Matter}

\author{David Curtin,}
\author{Shayne Gryba}

\affiliation{Department of Physics, University of Toronto, Toronto, ON M5S 1A7, Canada}

\emailAdd{dcurtin@physics.utoronto.ca}
\emailAdd{sgryba@physics.utoronto.ca}

\abstract{
Many minimal models of dark matter (DM) or canonical solutions to the hierarchy problem are either excluded or severely constrained by LHC and direct detection null results. 
In particular, Higgs Portal Dark Matter (HPDM) features a scalar coupling to the Higgs via a quartic interaction, and obtaining the measured relic density via thermal freeze-out gives definite direct detection predictions which are now almost entirely excluded. 
The Twin Higgs solves the little hierarchy problem without coloured top partners by introducing a twin sector related to the Standard Model (SM) by a discrete symmetry. 
We generalize HPDM to arbitrary Twin Higgs models and introduce \emph{Twin Higgs Portal Dark Matter} (THPDM), which features a DM candidate with an $SU(4)$-invariant quartic coupling to the Twin Higgs scalar sector. 
Given the size of quadratic corrections to the DM mass, its most motivated scale is near the mass of the radial mode. In that case, DM annihilation proceeds with the full Twin Higgs portal coupling, while direct detection is suppressed by the pNGB nature of the 125 GeV Higgs. 
For a standard cosmological history, this results in a predicted direct detection signal for THPDM that is orders of magnitude below that of HPDM with very little dependence on the precise details of the twin sector, evading current bounds but predicting possible signals at next generation experiments.
In many Twin Higgs models, twin radiation contributions to $\Delta N_\mathrm{eff}$ are suppressed by an asymmetric reheating mechanism. 
We study this by extending the $\nu$MTH and $X$MTH models to include THPDM and compute the viable parameter space according to the latest CMB bounds.
The injected entropy dilutes the DM abundance as well, resulting in additional suppression of direct detection below the neutrino floor. 
 }

\begin{document}

\begin{flushright}
\small{.}
\end{flushright}

\maketitle

\section{Introduction}

The electroweak hierarchy problem was the subject of an intense research program long before the discovery of the Higgs boson at the Large Hadron Collider (LHC) in 2012, and to this day the apparent instability of the Higgs mass to large radiative corrections continues to strongly motivate the existence of new physics near the TeV scale.
Several theories could explain the observed Higgs mass without excessive fine-tuning, both in the form of UV-complete models that guarantee naturalness up to the Planck scale, such as Natural SUSY \cite{Batra:2003nj, Ellwanger:2009dp, Hall:2011aa, Curtin:2014zua, Casas:2014eca} or Composite Higgs \cite{Kaplan:1983sm}, as well as proposals like the Little Higgs \cite{ArkaniHamed:2001nc, Schmaltz:2005ky} that seek only to stabilize the Higgs sector up to an experimental cutoff of 5 - 10 TeV, where additional physics can make the theory natural to higher scales.
However, all of these solutions predict TeV-scale particles with SM quantum numbers, in particular  QCD colour, which have not yet been observed at the Large Hadron Collider (LHC) contrary to earlier expectation ~\cite{Aad:2019ftg,Aaboud:2018ujj,Aaboud:2018kya,Aaboud:2017vwy,Aaboud:2017aeu,Aaboud:2018zjf,Aaboud:2017phn,Sirunyan:2019ctn,Sirunyan:2019zyu,Sirunyan:2019hzr,Sirunyan:2018psa}.
Canonical solutions to the hierarchy problem with new states heavy enough to satisfy LHC bounds would therefore exhibit significant fine-tuning, giving rise to the so-called ``little hierarchy problem.” 

The most important contributions to the Higgs potential come from top quark loops. These dangerous contributions can be canceled by introducing top partners related to the top by a symmetry, predicting top partners charged under SM colour that should be produced by the LHC at high rates.
The growing tension with experimental null results has led to the creation of a new class of models, called neutral naturalness models \cite{Chacko:2005pe,Craig:2015pha,Craig:2014aea,Barbieri:2016zxn,Chacko:2005un,PhysRevD.97.035017,PhysRevD.101.095014,Burdman:2006tz}, whose partner particles are SM singlets, or at least not charged under SM colour. 
 This is possible if the symmetry which relates the top and the top partner does not commute with SM colour. 
 These models solve the little hierarchy problem by providing an explanation for how TeV-scale top partners may have so far evaded our searches. Most notable among these models are the Twin Higgs (TH) models \cite{Chacko:2005pe,Craig:2015pha,Craig:2014aea,Barbieri:2016zxn,PhysRevD.97.035017,PhysRevD.101.095014}, which cancel quadratically divergent loop corrections to the Higgs potential through the introduction of a twin sector, related via a discrete $\mathbb{Z}_2$ symmetry to the SM. The Higgs sector of the theory transforms under an approximate global $SU(4)$ symmetry, which is broken by the gauge and Yukawa couplings of each sector, and the form of the potential induces spontaneous breaking from $SU(4)$ to $SU(3)$. After gauging an $SU(2)$ $\times$ $SU(2)$ subgroup of the $SU(4)$, under which $H$ transforms as $(H_A, H_B)$, only 1 of the 7 pseudo-Nambu Goldstone bosons (pNGBs) remains in the spectrum -- this can be identified as the 125 GeV Higgs. 
The most dangerous quadratically divergent one-loop corrections are always found in $\mathbb{Z}_2$-symmetric pairs, so these corrections to the potential exhibit an accidental $SU(4)$ symmetry.
Since the 125 GeV Higgs is a pNGB of $SU(4)$ breaking, its mass is necessarily protected until two-loop effects become important, typically up to scales of 5-10 TeV.

Phenomenological constraints force us to softly break the $\mathbb{Z}_2$ symmetry in the Higgs sector, analogous to soft SUSY breaking terms in e.g. the MSSM \cite{Martin:1997ns}. The completely $\mathbb{Z}_2$-symmetric TH model predicts large couplings between the Higgs and twin matter and a significant reduction in Higgs couplings to visible matter compared to the SM prediction. However, these couplings are tightly constrained by precision electroweak measurements~\cite{Burdman:2014zta, Chacko:2017xpd}, and couplings to the mirror sector are constrained by invisible branching ratio measurements. The latest results from ATLAS and CMS indicate $\text{Br}(h \to \text{invisible}) \lesssim 0.13$ at 95\% CL~\cite{ATLAS:2020cjb,Sirunyan:2018owy}, and the high-luminosity (HL) upgrade is expected to tighten this to $\sim$ 4\% at 95\% CL~\cite{CMS:2018tip}. To satisfy these bounds, soft $\mathbb{Z}_2$-breaking terms are introduced to misalign the vevs of the two sectors such that $v/f \lesssim 1/3$, where $v$ is the electroweak vev and $f$ is the vev of the full Higgs sector. This necessity comes at the price of a very modest $2 v^2/f^2 \lesssim 20\%$ tuning in minimal models~\cite{Craig:2015pha,Barbieri:2015lqa}, with further relaxation possible in certain constructions \cite{Beauchesne:2015lva,Harnik:2016koz,Goh:2007dh,Katz:2016wtw}.

If the $SU(4)$ symmetry is linearly realized, then the radial mode of the breaking is present in the spectrum, hereafter called the ``heavy Higgs" or ``twin Higgs." Current and future measurements are expected to be sensitive to this signature, in particular through di-boson production at the HL-LHC~\cite{Katz:2016wtw,Chacko:2017xpd,Ahmed:2017psb,Buttazzo:2015bka,Aad:2019zwb,Kilic:2018sew,Alipour-fard:2018mre}. The MTH also admits a number of ultra-violet (UV) completions, including supersymmetric models~\cite{Falkowski:2006qq,Chang:2006ra,Craig:2013fga,Katz:2016wtw,Badziak:2017syq,Badziak:2017kjk,Badziak:2017wxn,Asadi:2018abu}, holographic models~\cite{Geller:2014kta}, composite Higgs scenarios~\cite{Batra:2008jy,Geller:2014kta,Barbieri:2015lqa,Low:2015nqa}, and extra-dimensional orbifold gauge theories~\cite{Craig:2014aea,Craig:2014roa}. Many of these models predict heavy, exotic states that carry both visible and twin sector quantum numbers that can be searched for at colliders~\cite{Cheng:2015buv,Cheng:2016uqk,Li:2017xyf}.

It is interesting to ask whether Twin Higgs solutions to the little hierarchy problems are capable of providing plausible dark matter (DM) candidates as well. Indeed, the twin tau lepton and neutrino, twin neutralino, twin baryons, twin $W^{\pm}$, twin mesons, and the twin neutron have all been proposed as potential sources for DM in various TH constructions \cite{Prilepina:2016rlq,Craig:2015xla,Farina:2015uea,Garcia:2015loa,Garcia:2015toa,Cheng:2018vaj,Beauchesne:2020mih,PhysRevD.99.015005,Badziak:2019zys}, but some are already significantly constrained by the latest direct detection experiments \cite{Aprile:2015uzo,Akerib:2016vxi,Tan:2016zwf}.

In this paper, we examine the simple possibility of Scalar Higgs Portal Dark Matter (HPDM)~\cite{SILVEIRA1985136,McDonald:1993ex,Burgess:2000yq,Casas:2017jjg}, which is now essentially excluded \cite{Escudero:2016gzx,Cline:2013gha,Feng:2014vea}, in a Twin Higgs context. 
We define the Twin Higgs Portal Dark Matter (THPDM) scenario by adding a stable scalar $S$ coupled to the extended Twin Higgs sector through the $SU(4)$-symmetric quartic interaction $\lambda_{HS} S^2 ( H_A^2 + H_B^2 )$.
This interaction provides a portal between the DM scalar and the visible and twin matter sectors, allowing for both thermal annihilation in the early universe and direct detection. 
Just as for HPDM, for a given cosmological history and DM mass, reproducing the measured DM relic abundance yields a definite prediction for the coupling constant $\lambda_{HS}$,  and therefore for direct and indirect detection observables. However, unlike simple HPDM, the twin nature of the Higgs sector significantly changes the phenomenology and parameter space of the THPDM scenario.
In this paper, we carefully explore the direct detection parameter space of THPDM, for both standard and motivated non-standard cosmological histories.

The crucial difference between the HPDM and THPDM models arises from the pNGB nature of the 125 GeV Higgs. At low energies far below the mass scale of the radial mode, its couplings are dominated by small $SU(4)$- and $\mathbb{Z}_2$ breaking interactions in the Twin Higgs potential, while at higher energies these couplings are highly momentum dependent and unsuppressed.
Therefore, if the DM scalar  in THPDM has mass near or below the 125 GeV Higgs mass, it reproduces the phenomenology of HPDM, predicting direct detection signals that are already excluded.
On the other hand, if the DM mass is of a similar scale as the mass of the Twin Higgs radial mode, then it annihilates with the unsuppressed $\lambda_{HS}$ coupling but scatters in direct detection experiments with the suppressions arising from the pNGB nature of the 125 GeV Higgs. 

Compared to the HPDM scenario, this leads to a drastic reduction in the expected direct detection signal. 
Crucially, it is this latter scenario that is strongly favoured by naturalness considerations within the Twin Higgs setup, since loop corrections drive the DM scalar mass towards the mass scale of the radial mode.
Within the THPDM scenario, we therefore do not expect direct detection signals in current experiments. On the other hand, a signal in next generation experiments such as LUX-Zeplin (LZ)~\cite{Akerib:2015cja}, XENONnT \cite{Aprile:2015uzo}, and DARWIN \cite{Aalbers:2016jon} is predicted across large regions of THPDM parameter space.
This suppression of direct detection signals is also completely general for any DM candidate that couples to the Twin Higgs scalar sector in an $SU(4)$ invariant way, since it arises from the pNGB nature of the Higgs.
\footnote{This suppression of direct detection signatures due to pNGB dynamics is reminiscent of pNGB-DM models, which typically feature either the Higgs and the DM both as pNGBs as in \cite{Fonseca:2015gva,Balkin:2018tma,Ballesteros:2017xeg,Balkin:2017aep,Alanne:2018xli,Ahmed:2020hiw,Cai:2020njb}, or with only the DM as a pNGB as in \cite{Abe:2020iph,Alanne:2018zjm,Huitu:2018gbc,Karamitros:2019ewv,Jiang:2019soj,Arina:2019tib,Bandyopadhyay:2020otm}. To the best of our knowledge, our study is the first to encounter this phenomenon due to the Higgs but not the DM being a pNGB (although simply adding a Higgs-portal DM candidate to any composite Higgs model would achieve the same effect). Furthermore, the general idea of suppressing direct detection by enhancing the relative annihilation rate at early times has also been studied before, see e.g.~\cite{Schmaltz:2005ky, Batra:2003nj, Berger:2020maa}.}

The exact predictions for direct detection depend on the cosmological history of the Twin Higgs. The minimal Mirror Twin Higgs  (MTH) predicts light degrees of freedom present during Big Bang Nucleosynthesis (BBN) that are completely excluded by measurements of $\Delta N_{\text{eff}} \leq 0.23$ (at 2$\sigma$) \cite{Aghanim:2018eyx}. A class of models that includes the Fraternal Twin Higgs \cite{Craig:2015pha,Barbieri:2016zxn} solves this with hard $\mathbb{Z}_2$-breaking to remove light degrees of freedom while preserving a standard cosmological history, meaning direct detection signals of THPDM are uniquely determined by the DM mass. On the other hand, in the asymmetrically reheated Mirror Twin Higgs \cite{Chacko:2016hvu,Craig:2016lyx,Chacko:2018vss}, the twin sector contribution to $\Delta N_{\text{eff}}$ is reduced by some massive, long-lived particle species that decays preferentially to the visible sector at late times, reducing the number density of twin radiation. In this departure from standard cosmology the DM relic abundance is diluted alongside the twin sector, implying that a smaller coupling during freeze out is required to match the observed density today.\footnote{Alternatively, if dark matter is frozen in rather than a thermal relic, asymmetric reheating may help with achieving the correct abundance \cite{Koren:2019iuv}.} This possibility further suppresses direct detection signatures of  THPDM, anywhere from within the expected range of next generation direct detection experiments to below the neutrino floor.
We review two explicit models of asymmetric reheating, the $\nu$MTH \cite{Chacko:2016hvu} and the $X$MTH \cite{Craig:2016lyx}, update bounds on their parameter space, and show how DM dilution is correlated with $\Delta N_{\text{eff}}$ signals.
 In such scenarios, direct detection of DM could be very challenging, but we would expect to detect the small but nonzero deviation in $\Delta N_\mathrm{eff}$ at upcoming CMB-S4 experiments \cite{Abazajian:2016yjj}.

The paper is structured as follows. In Sections \ref{sec:MTH_model} and  \ref{sec:SSHPDM}  we briefly review relevant details of Twin Higgs models and singlet Scalar Higgs Portal DM (HPDM). 
In Section \ref{sec:THPDM} we define the THPDM scenario, analytically demonstrate the suppression mechanism for direct detection, and compare predictions for direct detection experiments to HPDM.  
Section \ref{sec:AR} is devoted to THPDM models with asymmetric reheating. We conclude in Section \ref{sec:conclusion}.

\section{Twin Higgs Models}
\label{sec:MTH_model}

Twin Higgs models solve the little hierarchy problem through the introduction of a twin sector of new particles, whose structure is related by a discrete symmetry to the structure of the Standard Model. The original proposal for a Twin Higgs model was the Mirror Twin Higgs (MTH) \cite{Chacko:2005pe}, which
features an extended gauge symmetry $SU(3)_A \otimes SU(2)_A \otimes U(1)_A \otimes SU(3)_B \otimes SU(2)_B \otimes U(1)_B$, where the SM fields are charged under SM$_A$ gauge groups and an exact $\mathbb{Z}_2$-symmetric copy of the SM matter sector is charged under the SM$_B$ gauge groups. Other varieties, mostly with  varying degrees of hard $\mathbb{Z}_2$ breaking to modify the twin matter spectrum, have been proposed \cite{Craig:2015pha,Barbieri:2016zxn,Barbieri:2017opf,Craig:2016kue} to solve the cosmological problems of the original setup. 
Here we review the basic Twin Higgs setup and the properties of the scalar sector that are common to most of these constructions and instrumental to the Twin Higgs Portal Dark Matter mechanism. We then comment on two benchmark scenarios for realistic Twin Higgs models, the Fraternal Twin Higgs \cite{Craig:2015pha} and the Asymmetrically Reheated Mirror Twin Higgs \cite{Chacko:2016hvu,Craig:2016lyx}, that we use to explore THPDM scenarios in this paper.

We follow mainly the exposition in Ref.~\cite{Ahmed:2017psb}, taking the extended Higgs sector to transform linearly as a global $SU(4)$ fundamental
\[
H = \begin{pmatrix} H_A \\ H_B \\ \end{pmatrix},
\]
where A denotes the SM or visible sector and B denotes the twin sector. We denote the vacuum expectation values of these fields $\frac{1}{\sqrt{2}} v_A,\frac{1}{\sqrt{2}} v_B$. The full potential includes terms that both spontaneously and explicitly break the $SU(4)$, required to satisfy current bounds on Higgs phenomenology. It is given by 
\beq
\label{eqn:V_TH}
V = \lambda \left(H_A^\dagger H_A + H_B^\dagger H_B -\frac{f_0^2}{2} \right)^2 + \kappa \left( \left(H_A^\dagger H_A \right)^2 + \left(H_B^\dagger H_B \right)^2 \right) + \sigma f_0^2 H_A^\dagger H_A.
\eeq
Note the soft $\mathbb{Z}_2$ and $SU(4)$ breaking term $\sigma$, which we discuss further below. 
The $\lambda$ term spontaneously breaks $SU(4)$ to $SU(3)$, generating 7 Nambu-Goldstone bosons (NGBs) in the process. Six of these are eaten by the electroweak bosons $W^{\pm}_{A,B}$, $Z_{A,B}$ of each sector, while the seventh remains uneaten and will eventually be identified as the observed 125 GeV Higgs. The spontaneous breaking of $SU(4)$ will also give rise to a massive radial mode with $m = \sqrt{2 \lambda} f_0$ for $\sigma = \kappa = 0$, which can be integrated out in a non-linear sigma model (NLSM) effective description far below the symmetry breaking scale. We work in the linear sigma model (LSM) description where the radial mode is included in the spectrum, in order to understand its effect on THPDM freeze--out in a perturbative Twin Higgs construction. 
However, the NLSM picture will be helpful to understand key features of THPDM phenomenology, making manifest the pseudo-Goldstone nature of the Higgs (see Section \ref{sec:THPDM}).

If we are going to identify the NGB as the 125 GeV Higgs then the $SU(4)$ symmetry must be explicitly broken. This is done by introducing the $\kappa$ and $\sigma$ terms, which softly break both the $\mathbb{Z}_2$ and $SU(4)$ symmetries. In order to avoid unreasonable levels of fine-tuning, we demand that the symmetry-breaking coefficients always be smaller than the symmetry-preserving ones, $\kappa, \sigma \ll \lambda$.
The $\kappa$ term in the potential explicitly breaks $SU(4)$ but preserves the $\mathbb{Z}_2$ symmetry. This generates a small mass for the surviving NGB, rendering it instead a pseudo-Nambu Goldstone boson (pNGB) and allowing us to identify it as the 125 GeV Higgs. 
One-loop corrections to the Higgs potential must have the form
\beq
\Delta V_{\text{top}} = \frac{3 \Lambda^2}{8 \pi^2} \left(y_A^2 H_A^\dagger H_A + y_B^2 H_B^\dagger H_B \right)
\eeq
where the $\mathbb{Z}_2$ symmetry enforces $y_A = y_B$, which is insensitive to soft $\mathbb{Z}_2$ breaking. These quadratic corrections therefore have an \emph{accidental} $SU(4)$ symmetry and cannot affect the light Higgs degree of freedom, making it a pNGB of an accidental global symmetry. Note that this effect operates only at one-loop, so a UV completion to solve the big Hierarchy Problem is needed above scales of 5-10 TeV where two-loop effects become relevant; see e.g.~\cite{Falkowski:2006qq,Chang:2006ra,Craig:2013fga,Katz:2016wtw,Badziak:2017syq,Badziak:2017kjk,Badziak:2017wxn,Asadi:2018abu,Geller:2014kta,Batra:2008jy,Geller:2014kta,Barbieri:2015lqa,Low:2015nqa,Craig:2014aea,Craig:2014roa}.

The $\sigma$ term breaks both $\mathbb{Z}_2$ and $SU(4)$, and its primary role is to misalign the vacuum. Without this term, the potential would be $\mathbb{Z}_2$ symmetric and the light pNGB would be shared equally between both sectors in field space, with $v_A = v_B$. In this case we would expect two related effects to occur: the couplings between the SM Higgs and the matter content of both sectors would become equal, and the size of the Higgs couplings to the SM would be reduced by a factor $1/\sqrt{2}$ relative to the SM prediction. However, the Higgs-SM couplings are already tightly constrained by precision electroweak measurements \cite{Burdman:2014zta, Chacko:2017xpd}, and current measurements find $\text{Br}(h \to \text{invisible}) \lesssim 0.13$ at 95\% CL~\cite{ATLAS:2020cjb,Sirunyan:2018owy}, limiting the possible size of Higgs couplings to the mirror sector. It is therefore necessary to break $\mathbb{Z}_2$ such that $\braket{H}$ lies mostly along the B direction, $v_B \gg v_A$. Dynamical methods of $\mathbb{Z}_2$ breaking are explored for example in \cite{Batell:2019ptb,Beauchesne:2015lva,Harnik:2016koz,Yu:2016swa,Jung:2019fsp,Yu:2016bku}, and these could easily be incorporated into our discussion, but we simply parameterize the soft $\mathbb{Z}_2$ breaking by adding the $\sigma$ term. Regardless of the $\mathbb{Z}_2$-breaking origin, the result is that the couplings between the SM Higgs and the mirror sector are reduced, and the Higgs-SM couplings approach the SM prediction. Satisfying experimental constraints requires $v_A/v_B \lesssim 1/3$, while Refs.~\cite{Craig:2015pha,Barbieri:2015lqa} show that the tuning of the theory is roughly $2v^2/f^2$, where $f^2 = v_A^2 + v_B^2$ and $v \equiv v_A = 246~\text{GeV}$. With this in mind we consider a range of $f/v$ from $3$ to roughly $7$ as being most motivated, which is within experimental limits and corresponds roughly to a tuning range of $4-20$\%.

We next examine the mass eigenstates of the theory. Minimizing the potential as in \cite{Ahmed:2017psb} leads to nonzero vevs in both sectors:
\beq
v_A = \sqrt{2} \braket{H_A} = f_0 \sqrt{ \frac{\kappa \lambda - (\kappa + \lambda) \sigma}{\kappa \left( \kappa + 2 \lambda \right)}}
\eeq

\beq
v_B = \sqrt{2} \braket{H_B} = f_0 \sqrt{ \frac{\lambda (\kappa + \sigma)}{\kappa \left( \kappa + 2 \lambda \right)}}
\eeq
and the fluctuations about the vacuum can be parameterized in the unitary gauge by
 \beq
 H_A = \frac{1}{\sqrt{2}} \begin{bmatrix} 0 \\ v_A + h_A \\ \end{bmatrix} ~~~~~,~~~~~  H_B = \frac{1}{\sqrt{2}} \begin{bmatrix} 0 \\ v_B + h_B \\ \end{bmatrix}.
 \eeq
To find the resulting mass eigenstates we diagonalize the potential through a rotation angle $\theta$,
\beq
\label{eqn:diag}
 \begin{pmatrix} h_- \\ h_+ \\ \end{pmatrix} = \begin{pmatrix} \cos \theta &- \sin \theta \\ \sin \theta &\cos \theta \\ \end{pmatrix} \begin{pmatrix} h_A \\ h_B \\ \end{pmatrix}
\eeq
which yields two mass eigenstates with masses given by
\beq
\label{eqn:mass_theory}
m_{\pm}^2 = f^2 \left(\kappa + \lambda \right) \pm \sqrt{\left(v_B^2 - v_A^2\right)^2 \left(\kappa + \lambda \right)^2 + 4 \lambda^2 v_A^2 v_B^2}.
\eeq
Note that in the $SU(4)$-symmetric limit with $\kappa = \sigma = 0$ we recover a massless mode $m_- = 0$ and a radial mode with $m_+ = 2 \lambda f^2$. The mixing angle $\theta$ is conveniently written as
\beq 
\label{eqn:angle_theory}
\tan(2 \theta) = \frac{2 v_A v_B}{\left(v_B^2 - v_A^2 \right) \left(1 + \kappa/\lambda \right)} =  \frac{ \left( 2 v/f \right) \sqrt{1- v^2/f^2}}{\left(1 - 2 v^2/f^2 \right) \left( 1 + \kappa/ \lambda \right)}.
\eeq
In the limit that $\kappa \ll \lambda$ and $ v/f \ll 1$, the mixing angle becomes $\theta \approx v/f$ and the mass eigenstates of the theory take the simple form 
\begin{align}
h_- &\equiv h = \cos (v/f) h_A - \sin(v/f)h_B \\
h_+ &\equiv \hat{h} = \sin(v/f) h_A + \cos(v/f) h_B
\end{align}
From here it can be seen that couplings between $h$ and the visible sector will be modified by a factor of $\cos (v/f)$. The implication is that in this model the 125 GeV Higgs is in fact \textit{mostly} the SM Higgs, but with a small twin Higgs component. As mentioned above, this mixing leads to predictions for invisible Higgs branching ratios and precision electroweak measurements. 

It is well known that the original Mirror Twin Higgs model is thoroughly excluded by cosmological measurements (see e.g.~\cite{Chacko:2016hvu,Craig:2016lyx,Chacko:2018vss}), since light mirror neutrinos and mirror photons produce very large contributions to $\Delta N_\mathrm{eff}$. 
In realistic Twin Higgs models, this is addressed in one of two ways: either by removing or lifting the light degrees of freedom in the mirror sector, or by invoking a non-standard cosmological history which reduces mirror sector contributions to $\Delta N_\mathrm{eff}$.

Hard  $\mathbb{Z}_2$ breaking, mostly in the fermion sector, can lift or eliminate light degrees of freedom. 
Since cancellations of the gauge and top loops corrections to the Higgs potential are the most important for naturalness, the $\mathbb{Z}_2$ symmetry can be broken for any of the lighter degrees of freedom in the twin spectrum. They can be removed or made heavier, which eliminates the problematic cosmology of the Mirror Twin Higgs. 
The Fraternal Twin Higgs (FTH) model \cite{Craig:2015pha} is in a sense the most minimal or extreme example of such an approach, eliminating the mirror photon and the first two mirror matter generations, but less extreme solutions with many mirror states at collider scales have also been considered \cite{Barbieri:2016zxn,Barbieri:2005ri,Farina:2015uea}.
Amongst the remaining states, only the mirror top mass has to be in agreement with the $\mathbb{Z}_2$ symmetry -- the mirror bottom, mirror tau and mirror tau neutrino masses are left as free parameters. 
Thus there may be no sources of twin radiation during BBN, and bounds on $\Delta N_{\text{eff}}$ can be trivially satisfied. 
The collider phenomenology of TH models without cosmological problems can be radically different from the MTH. In particular, the absence of light twin degrees of freedom can lead to the presence of long-lived particles in the twin spectrum, such as twin glueballs or twin bottomonium in the FTH \cite{Craig:2015pha}. These states can only decay through the small Higgs portal coupling that is required by naturalness, making them meta-stable. 
It has been shown~\cite{Curtin:2015fna,Cheng:2015buv}  that displaced vertex searches in ATLAS, CMS, or LHCb \cite{Aad:2020srt,Aad:2019xav,CMS:2020idp,Sirunyan:2019nfw,LHCb:2018hiv}, or in external transverse LLP detectors for the HL-LHC like MATHUSLA and CODEX-b \cite{Beacham:2019nyx,Lubatti:2019vkf,Curtin:2018mvb,Gligorov:2017nwh},
can discover these states for FTH-type models with top partner masses in excess of a  TeV.
Other TH constructions with hard $\mathbb{Z}_2$ breaking can also give rise to a variety of LLP signatures, including decays via the twin photon portal \cite{Cheng:2015buv}.

The cosmological history in Mirror Twin Higgs models can be easily modified by including a source of asymmetric reheating \cite{Chacko:2016hvu,Craig:2016lyx} while maintaining a $\mathbb{Z}_2$ symmetric twin spectrum. As mentioned in the introduction, these models feature late-decaying particles that can preferentially reheat the visible sector relative to the twin sector after the Higgs portal freezes out, thereby reducing the twin sector's contribution to $\Delta N_{\text{eff}}$. 
If there is a small twin baryon asymmetry (motivated by the $\mathbb{Z}_2$ symmetry), these twin baryons could constitute a fraction of dark matter~\cite{Markevitch:2003at,Kahlhoefer:2013dca,Harvey:2015hha,Robertson:2016xjh,Wittman:2017gxn,Harvey:2018uwf,Garcia:2015toa,Farina:2015uea}, though self-interaction bounds usually preclude them from making up its entirety. This leads to a rich array of cosmological signatures like mirror baryonic oscillations and their imprints in Large Scale Structure and the CMB~\cite{Chacko:2018vss}, as well as potentially spectacular astrophysical predictions like the existence and observability of mirror stars~\cite{Curtin:2019ngc,Curtin:2019lhm}. 
The asymmetric reheating also has important direct implications for dark matter direct detection, since any existing dark matter population is diluted alongside the twin sector. 
This can have important implications for direct detection experiments, since in many prominent dark matter models the coupling that controls the freeze-out abundance is the same one that is probed in the laboratory. 
Dilution of dark matter is a major point of interest in our analysis.

The Twin Higgs Portal Dark Matter mechanism is a general consequence of the Twin Higgs setup. After briefly reviewing scalar Higgs Portal Dark Matter in Section \ref{sec:SSHPDM}, we therefore study the phenomenology of THPDM first by assuming standard cosmological history in Section \ref{sec:THPDM} for two extreme cases of twin sectors, FTH-like and MTH-like. The former is the most minimal twin sector that can solve the little hierarchy problem, while the latter is a stand-in for a TH construction with the smallest hard $\mathbb{Z}_2$-breaking necessary to solve the $\Delta N_\mathrm{eff}$ problem. The results will depend only very weakly on details of the twin spectrum. 
In Section \ref{sec:AR} we study the asymmetrically reheated scenario in detail, reviewing and extending calculations in the literature for the $\nu$MTH \cite{Chacko:2016hvu} and $X$MTH \cite{Craig:2016lyx} and demonstrating the additional effect of DM dilution on the THPDM scenario.

\section{Scalar Higgs Portal Dark Matter}
\label{sec:SSHPDM}

The singlet-scalar Higgs portal DM (HPDM) \cite{SILVEIRA1985136,McDonald:1993ex,Burgess:2000yq,Casas:2017jjg} scenario is one of the simplest models of dark matter.\footnote{Note that we do not study fermionic HPDM that couples to the Higgs via a non-renormalizable operator $\frac{1}{\Lambda} |H|^2 \bar \psi \psi$ (see e.g.~\cite{LopezHonorez:2012kv, deSimone:2014pda, Fedderke:2014wda}), but we expect the THPDM mechanism to work similarly for this case as well.} The only addition to the SM is a minimal coupling of a stable scalar particle $S$ to the Higgs through the potential 
\beq
V = \frac{1}{2} m_S^2 S^2 + \frac{1}{4!} \lambda_S S^4 + \frac{1}{2} \lambda_{HS} H^\dagger H S^2.
\eeq
Electroweak symmetry breaking then induces a 3-point interaction term between the Higgs and the scalar, $\left(\frac{1}{2} \lambda_{HS} v\right) h S^2$, which enables it to freeze out in the early universe via Higgs-mediated annihilation into SM particles. This gives rise to a relic density today of 
\beq
\label{eqn:omega_gen}
\Omega_0 = \frac{1}{3 M_{Pl}^2 H_0^2} m Y_0 s_0
\eeq
where $s_0$ is the current entropy density, $H_0$ is the current value of the Hubble parameter, and $Y_0$ is the present number of dark matter particles per comoving volume, defined $Y_0 \equiv \frac{n_0}{s_0}$. When comoving entropy is conserved, $Y$ remains constant as the universe evolves. If comoving entropy is conserved at all times between freeze out and today, then $Y_0 = Y_f$ (where $f$ stands for freeze-out) and 
\beq
\label{eqn:omega_SHP}
\Omega_0 = \frac{1}{3 M_{Pl}^2 H_0^2} m Y_f s_0.
\eeq
In the freeze-out paradigm the relic density of dark matter is directly related to its annihilation cross section, $\Omega_0 \sim Y_f \sim 1/\braket{\sigma v}$. The thermally averaged cross section is  \cite{Gondolo:1990dk}
\beq
\braket {\sigma v} = \frac{1}{8 m_S^4 T K_2^2\left(m_S/T\right) } \int_{4 m_S^2}^\infty \mathrm{d}s~ \sigma(s) \left(s - 4 m_S^2 \right) \sqrt{s} K_1 \left(\sqrt{s}/T \right) 
\eeq
where $s$ is the squared center-of-mass energy, $\sigma(s)$ is the center-of-mass annihilation cross section, $T$ is the temperature of the thermal bath, and the $K_1, K_2$ are modified Bessel functions of the second kind.  Its exact relation to $Y_f$ can then be found either by solving the Boltzmann equation numerically or through an analytical approximation, as done e.g. in \cite{Kolb:206230}. In this work we follow the latter approach and find good agreement with known results.

\begin{figure}
\begin{center}
\includegraphics[height=8cm]{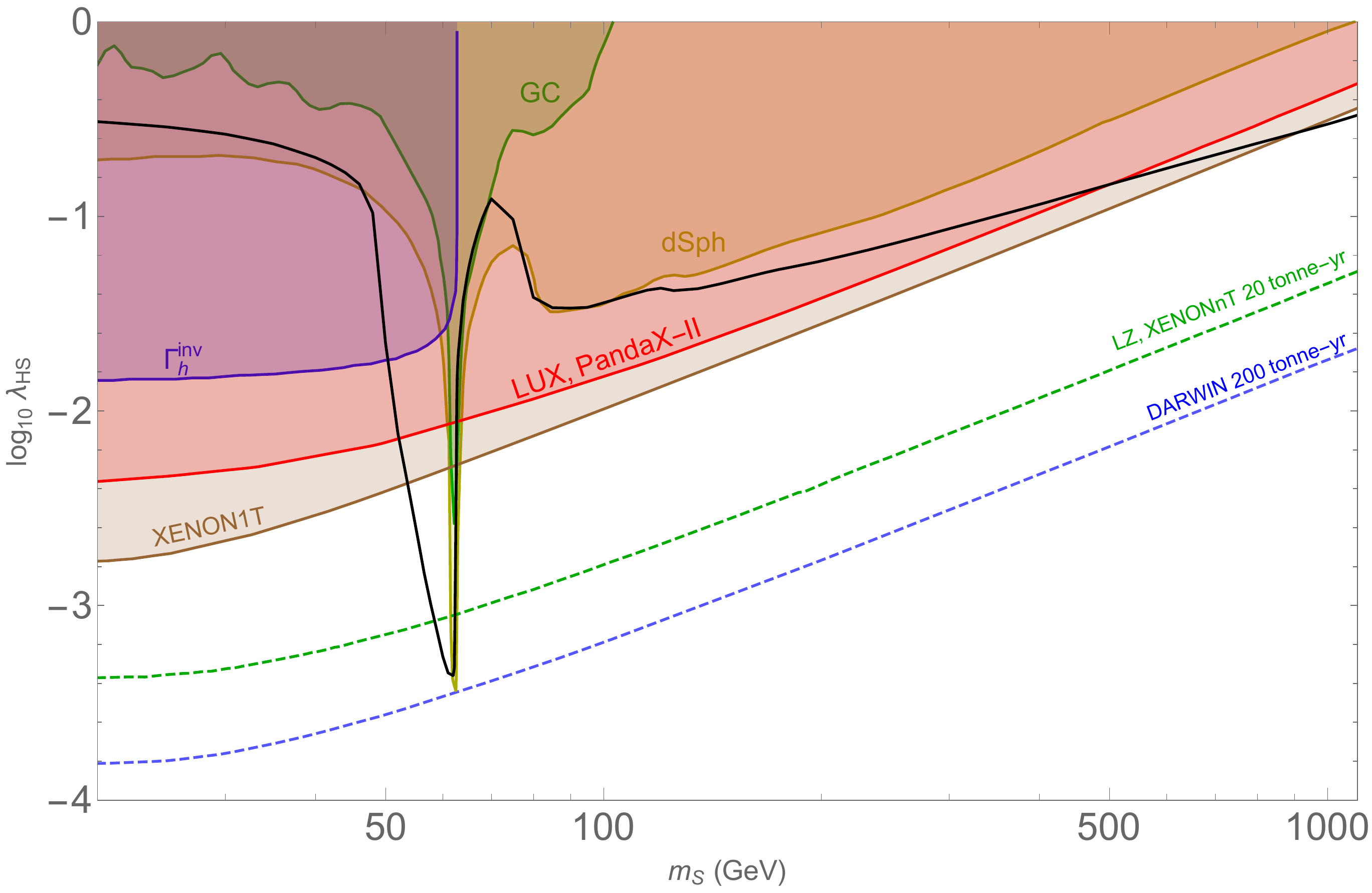}
\caption{Direct and indirect detection constraints for singlet scalar Higgs portal dark matter (HPDM). Current direct detection constraints include LUX, PandaX-II~\cite{Akerib:2016vxi,Tan:2016zwf} (red) and XENON1T \cite{Aprile:2018dbl} (brown); projected direct detection constraints from LUX-Zeplin (LZ)~\cite{Akerib:2015cja} and XENONnT \cite{Aprile:2015uzo} after 20 tonne-years are shown in dashed green, while 200 tonne-year projections from DARWIN are shown in dashed blue \cite{Aalbers:2016jon}. Indirect detection constraints from the Galactic Center (GC) and spheroidal dwarf galaxies (dSph) come from Fermi-LAT data \cite{Ahnen:2016qkx,Ackermann:2015lka} and are shown in dark green an orange respectively, while CMS constraints on the Higgs invisible decay width \cite{Khachatryan:2016whc} are shown in blue. Constraints from GC, dSph, LUX, LZ and $\Gamma_h^{\text{inv}}$ are derived in Ref. \cite{Casas:2017jjg}, while the XENON1T, XENONnT, and DARWIN bounds were added in the present work. The black curve indicates the expected coupling $\lambda_{HS}$ for varying DM mass consistent with the observed DM relic density $\Omega h^2 = 0.120 \pm .001$ \cite{Aghanim:2018eyx} (error bar on curve negligible), assuming that the scalar comprises all of DM. 
}
\label{fig:SHP}
\end{center}
\end{figure}

At tree level, the total thermally averaged cross section $\braket{\sigma v}$ is simply proportional to $\lambda_{HS}^2$. Since larger annihilation cross sections lead to a smaller relic abundance today, the measured relic abundance of $\Omega h^2 = 0.120 \pm .001$ \cite{Aghanim:2018eyx} leads to a prediction for the coupling $\lambda_{HS}$ as a function of mass $m_S$, shown in Figure \ref{fig:SHP} as the black curve. This prediction reproduces known results from the literature, e.g. \cite{Casas:2017jjg}. The curve was generated under the assumption that the scalar constitutes all of the dark matter; if the scalar dark matter fraction is $r < 1$, $\lambda_{HS}$ is larger  by a factor of $1/\sqrt{r}$. The cross section for scalar annihilation into visible matter was computed mainly following Ref. \cite{Cline:2013gha}, and includes all 2-2 tree level processes as well as QCD one-loop corrections for all quark final states.

The coupling $\lambda_{HS}$ can be probed by both direct and indirect detection experiments. Apart from the Higgs resonance, direct detection experiments constitute the dominant probe of the $\{\lambda_{HS}, m_S \}$ parameter space and there are many current and planned experiments aiming to constrain this space. The spin-independent direct detection cross section for scalar HPDM is \cite{Cline:2013gha}
\beq 
\label{eqn:sigma_SHP}
\sigma = \frac{1}{4 \pi} \frac{\lambda_{HS}^2 f_N^2}{m_h^4} \frac{\mu^2 m_N^2}{m_S^2}
\eeq
where $m_N$ is the nucleon mass, $\mu = \frac{m_N m_S}{m_N + m_S}$ is the reduced mass, and $f_N$ encodes the individual quark contributions in the Higgs-nucleon coupling, $\frac{f_N m_N}{v}$. It is defined in \cite{Cline:2013gha} as
 \beq 
 f_N = \sum_q \frac{m_q}{m_N} \bra N  \bar{q} q  \ket{N}.
 \eeq
As shown in Figure \ref{fig:SHP}, the parameter space of this model has been almost entirely excluded by Xenon1T, LUX, and PandaX-II \cite{Aprile:2015uzo,Akerib:2016vxi,Tan:2016zwf}, with DM masses above a TeV and the small window near the Higgs resonance expected to be covered by LZ, XENON and Darwin in the near future \cite{Akerib:2015cja,Aprile:2015uzo,Aalbers:2016jon}. 
We now show in the next section how the HPDM idea, when applied to Twin Higgs models, automatically leads to suppressed direct detection signatures due to the pNGB nature of the Higgs. This evades current bounds but could be discoverable in future experiments.

\section{Twin Higgs Portal Dark Matter}
\label{sec:THPDM}

We now construct a Twin Higgs version of the scalar Higgs Portal DM model by simply adding a real singlet scalar particle that couples to the Twin Higgs scalar sector in an $SU(4)$-invariant way. The full potential is
\beq
\label{eqn:VTHVS}
V = V_\mathrm{TH} + V_\mathrm{S}
\eeq
where $V_{\text{TH}}$ is the Twin Higgs potential given by Equation (\ref{eqn:V_TH}), and 
\beq
\label{eqn:VS}
V_S = \frac{1}{2} \mu_S^2 S^2 + \frac{1}{2} \lambda_{HS} S^2 \left( H_A^\dagger H_A + H_B^\dagger H_B \right).
\eeq
We expect that a UV complete Twin Higgs model could be extended to give rise to this Lagrangian, but leave explicit construction of such a model to future investigations.
To ensure $S$ is stable we require it to be odd under an unbroken $\mathbb{Z}_2$ symmetry $S \to -S$. In UV completions this could arise as a subgroup of a larger symmetry (like $U(1)$ if $S$ is complex), but this would not qualitatively change our analysis. 
We will consider this scalar potential for both an MTH-type model (where the mirror sector is a $\mathbb{Z}_2$ copy of the SM, except for the soft $\mathbb{Z}_2$ breaking term in $V_\mathrm{TH}$ necessary to ensure $v_B > v_A$) and an FTH-type model where the twin photon and first two twin matter generations have been removed. 
The physical mass of the singlet is 
\begin{equation}
\label{eq.mSsq}
m_S^2 = \mu_S^2 + \frac{1}{2} \lambda_{HS} f^2 
\end{equation}
and we require this to be positive to ensure $S$ does not acquire a vev and become unstable. 

\subsection{Pseudo-Goldstone Suppression of Direct Detection}

The pseudo-Goldstone nature of the 125 GeV Higgs boson radically changes the phenomenology of Higgs portal dark matter in our scenario. This is most apparent in the non-linear sigma model picture.
Since $S$ couples in an $SU(4)$-invariant way, any non-derivative couplings to the light Higgs can only arise due to small explicit symmetry-breaking interactions.
For $m_S \ll m_{\hat h}$, both $SS$ annihilation in the early universe and scattering off nuclei via Higgs exchange in direct detection are dictated entirely by these small interactions, and the ratio between annihilation and direct detection cross sections is  identical to HPDM. Therefore, the predictions of the direct detection signal for a given DM relic abundance are the same in THPDM as in HPDM for $m_S \ll m_{\hat h}$. 
On the other hand, for $m_S \gtrsim \frac{1}{2} m_{\hat h}$, annihilation of the singlets can now excite the radial mode, and the NLSM picture involving only exchange of  the pNGB via small explicit symmetry-breaking interactions breaks down. $SS$ annihilation is now dictated by the unsuppressed $SU(4)$-symmetric coupling $\lambda_{HS}$, while direct detection is still suppressed by the pseudo-Goldstone nature of the Higgs, since  nuclear scattering is always a low-energy process. Relative to the scattering cross section, the annihilation cross section is now drastically enhanced. Therefore, to reproduce the correct relic abundance, $\lambda_{HS}$ must be much smaller for  $m_S \gtrsim  \frac{1}{2} m_{\hat h}$  than for $m_S \ll  m_{\hat h}$, which reduces direct detection signatures. For $m_S > m_h$, THPDM can evade current bounds on direct detection while being potentially detectable in future experiments. 

We now demonstrate this mechanism analytically within the linear sigma model picture of the Twin Higgs scalar sector outlined in Section~\ref{sec:MTH_model}, and then present numerical predictions for direct detection of THPDM that confirm these arguments.

After electroweak symmetry breaking it is easy to show that the potential in Equation~(\ref{eqn:VS}) takes the form
\beq
\label{eqn:VS2}
V_S = \frac{1}{2} \lambda_{HS} S^2 \left[ h \left(v_A \cos \theta - v_B \sin \theta \right) + \hat{h} \left(v_A \sin \theta + v_B \cos \theta \right) + \frac{1}{2} h^2 + \frac{1}{2} \hat{h}^2 + \frac{1}{2} f^2 \right]
\eeq
where we use the results of Section \ref{sec:MTH_model} and note that $f^2 = v_A^2 + v_B^2$. Defining
\begin{align}
\lambda_{SSh} \equiv~ &\lambda_{HS} \times \left( v_A \cos \theta - v_B \sin \theta \right) \\
\lambda_{SS\hat{h}} \equiv~ &\lambda_{HS} \times \left( v_A \sin \theta + v_B \cos \theta \right) 
\end{align}
we can show via Equation (\ref{eqn:angle_theory}) that to first order in the $SU(4)$-violating coupling $\kappa$,
\begin{eqnarray}
\lambda_{SSh} &=& \lambda_{HS} \left( \frac{\kappa}{\lambda} \right) 
 v \left(1 - 2 v^2/f^2 \right) \left(1 - v^2/f^2 \right)^{1/2} + \mathcal{O}\left(\frac{\kappa^2}{\lambda^2} \right)\\
\lambda_{SS\hat{h}}  &=& \lambda_{HS}  f + \mathcal{O}\left(\frac{\kappa^2}{\lambda^2} \right)
\end{eqnarray}
where we recall $v \equiv v_A$ in the LSM. Note that the $\lambda_{SS\hat h}$ coupling of the DM scalar to the heavy Higgs $\hat h$ is unsuppressed, while the same coupling to the light Higgs, $\lambda_{SSh}$, is only generated if there is both explicit $SU(4)$ breaking (i.e. $\kappa \neq 0$) and explicit $\mathbb{Z}_2$ breaking (i.e. $v \neq f/\sqrt{2}$). In particular, $\lambda_{SSh}$ is suppressed by the small ratio $\kappa/\lambda$, and is simply $\lambda_{HS} v(\kappa /\lambda)$ to first order in $v/f$. 

\begin{figure}
\begin{center}
\includegraphics[height=6cm]{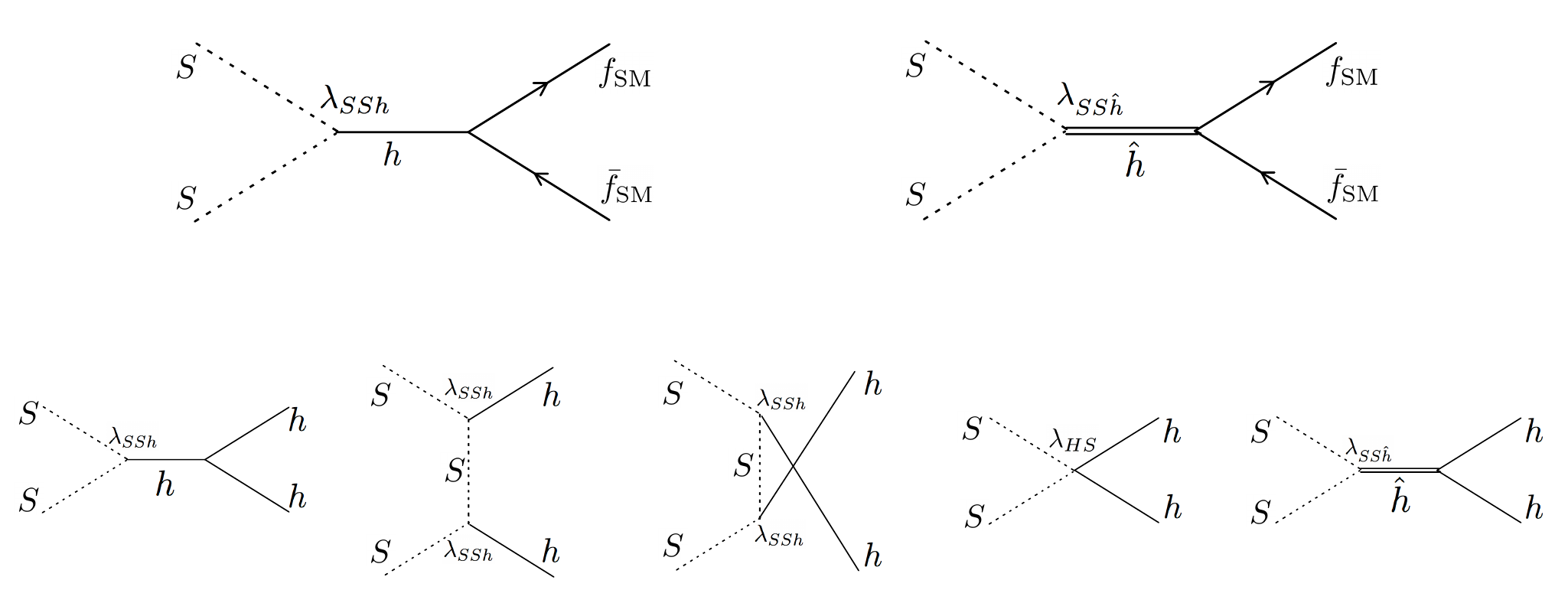}
\caption{\emph{Top}: contributions to $\mathcal{A}(SS \to \bar f_\mathrm{SM} f_\mathrm{SM})$. The first diagram dominates in the $v^2/f^2\ll 1$ limit but vanishes in the $SU(4)$ and $\mathbb{Z}_2$  symmetric limit. 
\emph{Bottom}: contributions to $\mathcal{A}(SS \to hh)$. In the $SU(4)$ and $\mathbb{Z}_2$ limit the first three diagrams are zero, while the last two cancel in the $SU(4)$ and $\mathbb{Z}_2$ limit with $s \to 0$.
}\label{fig:SShh}
\end{center}
\end{figure}

To elucidate the THPDM mechanism, let us first demonstrate how the HPDM phenomenology is recovered in the $m_S \ll m_{\hat h}$ limit. 
In that case, SS annihilation to the SM dominates, and the relevant diagrams are shown in Figure~\ref{fig:SShh}. 
The top two diagrams contribute to annihilation into SM fermions, and are (up to crossing symmetry) the same as the $h$- and $\hat h$-mediated diagrams that give the direct detection cross section. 
The $h$-mediated diagram is suppressed only by the small $\lambda_{SSh}$ coupling, while the $\hat h$-mediated diagram is suppressed by the higher mass of the heavy Higgs, as well as its smaller coupling to SM fermions. In the low-momentum limit, the ratio of the second to the first diagram is
\begin{equation}
\frac{
{\lambda_{SS\hat h} \sin \theta }/{m_{\hat h}^2}
}{
{\lambda_{SS h} \cos \theta }/{m_h^2}
}
\approx 
\lambda^2  \frac{v^2}{f^2} \ll 1
\end{equation}
if $v^2/f^2 \ll 1$, which is a reasonable approximation for our parametric arguments. We can therefore assume that both annihilation into fermions and nuclear scattering in direct detection is dominated by $h$-exchange. 

The bottom five diagrams in Figure~\ref{fig:SShh} contribute to annihilation into two Higgs bosons $SS \to hh$, which is relevant if $m_S > m_h$. The first three diagrams have exact equivalents in the HPDM model. The last two are unique to the Twin Higgs setup, but together are equivalent to the quartic $SShh$ diagram in HPDM. 
The appearance of the fifth diagram in the THPDM calculation can be seen as an artifact of using the Linear Sigma Model picture,\footnote{We thank Jack Setford for bringing this to our attention.} since in the NLSM formulation the $SS hh$ and $\hat h hh$ couplings are zero in the symmetric limit for a Goldstone $h$. 
In the LSM this requirement manifests instead as a cancellation between the last two diagrams. Explicitly, their combined amplitude is
\beq
\mathcal{A} = - i \lambda_{HS} \left[ 1 + \frac{2 f}{s - m_{\hat{h}}^2} \left( v_A \left[ \lambda \sin^3 \theta + 3 \kappa \sin \theta \cos^2 \theta \right] + v_B \left[ \lambda \cos^3 \theta + 3 \kappa \sin^2 \theta \cos \theta \right] \right) \right],
\eeq 
which, to first order in $\kappa/\lambda$ and $s \to 0$, reduces to
\beq
\label{eqn:goldstonecancel}
\mathcal{A} \approx - i \lambda_{HS} \frac{\kappa}{\lambda} \left( 1 - \frac{8 v^2}{f^2} + \frac{8 v^4}{f^4} \right)
\eeq
demonstrating that the 4-point interaction is generated by explicitly breaking the $SU(4)$.

The $s \to 0$ limit is relevant when $m_S \ll m_{\hat h}$. In this regime we can plainly see the connection between THPDM and HPDM: up to $\mathcal{O}\left(v^2/f^2 \right)$ corrections, the $SSh$ and $SShh$ couplings in THPDM are simply a factor of $\kappa/\lambda$ smaller than the corresponding HPDM couplings, since the 125 GeV Higgs is a pNGB of the approximate $SU(4)$. However, the two couplings behave differently in the $\mathbb{Z}_2$ limit, since in this limit $h = \frac{1}{\sqrt{2}} \left(h_A - h_B \right)$ is odd, meaning $SSh$ vanishes but $SShh$ does not. 

It is now easy to understand what happens when $m_S$ approaches or exceeds $\frac{1}{2} m_{\hat h}$. The Goldstone suppression demonstrated in Equation~(\ref{eqn:goldstonecancel}) breaks down, and annihilation to all kinematically accessible final state particles proceeds dominantly through the radial mode $\hat{h}$ via the unsuppressed coupling $\lambda_{HS}$.\footnote{Annihilation is of course further enhanced at high masses since $S$ can now annihilate to many more degrees of freedom in both the visible and twin sectors, but this effect is less important than the breakdown of the pNGB-suppression.} Direct detection, by contrast, is still only determined by the suppressed $\lambda_{SSh}$ coupling, and so to match the observed relic density $\lambda_{HS}$ must be significantly smaller, leading to reduced direct detection signatures. 

\subsection{Direct Detection Predictions for THPDM}

We now derive numerical predictions for direct detection of THPDM and compare to HPDM.
$SS$ annihilation is computed as outlined in Section~\ref{sec:SSHPDM}, as a function of $\lambda_{HS}, m_S, f/v, m_{\hat h}$, which yields a prediction for the coupling $\lambda_{HS}$ required to obtain the measured DM relic density. 
Making use of the fact that momentum exchange is negligible compared to the $h, \hat h$ masses, the direct detection cross section in THPDM is
\beq
\label{eqn:sigma_TH}
\sigma = \frac{1}{4 \pi} \frac{f_N^2}{v^2} \left[ \frac{\lambda_{SSh} \cos \theta}{m_h^2} + \frac{ \lambda_{SS\hat{h}} \sin \theta}{m_{\hat{h}}^2} \right]^2 \frac{\mu^2 m_N^2}{m_S^2}.
\eeq
In the language of effective field theory, both theories contain the same low energy operator, $S^2 \bar N N$, whose coefficient is constrained by experiment regardless of what UV physics gives rise to that operator. 
Comparing Equations (\ref{eqn:sigma_TH}) and (\ref{eqn:sigma_SHP}), we therefore see that THPDM predictions for direct detection can be shown in the same coupling-mass plane as for the HPDM if we  parameterize the $S^2 \bar N N$ operator via an an effective coupling to nucleons given by
\beq
\lambda_{\text{eff}}^2 \equiv \frac{m_h^4}{v^2} \left[ \frac{\lambda_{SSh} \cos \theta}{m_h^2} + \frac{ \lambda_{SS\hat{h}} \sin \theta}{m_{\hat{h}}^2} \right]^2 .
\eeq
For a given $m_S, f/v, m_{\hat h}$, this effective coupling is predicted by the observed DM relic density. 
This coupling reduces to $\lambda_{\text{HS}} \kappa/\lambda \approx \lambda_{SSh}/v$ in the $\theta \to 0$, or $f \gg v$, limit, consistent with the substitution required to recover the HPDM phenomenology for $m_S \ll m_{\hat h}$ explained above.

\begin{figure}
\begin{center}
\begin{tabular}{cc}
\includegraphics[height=5.1cm]{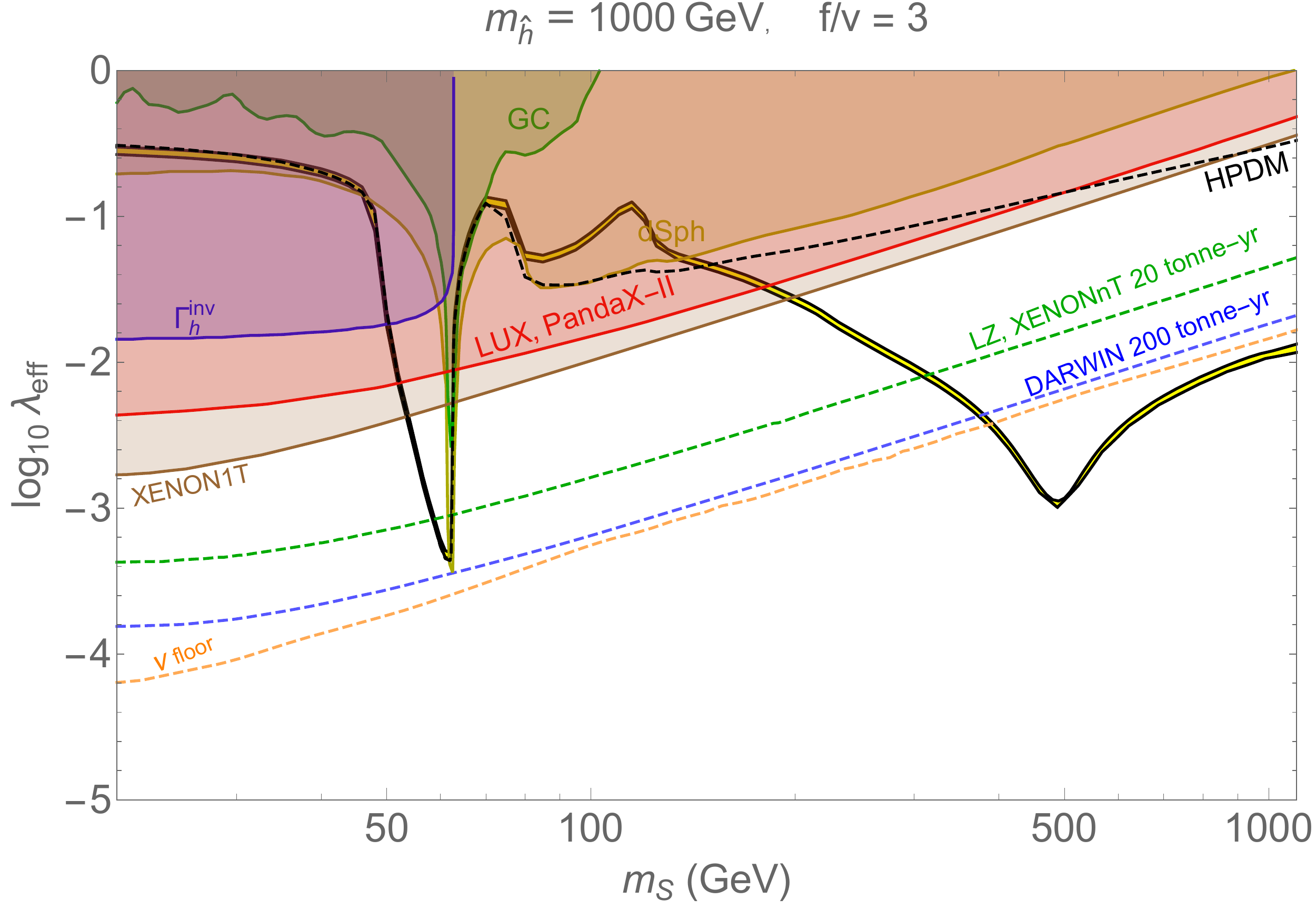} & \includegraphics[height=5.1cm]{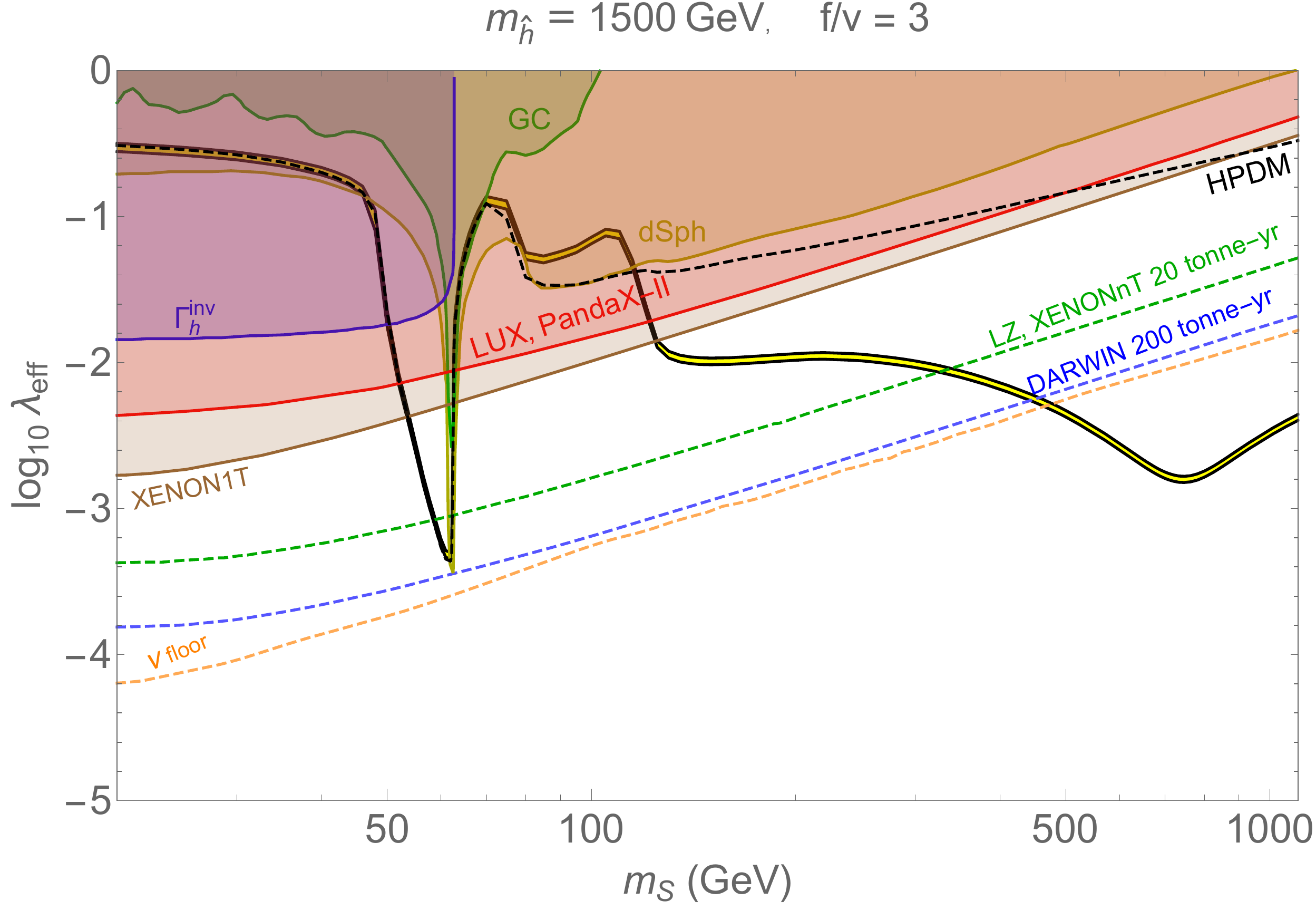} \\
\includegraphics[height=5.1cm]{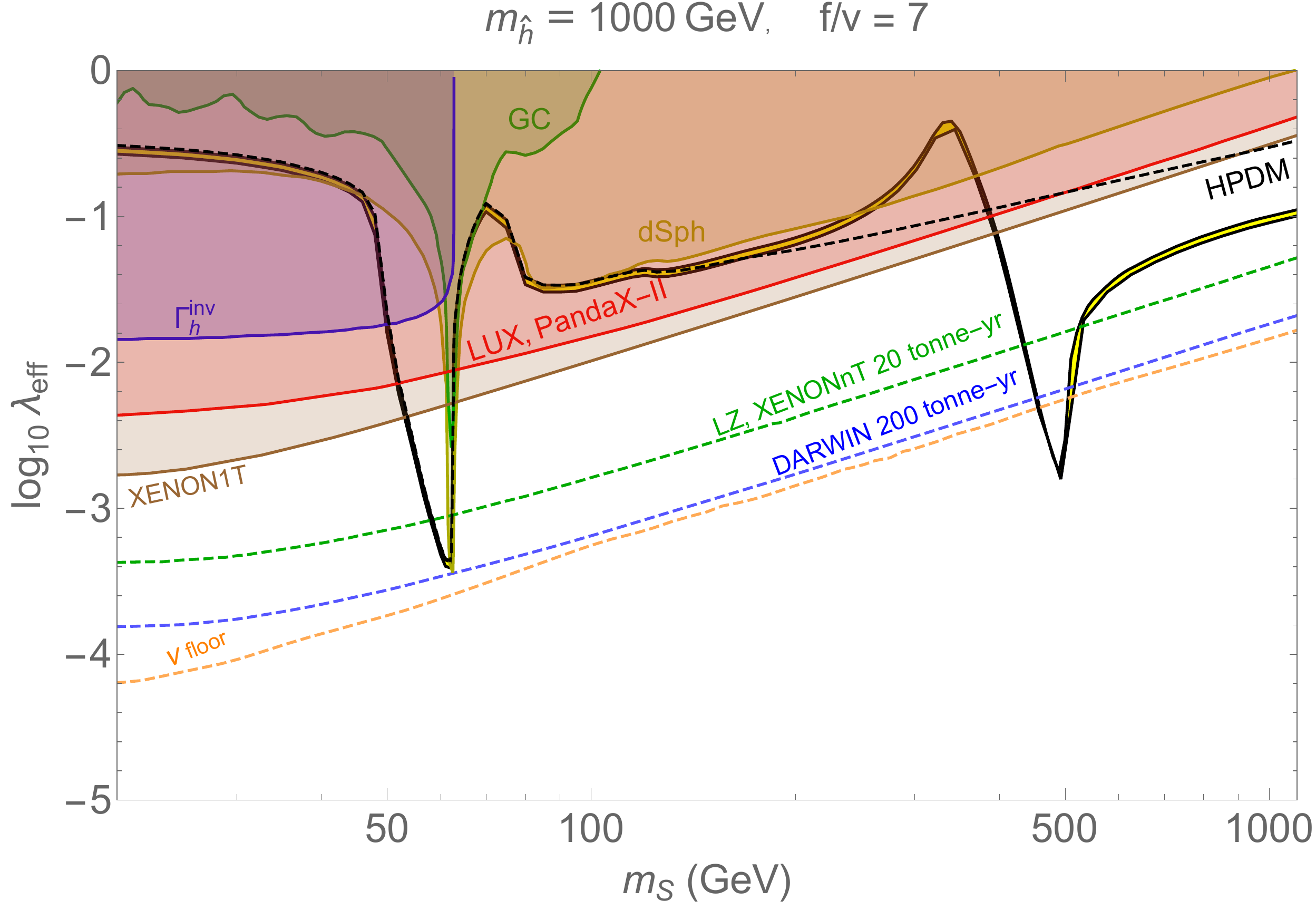} & \includegraphics[height=5.1cm]{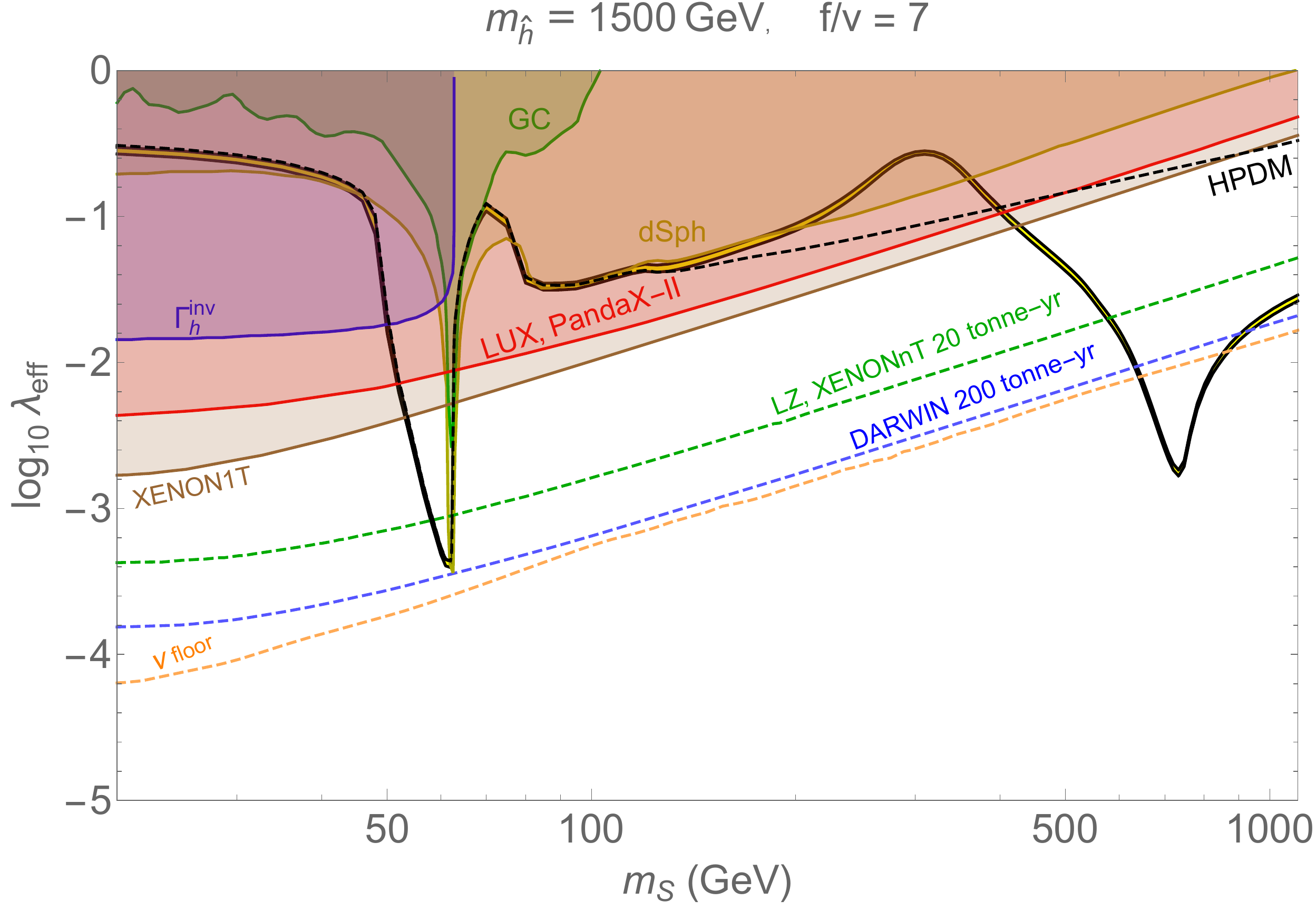} \\
\end{tabular}
\caption{Parameter space for the Twin Higgs Portal Dark Matter (THPDM). The yellow band represents the effective coupling between the dark matter $S$ and a nucleon that is consistent with the observed relic abundance of dark matter; the upper edge of the yellow band is for the reduced particle content of the Fraternal Twin Higgs (FTH) model, while the lower edge was computed for the full Mirror Twin Higgs (MTH). Experimental constraints are the same as those in Figure \ref{fig:SHP}. For comparison, the singlet-scalar Higgs Portal DM (HPDM) curve is shown in dashed black.}
\label{fig:THP_nodil}
\end{center}
\end{figure}

Our results are shown in Figure \ref{fig:THP_nodil}, for both MTH- and FTH-type mirror spectra. The breakdown of the NLSM picture in $SS$ annihilation for $m_{S}$ approaching or exceeding $\frac{1}{2} m_{\hat h}$ is clearly visible, resulting in a decrease in the effective predicted DM-nucleon coupling by one to two orders of magnitude compared to HPDM. 
This suppression is most pronounced near the $m_S \approx \frac{1}{2} m_{\hat h}$ resonance. 
It is interesting to note that THPDM becomes easier to discover in direct detection for larger $f/v$, providing an important complementarity to collider constraints on $f/v$ from Higgs coupling measurements and invisible or exotic decay searches, which lose sensitivity for larger $f/v$. 
The difference between MTH- and FTH-type mirror sectors is minimal, since the FTH contains the states most important for solving the hierarchy problem and hence having the largest coupling to the scalar sector. This makes the direct detection predictions of THPDM robust with respect to the details of the mirror spectrum. Future generations of experiments like LZ, XENONnT and DARWIN, together with collider searches, will be able to probe most of the THPDM parameter space in the coming years.

The calculations in this section assumed standard cosmology, as would be the case for Twin Higgs models with sufficient hard $\mathbb{Z}_2$ breaking to eliminate the problem of large $\Delta N_\mathrm{eff}$ contributions from light twin degrees of freedom. In Section \ref{sec:AR} we will examine Twin Higgs models that solve these cosmological issues with asymmetric reheating. We will see that this dilutes DM abundance and suppresses direct detection further -- additionally, in the event of a positive signal the magnitude of dilution could actually be measured via the reduced direct detection cross section, in combination with cosmological and collider measurements. This would serve as a probe of the non-standard cosmological history in Twin Higgs models.

\subsection{Natural Mass of Twin Higgs Portal Scalar Dark Matter}

Since the DM scalar $S$ couples to the Twin Higgs scalar sector in an $SU(4)$-invariant way, it does not spoil the accidental symmetry protection of the light 125 GeV Higgs mass. This is also evident from the NLSM picture, where the pNGB only couples to $SS$ via suppressed explicit symmetry-breaking interactions.\footnote{In the LSM picture, this is again manifest via a cancellation between quartic and $\hat h$-mediated contributions to 1-loop corrections of the light Higgs mass.}
Direct detection relies on the same interactions, and Figure~\ref{fig:THP_nodil} shows that the effective size of this explicit symmetry-breaking coupling required for $S$ to have the observed DM relic density is small, typically $\lesssim 10^{-2}$ over most of the parameter space of interest. Therefore, introducing $S$ as a DM candidate does not spoil the Twin Higgs solution of the little hierarchy problem.

However, it is also evident that the scalar $S$ itself enjoys no symmetry protection of its mass within the low-energy effective Twin Higgs model 
defined by Equations~(\ref{eqn:VTHVS}) and (\ref{eqn:VS}). 
In fact, it has an unsuppressed quartic coupling $\lambda_{HS}$ to the heavy radial mode $\hat h$.
The hierarchy
$m_S \ll m_{\hat h}$, which requires sizeable $\lambda_{HS}$ to overcome the pNGB suppression active during annihilation and achieve the observed relic abundance, is therefore unnatural for two reasons: first, 1-loop corrections to $m_S$ from $\hat h$ loops would be large, and second, such a small but positive $m_S^2$ would require a tuning of the tree-level parameters in Equation~(\ref{eq.mSsq}). These problems disappear for $m_S \sim m_{\hat h}$, both because there is no mismatch of mass-scales and because the required $\lambda_{HS}$ is smaller.
Even within the realm of applicability of the low-energy effective Twin Higgs model, this suggests that the masses of $\hat h$ and $S$ should not be too different from each other.
From a UV perspective, it therefore seems that $S$  with $m_S \sim m_{\hat h}$ should be interpreted as being part of the approximately $SU(4)$ invariant ``UV structure'' of the Twin Higgs scalar sector. The full hierarchy problem for $S$ would then be solved by whatever UV completion makes the masses of $\hat h$ and $h$ natural above 5-10 TeV.

\emph{All of these considerations strongly motivate the DM candidate $S$ to have  the same mass scale as the radial mode $\hat h$.} This in turn pushes THPDM into the regime where the pNGB-suppression of direct detection compared to DM annihilation is most pronounced. 
Null results in direct detection experiments to date are therefore entirely natural in our model. By the same token, our predictions in Figure~\ref{fig:THP_nodil} show that future detectors will have a good chance of detecting THPDM in the most relevant regions of parameter space.

\section{Asymmetrically Reheated Twin Higgs Portal Dark Matter}
\label{sec:AR}

In the previous section, we showed how the Twin Higgs mechanism leads to significantly suppressed direct detection rates for THPDM assuming standard cosmology.
This conclusion applies equally well to both MTH-type and FTH-type models, the latter being those that satisfy $\Delta N_\mathrm{eff}$ constraints by modifying the twin spectrum to remove light degrees of freedom. 
An alternative solution to the cosmological problems of the MTH is asymmetric reheating following an early period of matter domination~\cite{Chacko:2016hvu,Craig:2016lyx}. Additional entropy is injected into the visible sector after it decouples from the twin sector, diluting the twin contribution to $\Delta N_\mathrm{eff}$ during CMB times.
This dilution also reduces the energy density of any DM candidate that froze out before asymmetric reheating by a \emph{dilution factor} $D$: $\Omega \to \Omega/D$. 
In models with asymmetric reheating, the DM density at freeze-out must therefore be \emph{larger} by a factor of $D$ compared to the expectation from standard cosmology.
If the same diagrams are responsible for both annihilation in the early universe and direct detection, this corresponds to a \emph{decrease} in the direct detection rate by the same dilution factor $D$. 
This reduction in direct detection due to asymmetric reheating is universal and does not depend on the particular DM model.
In this section we carefully examine the effect of asymmetric reheating on DM in general and THPDM in particular.

In order for our discussion to be self-contained, we first review basic MTH cosmology in Section~\ref{sec:MTH_cosmo} to establish notation and re-derive the well-known discrepancy between $\Delta N_\mathrm{eff}$ in the $\mathbb{Z}_2$-symmetric MTH and measurements from the CMB. Then we review the asymmetric reheating mechanism in Section~\ref{sec:AR_mech}, largely following and updating the discussion in~\cite{Chacko:2016hvu}. Readers familiar with Twin Higgs cosmology are invited to skip these sections. 

In Section~\ref{sec:AR_signatures} we define the DM dilution factor $D$ and discuss its relation to the $\Delta N_\mathrm{eff}$ prediction  of asymmetrically reheated models. We will find that direct detection could potentially make the dilution a separate observable from $\Delta N_\mathrm{eff}$, allowing for an additional probe of Twin Higgs cosmology.

In Section~\ref{sec:nuMTHXMTH} we compute the dilution factors that are generated by two representative Twin Higgs models of asymmetric reheating, the $\nu$MTH~\cite{Chacko:2016hvu}  in Section~\ref{sec:nuMTH} featuring late-time decay of right-handed neutrinos, and the $X$MTH~\cite{Craig:2016lyx} in Section~\ref{sec:XMTH} where the reheating particle is a long-lived scalar $X$. 
We also re-examine the viability of these models in light of the updated $\Delta N_\mathrm{eff} < 0.23$ (at 2$\sigma$) constraints~\cite{Aghanim:2018eyx}, and show how predictions for $\Delta N_\mathrm{eff}$ and the DM dilution factor $D$ are correlated. 
Both models lead to a positive signal in future CMB-S4 measurements~\cite{Abazajian:2016yjj} for much of their respective parameter spaces.

Section~\ref{sec:results} contains predictions for THPDM direct detection for the range of dilution factors $D \sim 100 - 1000$ motivated in realistic $\nu$MTH and $X$MTH models.

\subsection{Review of Standard Mirror Twin Higgs Cosmological History}
\label{sec:MTH_cosmo}

\begin{figure}
\begin{center}
\includegraphics[height=6cm]{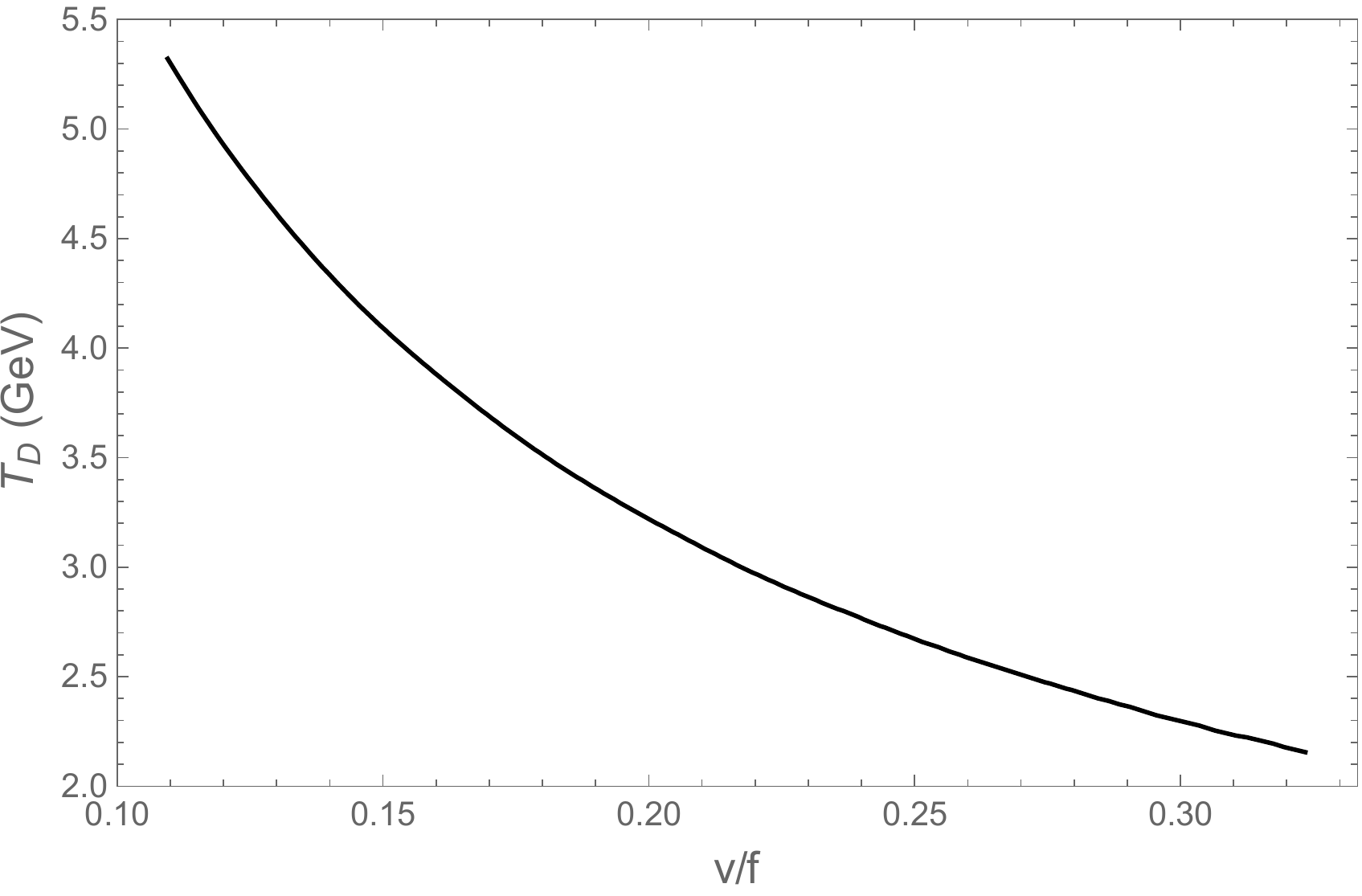}
\caption{The temperature of decoupling between the visible and mirror sectors as a function of the ratio $v/f$. This plot was generated from the solution of Equation (\ref{eqn:T_D}), where the Yukawa couplings $y_A^i$ and $y_B^j$ are those of the heaviest active fermions in the thermal bath of each sector. To reflect the continuous nature of particle freeze out, the Yukawas were linearly interpolated from the coupling of the lightest active particle at $T = m$ to the next lightest particle at $T = \frac{1}{6} m$, by which time approximately 80\% of the particles of mass $m$ have frozen out.}
\label{fig:T_D}
\end{center}
\end{figure}

It is helpful to review basic Mirror Twin Higgs cosmology, explicitly demonstrating how the unmodified model predicts $\Delta N_\mathrm{eff} \sim 6$. We follow the discussion in~\cite{Chacko:2016hvu}. 
The fact that the 125 GeV Higgs boson $h$ is an admixture of $h_A$ and $h_B$ enables the visible and mirror sectors to maintain thermal equilibrium in the early universe down to $\sim \mathcal{O}(\gev)$ temperatures through Higgs-mediated fermion scattering.
The thermally averaged interaction rate for these Higgs portal processes is of order
\beq
\braket {\sigma v}  \approx \left(y_A^i y_B^j \right)^2 \left(\frac{v}{f} \right)^2 \frac{T^2}{m_h^4}
\eeq
where $y_A^i$ and $y_B^j$ are the Yukawa couplings from the heaviest visible and mirror sector fermions still present in the thermal bath at temperature $T$.
This crude approximation is sufficient to estimate the decoupling temperature of the Higgs portal. 
Freeze out occurs when $n \Gamma = n \braket{ \sigma v} \lesssim H$, where $n$ is the fermion number density and $H$ is the Hubble parameter. The approximate temperature $T_D$ at which the two sectors thermally decouple can therefore be found by solving
\beq
\label{eqn:T_D}
\frac{3}{4} \left( \frac{\zeta(3)}{\pi^2}\right) g T_D^3 
\left(y_A^i y_B^j \right)^2 \left(\frac{v}{f} \right)^2 \frac{T_D^2}{m_h^4} \simeq \sqrt{\frac{\pi^2}{90}} \frac{T_D^2}{M_{\text{Pl}}} g_{*S}^{1/2}
\ ,
\eeq
where $g_{*S}$ is the total number of relativistic degrees of freedom in the bath at temperature $T_D$, and $g$ is the number of degrees of freedom of the target fermion. 
To estimate $T_D$, we note that roughly 80\% of the annihilations for a species leaving the bath happen in the interval $\frac{1}{6} m < T < m$~\cite{Choi:2018gho}. This means that to a good approximation we can construct continuous expressions for the effective Yukawa couplings $y_A^i$ and $y_B^j$ by linearly interpolating between the coupling of the lightest active particle at $T = m$ and the next lightest at $T = \frac{1}{6} m$, keeping the expression constant outside of these intermediary regions.
The decoupling temperature is shown as a function of $v/f$ in Figure \ref{fig:T_D}, coinciding with the $\sim$ few GeV expectation found in e.g.~\cite{Barbieri:2016zxn,Chacko:2016hvu,Craig:2016lyx}. Smaller $v/f$ leads to earlier decoupling.

Once the two sectors are decoupled, each thermal bath evolves separately and the total entropy of each sector is independently conserved. During this radiation-dominated epoch we therefore have
\beq
\label{eqn:entr_cons}
g_{*A} T_A^3 a^3 = \text{const} ~~~~~,~~~~~ g_{*B} T_B^3 a^3 = \text{const}
\eeq
where $g_{*A}$ is the number of relativistic degrees of freedom in the visible sector, $g_{*B}$ the mirror sector counterpart, and $a$ is the scale factor. Note here that the constants for A and B are generally not equal to each other, and we assume that all relativistic degrees of freedom in a given sector have the same temperature so that $g_{*} = g_{*S}$. Where relevant, the difference between $g_{*}$ and $g_{*S}$ is accounted for. 

Additionally, the energy density of a given sector during radiation domination is given by $\rho = \frac{\pi^2}{30} g_{*} T^4$.
Therefore, as long as comoving entropy is conserved, the energy density of each sector evolves as
\beq
\label{eqn:rho_evolve}
\rho_f = \left( \frac{g_{*i}}{g_{*f}} \right)^{1/3} \rho_i \left( \frac{a_i}{a_f} \right)^4
\eeq
where $i$ and $f$ denote some initial and final times. 
The decrease in energy density due the expansion of the universe is slowed by particles becoming non-relativistic and leaving the thermal bath.

\begin{figure}
\begin{center}
\includegraphics[height=8cm]{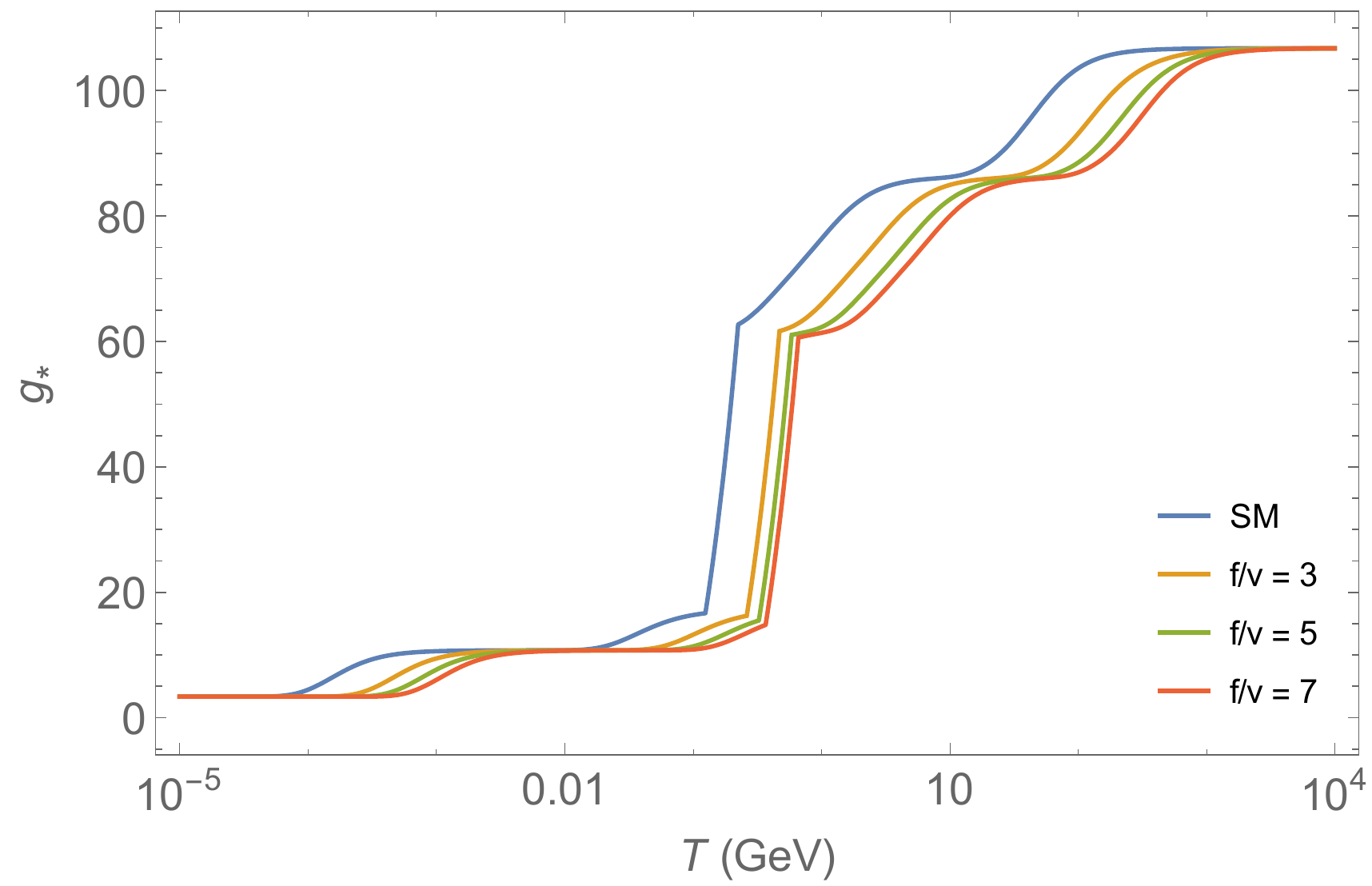}
\caption{Values of the effective relativistic degrees of freedom $g_{*}$ in the mirror sector for $f/v = 3, 5, 7$, shown in orange, green, and red respectively. The SM case is shown in blue for comparison (and corresponds to $f/v = 1$). In the region of the QCD phase transition we follow Ref. \cite{Craig:2016lyx} and linearly interpolate between known values before and after the transition. For the SM case we interpolate from a temperature of 125 MeV to 225 MeV, while for the MTH case the central value and width of the interpolated region is scaled by $(1 + \log (v/f))$. This reproduces and mildly extends the result in Ref. \cite{Craig:2016lyx}. The contribution to $g_*$ from neutrinos during their decoupling ($T \lesssim .01$ GeV) was calculated in this work following 
Ref. \cite{Husdal:2016haj}.}
\label{fig:gStar}
\end{center}
\end{figure}

We can now calculate the ratio of energy densities between the visible sector A and the mirror sector B at BBN. From Equation (\ref{eqn:rho_evolve}) we find that
\beq
\frac{\rho_B}{\rho_A} \eval_{\text{BBN}} = \left( \frac{g_{*A,\text{BBN}}}{g_{*B,\text{BBN}}} \right)^{1/3} \left( \frac{g_{*B,D}}{g_{*A,D}} \right)^{1/3}   \frac{\rho_B}{\rho_A} \eval_D
\eeq
where $g_{*A,\text{BBN}}$ and $g_{*A,D}$ denote the relativistic degrees of freedom in the visible sector at the time of BBN (in the visible sector) and mirror sector decoupling respectively, with analogous descriptions for  $g_{*B,\text{BBN}}$ and $g_{*B,D}$ in the mirror sector. Further, since the temperatures of the two sectors are equal at decoupling and the two sectors have roughly the same number of active degrees of freedom at BBN in the symmetric MTH model, the above reduces to

\beq
\frac{\rho_B}{\rho_A} \eval_{\text{BBN}} =  \left( \frac{g_{*B,D}}{g_{*A,D}} \right)^{4/3}.
\eeq
$\Delta N_{\text{eff}}$ can easily be written in terms of this ratio. Conventionally, $\Delta N_{\text{eff}}$ is defined as
\beq
\Delta N_{\text{eff}} \equiv \frac{\rho_B}{\rho_{\nu,i}} \approx 3 \frac{\rho_B}{\rho_{\nu}}
\eeq
where $\rho_B$ represents the dark radiation density and $\rho_{\nu,i}$ is the energy density of a single SM neutrino species. To the extent that we can ignore mass differences in the neutrino species\footnote{Since we are concerned with temperature scales much larger than the neutrino masses this is an excellent approximation. However see Ref. \cite{Craig:2016lyx} for a calculation of $\Delta N_{\text{eff}}$ with finite neutrino masses.} , the total neutrino energy density $\rho_{\nu}$ is simply $\rho_{\nu} \approx 3 \rho_{\nu,i}$. Since the neutrinos decouple from the bath prior to electron-positron annihilation, their temperature is reduced by a factor $\left(4/11 \right)^{1/3}$ relative to the photons during BBN. As a result the SM neutrinos comprise roughly 40.5\% of the total radiation in the SM bath. This implies
\begin{eqnarray}
\label{eqn:DNeff}
\Delta N_{\text{eff}} = 3~ \frac{\rho_B}{\rho_{\nu}} \eval_{BBN} &\approx &7.4~ \frac{\rho_B}{\rho_A} \eval_{\text{BBN}}   \ ,
\end{eqnarray}
which for the completely mirror symmetric Twin Higgs is well approximated by
\begin{eqnarray}
\Delta N_{\text{eff}}  &\approx& 7.4~ \left( \frac{g_{*B,D}}{g_{*A,D}} \right)^{4/3}.
\end{eqnarray}
Because of the $\mathbb{Z}_2$ breaking, the mirror sector particles will be a factor of $\frac{v_B}{v_A} \approx \frac{f}{v}$ heavier than their SM counterparts, which will in turn affect the size of $g_{*B}$ for a given temperature. This dependence is shown in Figure \ref{fig:gStar} for a variety of $f/v$. Regardless of the chosen $f/v$, the resulting number of effective neutrino species present during BBN is well outside the experimental bound of $\Delta N_{\text{eff}} \lesssim 0.23$~\cite{Aghanim:2018eyx}. Concretely, the MTH predicts
\begin{align}
\Delta N_{\text{eff}} &\approx 6.3 ~~~~~,~~~~~ f/v = 3 \\
\Delta N_{\text{eff}} &\approx5.8 ~~~~~,~~~~~ f/v = 7
\end{align}
indicating that without modification it is not a cosmologically viable model for any $f/v$.

\subsection{Review of Asymmetric Reheating Mechanism}
\label{sec:AR_mech}

Asymmetric reheating occurs when massive, relatively long-lived particles, labeled $Q$, freeze out relativistically due to their weak couplings. This leads to an early period of matter domination before they decay out-of-equilibrium, preferentially reheating the visible sector compared to the hidden sector. 
If the decay products quickly thermalize with the respective baths, the visible sector temperature and hence energy density is raised relative to the hidden sector, and $\Delta N_\mathrm{eff}$ is reduced -- see Equation (\ref{eqn:DNeff}).
Here we briefly review the model-independent description of the asymmetric reheating mechanism for Twin Higgs theories from~\cite{Chacko:2016hvu}.
We also take into account redshifting effects that are relevant when $Q$ freezes out at $T \gg M_Q$, discussed in~\cite{Craig:2016lyx}, and slightly update the discussion to include the latest bounds on $\Delta N_\mathrm{eff}$~\cite{Aghanim:2018eyx}.
The properties of the decaying particles are left as general as possible so that the results are easily applicable to the $\nu$MTH~\cite{Chacko:2016hvu} and $X$MTH~\cite{Craig:2016lyx} models discussed in Section~\ref{sec:nuMTHXMTH}.

In general, the mirror sector's contribution to $\Delta N_{\text{eff}}$ depends on the ratio $f/v$, the heavy Higgs mass $m_{\hat{h}}$, and the properties of the $Q$.
 $\Delta N_{\text{eff}}$ is most sensitive to the decay width $\Gamma_Q$, the mass $M_Q$, the temperature at which it decouples from the thermal bath $T_{Q,0}$, and the mirror sector branching ratio $\epsilon$, defined by
\beq
\label{eqn:epsilon}
\epsilon \equiv \frac{\Gamma_{Q \to B}}{\Gamma_Q} \ ,
\eeq
where asymmetric reheating requires $\epsilon \ll 1$. In practice, $\Delta N_\mathrm{eff}$ is relatively insensitive to $m_{\hat{h}}$, except insofar as it can determine whether $Q$ ever thermalizes with the SM/mirror sector bath early in the universe in scenarios where $Q$ is very weakly coupled.

There is a particular window of time during which the decay of the $Q$ must take place in order for the dilution mechanism to be effective. If the decay happens before the visible and twin sector (denoted A and B respectively) are decoupled from each other, then the entropy dumped into A can easily leak back into B through the Higgs portal and the dilution will not be sufficient to satisfy $\Delta N_{\text{eff}}$ bounds. It can be seen from Equation (\ref{eqn:DNeff}) that satisfying $\Delta N_{\text{eff}} < 0.23$ requires the energy density in the visible sector to be at least a factor of 30 larger than the mirror sector after reheating.
Since the mirror and visible sectors have similar numbers of degrees of freedom, it follows that no more than $\sim 1/15$ of the $Q$ can have decayed before this time.
This implies roughly that $\Gamma_Q \lesssim H/15$, where $H$ is evaluated at $T_D$.\footnote{Note that this is an updated prediction from \cite{Chacko:2016hvu} based on more recent $\Delta N_{\text{eff}}$ bounds.}. 
For a radiation dominated universe, we then find
\beq
\label{eq:GammaQMax}
\Gamma_Q \lesssim \sqrt{\frac{\pi^2}{90}} \frac{1}{15 M_{\text{Pl}}} g_{*,D}^{1/2} T_D^2
\eeq
which leads to an upper bound of $\Gamma_Q \lesssim 5 \times 10^{-19} ~(2 \times 10^{-18})$ GeV for $f/v = 3 ~(7)$, or a minimum lifetime of $\tau \gtrsim 1~(0.3)~\mu s$.

On the other hand, the asymmetric reheating cannot be allowed to disrupt BBN. The temperature of the visible sector after reheating, $T_{A, R}$, must therefore be larger than $\mathcal{O}(1-10 \mev)$~\cite{deSalas:2015glj}. A more precise constraint would require a dedicated analysis, but for the purposes of estimating the viable parameter space of asymmetrically reheated Twin Higgs models we will consider two lower bounds on $T_{A,R}$, 1 MeV and 10 MeV. 
This corresponds to
$\Gamma_Q \gtrsim 4 \times 10^{-25}$ GeV, or $\tau \lesssim 1s$ for 1 MeV and $\Gamma_Q \gtrsim 4 \times 10^{-23}$ GeV, $\tau \lesssim 10^{-2}s$ for 10 MeV -- see Equation~(\ref{eqn:rhoAR}) below.
Respecting the BBN bound also ensures that reheating does not occur after the visible sector neutrinos decouple from the SM bath at $T^0_{\text{SM}} \sim 1 \mev$. This avoids dilution of the SM neutrino energy density due to entropy injection into the active SM bath, which could reduce $\Delta N_\mathrm{eff}$ below the SM expectation.
The situation is potentially a bit more complicated for twin neutrinos, which decouple from twin electrons when the mirror sector roughly has temperature $T^0_{\text{Twin}} = \left(f/v\right)^{4/3} T^0_{\text{SM}}$. Recall that both sectors have very similar temperature before asymmetric reheating. 
Requiring $T_{A,R} > 10 \mev$ ensures that entropy injection takes place when twin neutrinos are still in thermal contact with the twin electron bath for most of our parameters of interest, and we can treat the surviving radiation components to be thermal, making estimation of $\Delta N_\mathrm{eff}$ straightforward. 
However, if the $T_{A,R}$ bound is relaxed to 1 MeV, then the twin neutrinos could receive a contribution to their number density from $Q$ decays (depending on the precise branching ratios within each sector) that never thermalizes but instead has a distribution dictated by $Q$ decays~\cite{Craig:2016lyx}. We leave a precise treatment of these non-thermal effects on $\Delta N_\mathrm{eff}$ for future investigation, and here simply present results for $T_{A, R} = 1 \mev$ as if the twin neutrino radiation component is thermal as a first crude estimate.

When the $Q$ become the dominant component of the universe's energy density, we find
\beq
\rho = 3 H^2 M_{\text{Pl}}^2 = M_Q n_Q
\eeq
where $H$ is the Hubble parameter, $M_{\text{Pl}}$ is the reduced Planck mass, and $n_Q$ is the number density of the $Q$. In the limit that all the decays happen at $H = \Gamma_Q$ we can then equate the energy density before and after to obtain
\beq
\label{eqn:rhoAR}
\rho_{A,R} = \frac{\pi^2}{30} g_{*A,R} T_{A,R}^4 = 3 (1- \epsilon) \Gamma_Q^2 M_{\text{Pl}}^2 \approx 3  \Gamma_Q^2 M_{\text{Pl}}^2 =  M_Q n_{Q,R}
\eeq
where $T_{A,R}$ is the temperature of the visible sector after reheating, $g_{*A,R}$ the associated degrees of freedom, $n_{Q,R}$ denotes the number density of the $Q$ after reheating, and we are implicitly assuming that the original energy density in SM radiation is dominated by the radiation produced from the $Q$ decay.
Equation~(\ref{eq:GammaQMax}) then requires the reheat temperature of the visible sector to satisfy $T_{A,R} \lsim 0.7 \ (1.3)$ GeV for $f/v = 3\ (7)$.

The mirror sector receives a much smaller portion of the total energy density from the $Q$ sector, so we have to take the pre-existing energy density into account. Within the instantaneous decay approximation, this gives
\beq
\label{eqn:rhoBR}
\rho_{B,R} = 3 \epsilon \Gamma_Q^2 M_{\text{Pl}}^2 + \left( \frac{g_{*B,D}}{g_{*B,R}} \right) ^{1/3}  \rho_{B,D} \left( \frac{a_D}{a_R} \right)^4.
\eeq
Dividing (\ref{eqn:rhoBR}) by (\ref{eqn:rhoAR}) then leads to
\beq
\label{eqn:ratio}
\frac{\rho_B}{\rho_A}\eval_{R} \equiv \frac{\epsilon + R_Q}{1-\epsilon} \approx \epsilon + R_Q
\eeq
where $R_Q$ is the ratio of the hidden sector energy density before reheating compared to the total entropy injected by $Q$ decays, defined
\beq 
\label{eqn:RQtemp}
R_Q = \frac{ \left( \frac{g_{*B,D}}{g_{*B,R}} \right) ^{1/3} \rho_{B,D} \left( \frac{a_D}{a_R} \right)^4}{M_Q n_{Q,R}}.
\eeq
Thus from Equations (\ref{eqn:ratio}) and (\ref{eqn:DNeff}) we find
\beq
\label{eqn:DNeffv3}
\Delta N_{\text{eff}} \approx 7.4~ \frac{\rho_B}{\rho_A} \eval_{\text{BBN}} = 7.4 \left( \frac{g_{*B,R}}{g_{*A,R}} \right)^{1/3}
(\epsilon + R_Q)
\eeq
where the $\left(g_{*B,R}/g_{*A,R} \right)^{1/3}$ factor accounts for the changing constituents of the thermal baths in both sectors between reheating and BBN.
Intuitively, $R_Q$ is a measure of how washed out the original energy density in the twin sector is after $Q$ decays. At the same time, $\epsilon$ is a measure of how much energy density is sent back into the twin sector through the same process. In order for $\Delta N_{\text{eff}}$ to be brought to within experimental limits, both of these quantities are required to be small. Effective asymmetric reheating requires that a large amount of entropy be injected into the visible sector, with very little leaking back into the twin sector.

To compute $\Delta N_\mathrm{eff}$ for a given Twin Higgs model of asymmetric reheating, our only remaining task is to compute $R_Q$. 
 At the time of decoupling between visible and mirror sectors, the energy density of the bath is split according to the relativistic degrees of freedom in either sector,
\beq
\label{eqn:rhoADBD}
\rho_{A,D} =  \frac{\pi^2 }{30} g_{*A,D} T_D^4~~~~~,~~~~~ \rho_{B,D} = \frac{\pi^2 }{30} g_{*B,D} T_D^4 \ .
\eeq
$Q$ is assumed to be non-relativistic and long decoupled when the hidden and visible sectors decouple from each other:
\beq
\label{eqn:n_DR}
n_{Q,R} \approx \left( \frac{a_D^3}{a_R^3} \right) n_{Q,D}.
\eeq
Thus from Equations (\ref{eqn:rhoAR},~\ref{eqn:RQtemp}-\ref{eqn:n_DR}) we find
\begin{equation}
\label{eqn:RQ}
R_Q =  \frac{\pi^2}{30} \left( \frac{g_{*B,D}}{g_{*B,R}} \right)^{1/3} g_{*B,D} \left(\frac{3 \Gamma_Q^2 M_{\text{Pl}}^2}{M_Q^4} \right)^{1/3} \left( \frac{T_D^4}{n_{Q,D}^{4/3}} \right) \ .
\end{equation}
If $Q$ freezes out at a temperature $T_{Q,0}$ not much higher than its mass, it becomes non-relativistic soon after and redshifting effects on its energy density can be neglected. In that case, we can write the number density as 
\begin{equation}
n_{Q,D} \approx \tilde{g}_Q \frac{\zeta(3)}{\pi^2} T_{Q,0}^3 \left( a_0^3/a_D^3 \right) \ ,
\end{equation}
where
\beq
\tilde{g} \equiv \sum_{\text{bosons}} g_i + \frac{3}{4} \sum_{\text{fermions}} g_i.
\eeq
$R_Q$ then takes the simplified form
\beq
\label{eqn:RQsimp}
R_Q = \frac{\pi^2}{90} \left( \frac{3 \pi^2}{\zeta(3)} \right)^{4/3} \left( \frac{ g_{*B,D}}{\tilde{g}_{Q}} \right)^{4/3} \left( \frac{g_{*0}}{g_{*D}} \right)^{4/3} \left( \frac{\Gamma_Q^2 M_{\text{Pl}}^2}{g_{*B,R} M_Q^4} \right)^{1/3}.
\eeq
On the other hand, if $Q$ is so weakly interacting that it freezes out at $T_{Q,0} \gg M_Q$, then it will redshift as decoupled radiation for a non-negligible period of time before becoming non-relativistic (which can be particularly important for the $X$MTH model).
In that case, the number density must be calculated using a redshifted Fermi-Dirac or Bose-Einstein distribution \cite{Jacques:2013xr}
\beq
n_{Q,D} = \frac{g}{2 \pi^2} \int_0^\infty \mathrm{d}p ~p^2 \left( \exp \left[ \frac{ \sqrt{ \left( \frac{T_{Q,0}}{T_{Q,D}} p \right)^2 + M_Q^2}}{T_{Q,0}} \right] \pm 1
\right)^{-1} \ ,
\eeq
where $\pm 1$ is for Fermi-Dirac or Bose-Einstein statistics respectively and $T_{Q,D} = \left( g_{*D}/g_{*0} \right)^{1/3} T_D$ is the temperature of the $Q$ at the time of visible-twin sector decoupling.

\subsection{Dark Matter Dilution}
\label{sec:AR_signatures}

Any DM abundance that freezes out before asymmetric reheating will be diluted similar to the mirror sector (assuming no DM is produced in $Q$ decays). 
The DM relic abundance  in asymmetrically reheated Twin Higgs theories is therefore computed analogously to standard freeze-out -- see Equation (\ref{eqn:omega_SHP}) -- but with two important differences. 
First, when the $Q$ decay they increase the total entropy of the universe by a factor $\Delta$,
\beq
s_0 a_0^3 = \Delta s_f a_f^3.
\eeq
This implies that
\beq
Y_0 = \frac{n_0}{s_0} = \Delta^{-1} \frac{n_f}{s_f} = \Delta^{-1} Y_f.
\eeq
Second, the entropy density today is the sum of the entropy densities in the two sectors, $s_{0,\text{MTH}}$ = $s_{0,A} + s_{0,B}$. Therefore the relic density is 
\beq
\Omega_0 = \frac{1}{3 M_{Pl}^2 H_0^2} m Y_f s_{0,A} \left[ \Delta^{-1} \left( 1 + \frac{s_{0,B}}{s_{0,A}} \right) \right]
\eeq
where we have factored out $s_{0,A}$ because this is the quantity we measure with CMB data. Comparing to Equation (\ref{eqn:omega_SHP}), we find that the observed relic abundance is given by 
\begin{equation}
\Omega_0 = \frac{ \Omega_{\Lambda \text{CDM}}}{D} \ ,
\end{equation}
where $\Omega_{\Lambda \text{CDM}}$ is the relic density formula in standard cosmology\footnote{Note that the value of $Y_f$ is model-dependent and will vary depending on both the thermally averaged annihilation cross section of the DM candidate, as well as the constituents of the thermal bath as it freezes out. In this definition we take $Y_f$ to be that of a Twin Higgs model in the absence of any asymmetric reheating. If one wishes to consider the dilution factor between an asymmetrically reheated Twin Higgs model and a more minimal model of dark matter (e.g. HPDM), then a factor $K = \frac{Y_{f,\text{CDM}}}{Y_{f,\text{MTH}}}$ should be included in the definition of $D$.}, and we have defined the \emph{dilution factor}
\beq
\label{eqn:D_def}
D \equiv \Delta \left( 1 + \frac{s_{0,B}}{s_{0,A}} \right)^{-1} \ ,
\eeq
which parameterizes the magnitude of dark matter dilution that arises from asymmetric reheating in Twin Higgs models.\footnote{This is a modified definition of the dilution factor compared to e.g. Ref. \cite{Hamdan:2017psw}, specific to the Twin Higgs since the entropy of the twin sector plays a significant role.} To better understand this quantity, we now write it in terms of relevant cosmological parameters. First, notice that 
\beq
\Delta = \frac{s_0 a_0^3}{s_f a_f^3} = \frac{s_R a_R^3}{s_D a_D^3} =\left( \frac{s_{A,R} + s_{B,R}}{s_D} \right) \left( \frac{a_R}{a_D} \right)^3
\eeq
since, by construction, all of the decays of the $Q$ happen after the mirror and visible sector decouple. We then find
\beq
\label{eqn:D1}
D =  \frac{s_{A,R}}{s_D} \left( \frac{a_R}{a_D} \right)^3
\eeq
since comoving entropy is conserved after reheating. This expression can be written in terms of energy densities as 
\beq
D = \frac{\rho_{A,R}^{3/4} ~ g_{*A,R}^{1/4} }{\rho_D^{3/4} g_{*D}^{1/4}} \left( \frac{a_R}{a_D} \right)^3,
\eeq
and using Equations (\ref{eqn:rhoAR}) and (\ref{eqn:RQtemp}), as well as the fact that $\rho_D = \left( \frac{g_{*D}}{g_{*B,D}} \right) \rho_{B,D}$, we find the relatively simple form
\beq
\label{eqn:D_simple}
D = \frac{1}{R_Q^{3/4}} \left( \frac{g_{*B,D}}{g_{*D}} \right) \left( \frac{g_{*A,R}}{g_{*B,R}} \right)^{1/4}.
\eeq
Ignoring $g_*$ factors, we see that $D$ depends only on $R_Q$ while $\Delta N_\mathrm{eff}$ depends on both $R_Q$ and $\epsilon$ (see Equation (\ref{eqn:DNeffv3})). 
This arises because the amount by which DM is diluted depends only on the total amount of entropy injected by $Q$-decay, not on the asymmetry of the reheating. $R_Q$ is defined as the twin sector energy before reheating compared to the total amount of energy injected, but since the former is straightforwardly related to the total energy density before reheating, the relative importance of the entropy injection is indeed specified by $R_Q$ and not $\epsilon$.
The dilution factor can be related to the properties of $Q$ using the simplified form of $R_Q$ in Equation~(\ref{eqn:RQsimp}) which is valid if $T_{Q,0}$ is not much larger than $M_Q$. Ignoring constants, this gives
\beq
D \sim M_Q \biggr(\frac{g_{*A,R}^{1/4}}{\Gamma_Q^{1/2} M_{Pl}^{1/2}} \biggr) \biggr( \frac{\tilde{g}_Q}{g_{*0}} \biggr)
\eeq
which shows that dilution increases if $Q$ is heavier, decays later, or has more degrees of freedom.

Dilution affects direct detection of a thermal freeze-out DM candidate $\chi$ in a very simple way. 
If 
$\Omega_0 \sim 1/\braket{\sigma v}$
is reduced by a factor of $D$ due to the reheating, and if the direct detection cross section $\sigma_{\chi N}$ is given by the same couplings that drive annihilation, $\sigma_{\chi N} \propto \langle \sigma v \rangle$, then both cross sections must decrease by a factor of $D$ compared to standard cosmology to overproduce DM before reheating and therefore produce the correct DM relic abundance today. 
This effect of reheating and decreasing direct detection signatures compared to the predictions of standard cosmology applies generally to any DM candidate -- see e.g. \cite{Evans:2019jcs,PhysRevD.92.103509,Arias:2019uol,Arias:2020qty,Bernal:2018ins}.
In THPDM, $\langle \sigma v \rangle \propto \sigma_{S N} \propto \lambda_{eff}^2$. Compared to the predictions of Section~\ref{sec:THPDM},  direct detection in Twin Higgs models with asymmetric reheating will therefore be additionally suppressed, with $D > 1$ reducing the expected effective coupling by
\beq
\lambda_{\text{eff}} \to \frac{\lambda_{\text{eff}}}{\sqrt{D}}.
\eeq
Further predictions require some information about the plausible size of the dilution factor $D$, and how $D$ is correlated with $\Delta N_\mathrm{eff}$. We study this for two explicit Twin Higgs models with asymmetric reheating in Section~\ref{sec:nuMTHXMTH} below, before showing predictions for direct detection in THPDM with asymmetric reheating in 
Section~\ref{sec:results}.

We point out that within a given DM framework, the dilution factor is observable in the event of a DM discovery by the lower direct detection rate compared to the standard cosmology expectation. $D$ therefore provides a cosmological probe of the properties of $Q$ that is complementary to $\Delta N_\mathrm{eff}$, allowing in principle independent probes of $\epsilon$ and $R_Q$. Together with other cosmological~\cite{Chacko:2018vss} and collider measurements~\cite{Burdman:2014zta, Chacko:2017xpd,Chacko:2019jgi}, this can allow for detailed examination of the twin protection and asymmetric reheating mechanisms. 

\subsection{Asymmetrically Reheated Twin Higgs Portal Dark Matter}
\label{sec:nuMTHXMTH}

In this section we examine two asymmetrically reheated MTH models, the $\nu$MTH~\cite{Chacko:2016hvu} and $X$MTH~\cite{Craig:2016lyx}. 
We review details of the models and their cosmological history, illustrate the viable parameter space in light of updated $\Delta N_\mathrm{eff}$ bounds, and show the range of correlated predictions for $\Delta N_\mathrm{eff}$ and DM dilution factor $D$, which allows us to make direct detection projections for THPDM with asymmetric reheating  in Section~\ref{sec:results}.

\subsubsection{$\nu$MTH}
\label{sec:nuMTH}

A simple way of incorporating the asymmetric reheating mechanism into the Twin Higgs framework is by adding right-handed neutrinos (RHN), proposed in Ref. \cite{Chacko:2016hvu}.
This is particularly elegant since RHNs are highly motivated to exist for other reasons, most importantly to explain non-zero neutrino masses \cite{Drewes:2013gca}. Ref.~\cite{Chacko:2016hvu} examines the case of 3 generations of RHNs in both the visible and twin sectors that mix with each other as well as the active neutrinos of their respective sectors, generating small active masses through the familiar type-I seesaw mechanism. In specific scenarios the width of each RHN generation is proportional to one active neutrino mass, and the known differences in the neutrino masses can then be translated into differences in decay widths of the RHNs.\footnote{For concreteness, here and in Ref.~\cite{Chacko:2016hvu} the normal ordering is assumed.}

Since the energy density of the RHNs continues to grow relative to the energy density of the thermal baths as the universe cools, later decays have a more significant effect on the subsequent asymmetric reheating and dilution. For this reason, many of the essential features of the more realistic 3 generation model can be captured with a 1 generation toy model, also covered in Ref.~\cite{Chacko:2016hvu}. In what follows we review the details of this simpler model and address its implications for asymmetric reheating and dilution, pointing out expected behaviour for the full 3 generation model where appropriate.

Consider the extension of the Twin Higgs by a single generation of RHNs that respect the $\mathbb{Z}_2$ symmetry,
\beq
\label{eqn:nuMTH_lagrange}
\mathcal{L} \supset - y \left( L_A H_A N_A + L_B H_B N_B \right) - \frac{1}{2} M_N \left(N_A^2 + N_B^2 \right) - M_{AB} N_A N_B + \text{h.c.}
\eeq
where for simplicity we invoke a hierarchy in the parameters $y \braket{H} \ll M_{AB} \ll M_N$ for both $H_A$ and $H_B$. Prior to electroweak symmetry breaking, the RHNs combine to form $\mathbb{Z}_2$-symmetric mass eigenstates $N_{\pm} = \frac{1}{\sqrt{2}} \left(N_A \pm N_B \right)$ with nearly degenerate masses $M_N \pm M_{AB}$. After electroweak symmetry breaking these states mix with the active neutrinos, which generates masses for the active neutrinos
\begin{align}
m_{\nu,A} &= \frac{y^2 v_A^2}{2 M_N} \left(1 + \mathcal{O}\left(\frac{M_{AB}}{M_N} \right) \right) \\
m_{\nu,B} &= \frac{y^2 v_B^2}{2 M_N} \left(1 + \mathcal{O}\left(\frac{M_{AB}}{M_N} \right) \right)
\end{align}
and allows the RHNs to decay to leptons through both neutral and charged currents. The decay width for the RHNs into the visible sector is proportional to  
\beq
\label{eqn:GammaN}
\Gamma_{N \to A} \propto \left( \frac{m_{\nu,A}}{M_N} \right) G_{F,A}^2 M_N^5
\eeq
where $G_{F,A}$ is the Fermi constant with visible sector vector bosons, and we neglect final state masses as well as $\mathcal{O}(m_{\nu}/M_N)$ and $\mathcal{O}(M_{AB}/M_N)$ corrections to the sterile neutrino masses. For actual computations we again follow Ref. \cite{Chacko:2016hvu}, which derives partial widths of the RHNs in Fermi theory, valid for  $M_N \lesssim M_W$. The RHN decays into the mirror sector are similar, 
\beq
\Gamma_{N \to B} \propto \left( \frac{m_{\nu,B}}{M_N} \right) G_{F,B}^2 M_N^5 \propto \left( \frac{v^2}{f^2} \right) \Gamma_{N \to A}
\eeq
but the net result is an overall suppression of $v^2/f^2$ due to the heavier neutrino and vector bosons in the mirror sector. This leads to asymmetric reheating of the visible sector
\beq
\epsilon \approx \frac{v^2}{f^2},
\eeq
but note that this ratio can be adjusted through, for example, explicit $\mathbb{Z}_2$ breaking in the Lagrangian of Equation (\ref{eqn:nuMTH_lagrange}), as discussed in Ref. \cite{Chacko:2016hvu}. In what follows we pay particular attention to the case of $\epsilon = v^2/f^2$, but ultimately treat it as a free parameter of the model.

Lastly, asymmetric reheating is effective only if the RHNs are able to dominate the cosmology before decaying. It is therefore crucial that they reach equilibrium abundance within the bath before freezing out — otherwise they will not carry away enough energy to reheat the visible sector at late times. In a 3 generation model this is simple to achieve. Neutrino oscillation measurements indicate that at least one neutrino has mass $m_{\nu} \gtrsim 0.05$ eV \cite{PhysRevD.98.030001}, corresponding to a Yukawa coupling of $\mathcal{O}(10^{-7})$, sufficient to guarantee that the corresponding RHN is in thermal contact with the SM bath in the early universe. The observation of large mixing angles in the neutrino sector further suggests that all the RHN Yukawas are of comparable size, meaning all three RHN generations are in thermal equilibrium with the visible sector prior to freeze-out.
In this work, we focus on the single-generation toy model to parametrically understand the correlation between $\Delta N_{eff}$ and dilution factor $D$, keeping in mind that a three-generation set up is likely more realistic. We therefore assume that regardless of the mass of our single RHN, it is thermalized at early times and freezes out with an equilibrium abundance at $T_0$, where
\beq
\label{eqn:freezeout}
T_0 \approx \left(\frac{M_N}{m_{\nu}} \right)^{1/3} \left(1~\text{MeV} \right) 
\eeq
since the interaction rate of the RHNs with the bath is a factor $\sim m_\nu/M_N$ weaker than the active neutrino interaction rate.

\begin{figure}
\begin{center}
\begin{tabular}{cc}
\includegraphics[width=7.6cm]{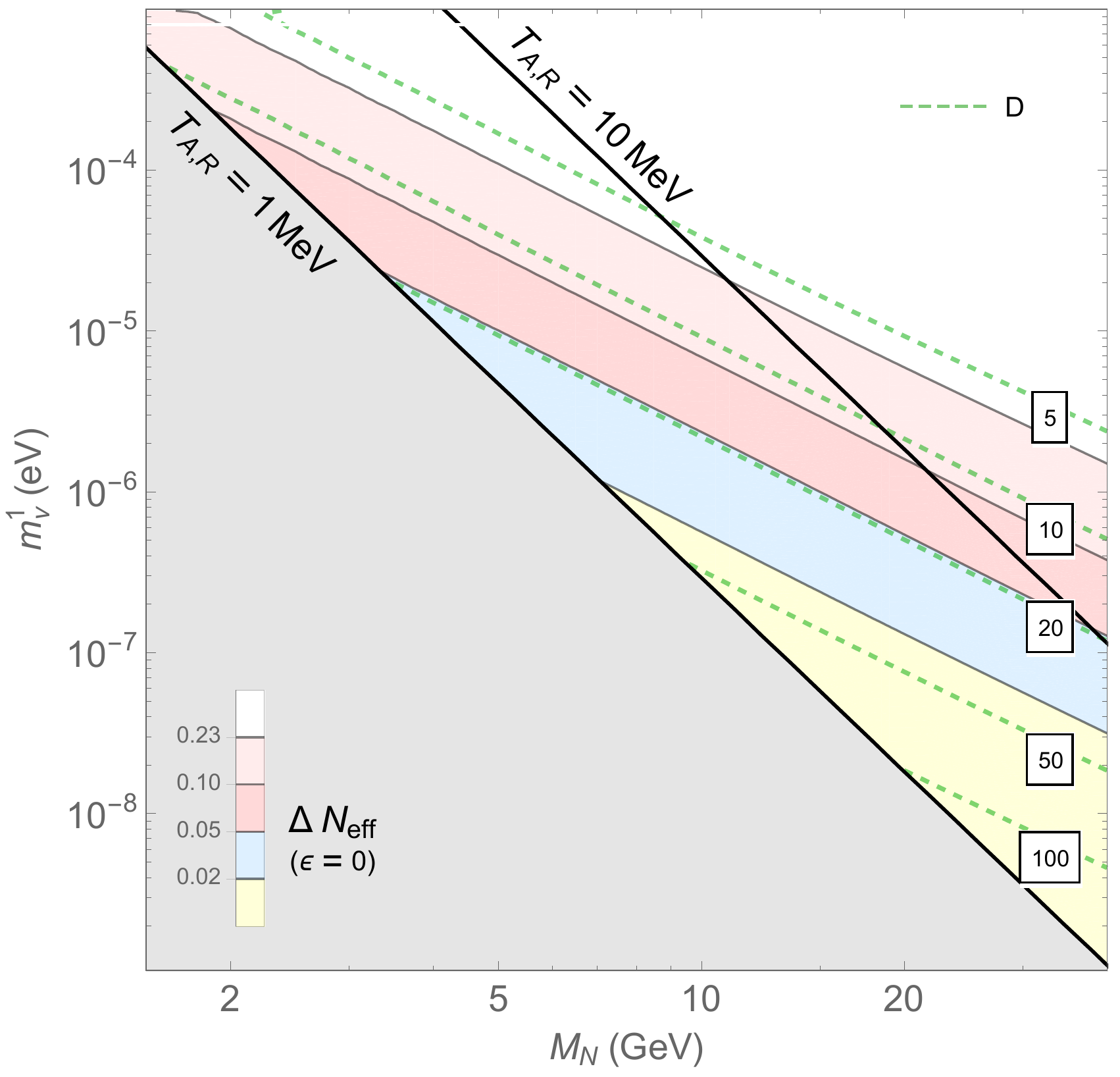} & \includegraphics[width=7.6cm]{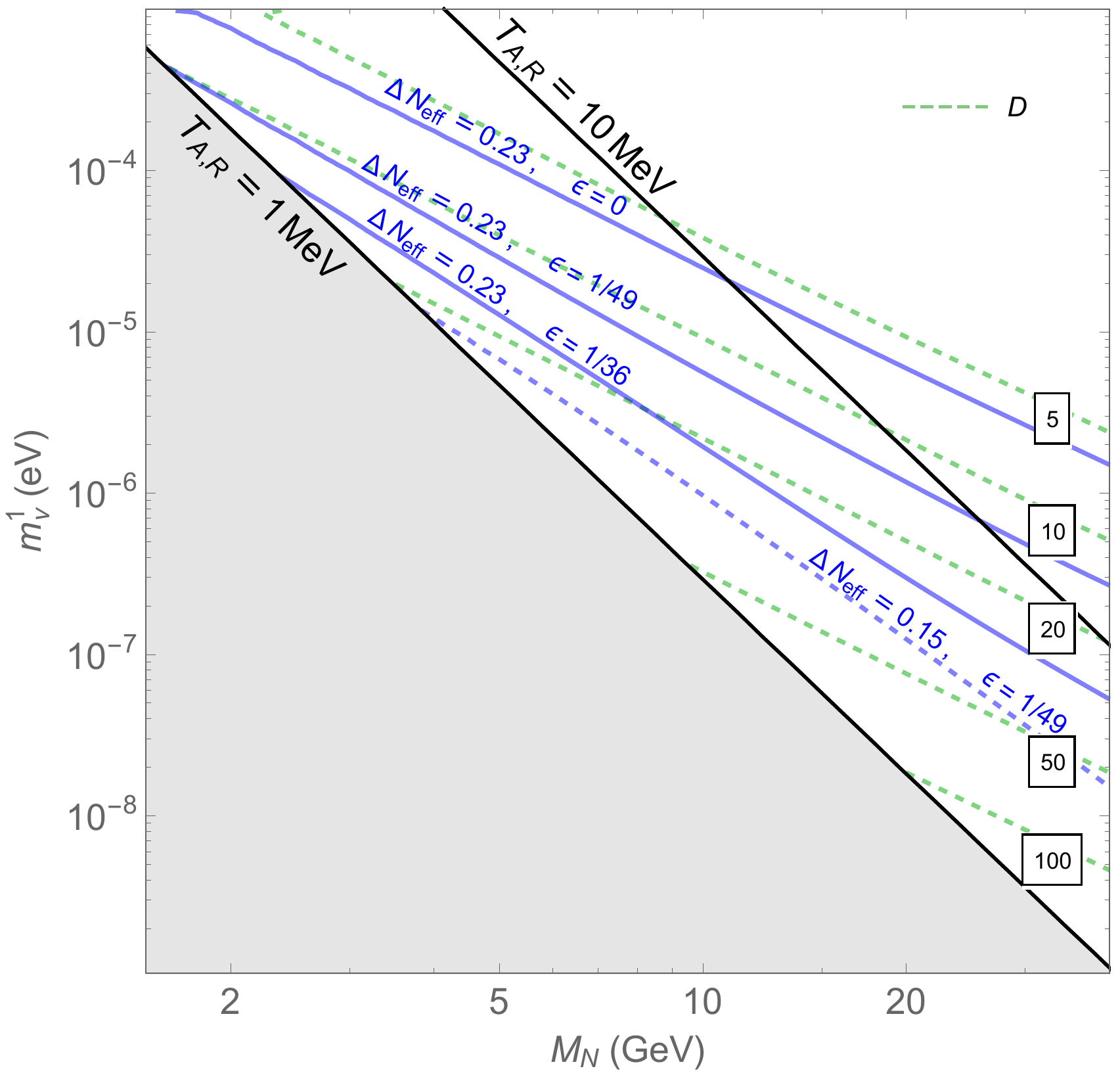}  \\
\end{tabular}
\caption{
The parameter space for the universal 1 generation $\nu$MTH model, written in terms of the lightest active neutrino mass $m_{\nu}^1$. 
Left: coloured contours show $\Delta N_{eff}$ for $\epsilon = 0$. 
Right: blue contours indicate $\Delta N_{eff} = 0.23$ or $0.15$ for various choices of $\epsilon$ inspired by the minimal $\nu$MTH prediction $\epsilon = v^2/f^2$. 
In both plots, green dashed contours show the dilution factor $D$. 
The 1 MeV and 10 MeV lines are the optimistic and conservative estimates for bounds on $T_{A,R}$ respectively, due to BBN constraints.}
\label{fig:nuMTHparam}
\end{center}
\end{figure}

When the RHNs decay, they give rise to a suppressed $\Delta N_{\text{eff}}$ signature according to Equation (\ref{eqn:DNeffv3}) and a dark matter dilution factor given by Equation (\ref{eqn:D_simple}). Note that computations were done accounting for the redshift of the RHNs, using Equation (\ref{eqn:RQ}) for $R_N$. The available parameter space for this toy model is shown in Figure \ref{fig:nuMTHparam}, where we consider reheating temperatures $T_{A,R}$ ranging from 1 to 10 MeV. Results are shown both for $\epsilon = 0$ and for $\epsilon = v^2/f^2$ at $f/v = 5,6,7$.

The results of this one-generation toy model can generally be regarded as conservative estimates for results expected in a more realistic, 3 generation model. For example, consider the $\Delta N_{\text{eff}}$ signature for the two models in terms of the branching ratio $\epsilon$. If, in the 3 generation model, the branching ratio for the last decay is large enough that the injected energy density into the twin sector dominates over the preexisting energy density, then this new radiation is what sets the ratio $T_B/T_A$, and the cosmological history of the previous two decays is essentially irrelevant. In this case the $\Delta N_{\text{eff}}$ signature for the 1 and 3 generation models should be similar. On the other hand, when $\epsilon = 0$ each successive decay will reheat the visible sector but not the twin sector, such that $T_B/T_A$ is further reduced after each decay takes place. In this case we expect that $\Delta N_{\text{eff}}$ will be smaller by a factor of a few in the 3 generation case.

Turning to the dilution factor, we note that the dark matter will be successively diluted relative to the visible sector for each RHN decay regardless of $\epsilon$, since the branching ratio of the RHNs into dark matter is by definition zero. We therefore expect $D$ to be a factor of a few larger for the 3 generation model in addition to a potential decrease in $\Delta N_{\text{eff}}$.  In either case, dilution factors up to $D \sim \mathcal{O}(100)$ are easily generated in the $\nu$MTH setup.

\subsubsection{$X$MTH}
\label{sec:XMTH}

The $X$MTH is another asymmetrically reheated Twin Higgs model. 
This model was first proposed in Ref. \cite{Craig:2016lyx}, and below we review the main features of the model and update some derivations and bounds.

The $X$MTH couples a scalar particle $X$ to the Higgs sector in much the same way that the dark matter $S$ is coupled. However, the $X$ is made unstable by introducing $x$, which may be identified as the vev of $X$ in some UV completion of the theory or a small linear coupling with the Twin Higgs multiplet:
\beq
\label{eqn:V_XMTH}
V \supset \lambda_X X (X + x) (H_A^\dagger H_A + H_B^\dagger H_B) + \frac{1}{2} m_X^2 X^2 
\eeq
 Since our focus in this work is to solve the little hierarchy problem, we will simply treat this is as a low-energy effective theory and remain agnostic to the possible UV origins of this framework.\footnote{Both $\lambda_X$ and $x/m_X$ will be tiny in our regions of interest. Therefore, $X$ itself does not inherit a hierarchy problem from its couplings to the Twin Higgs states within this effective IR description.}
 The small coupling $\lambda_X x$ allows $X$ to decay to both visible and twin matter through mixing with the Higgs in each respective sector. As we will see, for $m_X \sim \mathcal{O}(10 - 1000)$ GeV the branching ratio into the visible sector will dominate over the mirror sector branching ratio, primarily due to the lower mass thresholds in the visible sector.

The form of the potential admits mass mixing between $h_A$, $h_B$, and $X$. After expanding $H_A$ and $H_B$ around their vevs the mass matrix can be read off directly via Equations (\ref{eqn:V_TH}) and (\ref{eqn:V_XMTH}):

\beq
\label{eqn:mass_matrix}
M_{\text{gauge}}^2 \equiv \begin{pmatrix}  \lambda v_A^2 +  \frac{1}{2} \lambda (f^2 - f_0^2) + \frac{3}{2} \kappa v_A^2 + \frac{1}{2} \sigma f_0^2 && \lambda v_A v_B && \frac{1}{2} \lambda_X x v_A \\ \lambda v_A v_B && \lambda v_B^2 + \frac{1}{2} \lambda (f^2 - f_0^2) + \frac{3}{2} \kappa v_B^2  && \frac{1}{2} \lambda_X x v_B \\  \frac{1}{2} \lambda_X x v_A && \frac{1}{2} \lambda_X x v_B && \frac{1}{2} m_X^2 + \frac{1}{2} \lambda_X f^2 \\ \end{pmatrix}
\eeq

To understand the mass eigenstates in the theory, it is instructive to diagonalize the mass matrix in two steps. First, we diagonalize the $2 \times 2$ $(H_A, H_B)$ submatrix in exactly the same way as was done in Section \ref{sec:MTH_model}. This is done by rotation through angle $\theta$, and takes us to the $\{h, \hat{h}, X\}$ basis with mass matrix 

\beq
\label{eqn:mass_matrix_hH}
M_{\text{partial}}^2 \equiv \begin{pmatrix} \frac{1}{2} m_h^2 && 0 && \frac{1}{2} \lambda_X x \left(v_A \cos \theta - v_B \sin \theta \right) \\ 0 && \frac{1}{2} m_{\hat{h}}^2 &&  \frac{1}{2} \lambda_X x \left(v_A \sin \theta + v_B \cos \theta \right) \\ \frac{1}{2} \lambda_X x \left(v_A \cos \theta - v_B \sin \theta \right) &&  \frac{1}{2} \lambda_X x \left(v_A \sin \theta + v_B \cos \theta \right) && \frac{1}{2} m_X^2 + \frac{1}{2} \lambda_X f^2 
\end{pmatrix}.
\eeq

\begin{figure}
\begin{center}
\includegraphics[height=2.4cm]{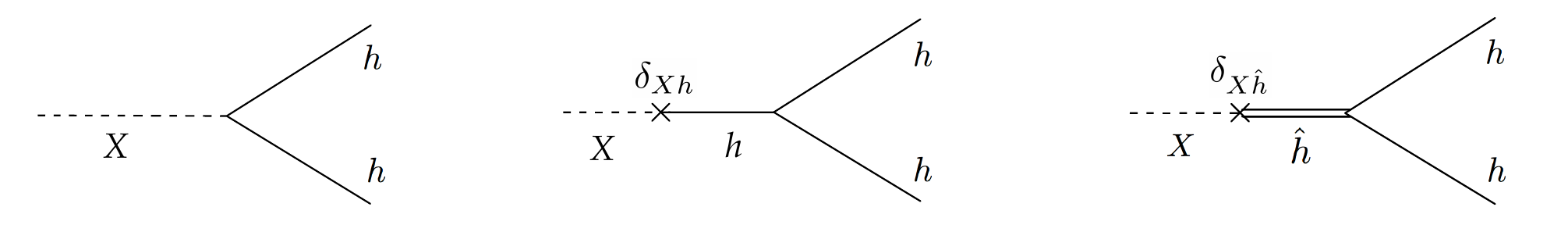}
\caption{Contributions to $\Gamma(X \to hh)$.}
\label{fig:Xhh}
\end{center}
\end{figure}

Since $X$ has to be long-lived, we take $\lambda_X x \ll m_X, m_h, m_{\hat h}$. This means that $M_{\text{partial}}^2$ is nearly diagonal, and so the states $\{h,\hat{h},X\}$ are already approximate mass eigenstates. While the exact Higgs mass eigenvalues will be shifted at $\mathcal{O}(\lambda_X x)$ from $m_h$ and $m_{\hat{h}}$, and the $X$ mass eigenvalue has already been shifted at $\mathcal{O}(\lambda_X)$, these corrections are small enough that throughout the rest of this paper we will refer to $\{h,\hat{h},X\}$ as mass eigenstates, with the following mixing angles between $X$ and $h, \hat h$:\footnote{This corrects a minor mistake in the derivation of these angles in~\cite{Craig:2016lyx}.} 
\begin{align}
\label{eqn:DeltaXH}
\delta_{X\hat{h}} &\equiv \frac{ \lambda_X x \left(v_A \sin\theta + v_B \cos\theta \right)}{m_X^2 - m_{\hat{h}}^2} + \mathcal{O}(\lambda_X^2)\\
\label{eqn:DeltaXh}
\delta_{Xh} &\equiv \frac{ \lambda_X x \left(v_A \cos\theta - v_B \sin\theta \right)}{m_X^2 - m_h^2} + \mathcal{O}(\lambda_X^2)
\end{align}
Note that $\tan \theta = v_A/v_B$ in the $SU(4)$-symmetric limit with $\kappa, \sigma = 0$, and hence $\delta_{Xh} \to 0$, consistent with the pNGB nature of $h$. 

It is also useful to translate Equations (\ref{eqn:DeltaXH} - \ref{eqn:DeltaXh}) into mixing angles between $X$ and $h_A$, $h_B$, since these give decay rates into the visible and mirror sectors respectively via off-shell Higgs bosons:
\begin{align}
\label{eqn:DeltaXA}
\delta_{XA} &= \delta_{Xh} \cos \theta + \delta_{X\hat{h}} \sin \theta \\
\label{eqn:DeltaXB}
\delta_{XB} &= \delta_{X\hat{h}} \cos \theta - \delta_{Xh} \sin \theta
\end{align}
To first order in $v/f$ and $m_X \gtrsim m_{\hat{h}}$ these take the simple form
\begin{align}
\label{eqn:DeltaXAlarge}
\delta_{XA} &\approx \frac{\lambda_X x}{m_X^2-m_{\hat{h}}^2} ~v ~~~~~~~~~~(m_X \gtrsim m_{\hat{h}}) \\
\label{eqn:DeltaXBlarge}
\delta_{XB} &\approx \frac{\lambda_X x}{m_X^2-m_{\hat{h}}^2} ~f ~~~~~~~~~~(m_X \gtrsim m_{\hat{h}}).
\end{align}
while for $m_X \ll m_h$ we expand to $\mathcal{O}\left(\kappa/\lambda\right)$:\begin{align}
\label{eqn:DeltaXAsmall}
\delta_{XA} &\approx - \left(\frac{\kappa}{\lambda}\right) \times \frac{\lambda_X x}{4 \kappa v}  ~~~~~~~~~~(m_X \ll m_h) \\
\label{eqn:DeltaXBsmall}
\delta_{XB} &\approx - \left(\frac{\kappa}{\lambda}\right) \times \frac{\lambda_X x}{4 \kappa v}~\frac{v}{f}  ~~~~~~~~~~(m_X \ll m_h).
\end{align}
Note however that both $\delta_{XA}$ and $\delta_{XB}$ generally have zeroes in the intermediate regime where $m_X$ ranges from $\sim m_h$ to $\sim f$ due to cancellations between the two terms in Equations (\ref{eqn:DeltaXA}, \ref{eqn:DeltaXB}) which are not captured by these simple approximations. Any numerical calculations should be performed using Equations (\ref{eqn:DeltaXH} - \ref{eqn:DeltaXB}).

\begin{figure}
\begin{center}
\includegraphics[height=7cm]{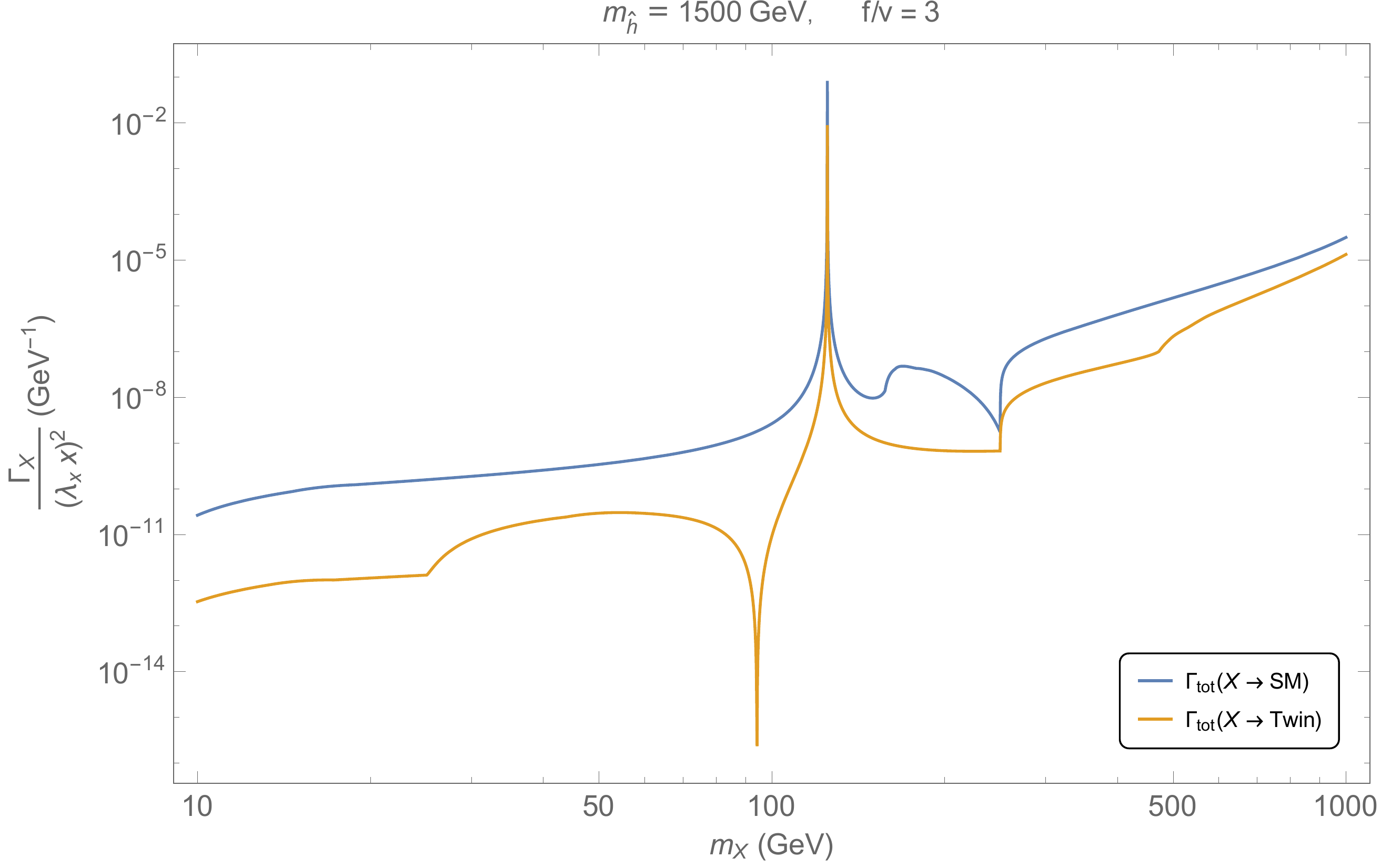}
\caption{Complete decay width of $X$ into visible (SM) and mirror (twin) sectors, normalized by $\left( \lambda_X x \right)^2$ which is common to all partial widths. In this figure we take $f/v = 3$ and $m_{\hat{h}} = 1500$ GeV.}
\label{fig:Xwidth}
\end{center}
\end{figure}

\subsubsection*{\underline{X decays}}

\indent \indent Since the $X$ only couples directly to the Higgs bosons of Twin Higgs theories, we divide discussion of the decay into two regimes, $m_X < 2m_h$, and $m_X \geq 2m_h$.

For $m_X < 2m_h$, all decays take place either through mixing with the Higgs bosons or to 3-body or higher final states via $X \to hh^*$. In all cases we find that $hh^*$ final states are subdominant to other processes and are thus not included in what follows. As discussed in Ref. \cite{Craig:2016lyx}, the partial widths through mass mixing at tree level for all non-Higgs final states are simply
\begin{align}
\label{eqn:GammaMixSM}
\Gamma_{\text{mix}}(X \to \text{SM}) &= |\delta_{XA}|^2 ~\Gamma(h_A \to \text{SM}) \Bigr\rvert_{m_{h_A} = m_X} \\
\label{eqn:GammaMixTW}
\Gamma_{\text{mix}}(X \to \text{Twin}) &= |\delta_{XB}|^2 ~\Gamma(h_B \to \text{Twin}) \Bigr\rvert_{m_{h_B} = m_X}
\end{align}
where we take $m_{h_A}$ and $m_{h_B}$ to mean the mass of a Higgs boson particle in either the visible or twin sector if we were to consider only that sector in isolation. That is, $\Gamma(h_B \to \text{Twin}) \rvert_{m_{h_B} = m_X}$ can be thought of as just the regular decay width of a Higgs of mass $m_X$ to all non-Higgs matter with masses determined by $f/v$. While decays of the SM Higgs are well-documented across a wide range of masses (see e.g. Ref. \cite{Dittmaier:2011ti}), for consistency we compute both $\Gamma(h_A \to \text{SM})$ and $\Gamma(h_B \to \text{Twin})$ by closely following the work of Ref. \cite{Djouadi:2005gi}. All processes relevant at the percent level or higher are included, and our results are in good agreement with Ref. \cite{Djouadi:2005gi}.

For $m_X > 2m_h$, the $X$ can decay directly into two light Higgs bosons $h$. Note that in addition to the direct $Xhh$ vertex, this process will receive two extra contributions from mass mixing\footnote{These contributions were amongst those that were neglected in~\cite{Craig:2016lyx}, which significantly modifies some numerical results but does not change the important qualitative behaviour of the model.} 
 that are shown in Figure \ref{fig:Xhh}. %
These contributions are important due to the pNGB nature of $h$, similar to the discussion in  Section \ref{sec:THPDM}. The decay width of $X \to hh$ is suppressed by $\mathcal{O}(\kappa/\lambda)$ and $\mathcal{O}(\sigma/\lambda)$ factors, particularly for $m_X \gtrsim 2 m_h$, which in our Linear Sigma Model picture is enforced by cancellations between these three diagrams. With this included, the decay $X \to hh$ vanishes in the $SU(4)$-symmetric limit. 
Above $m_X = m_h + m_{\hat{h}}$, the $X$ can decay into an $h\hat{h}$ pair. However, there is no direct coupling between these three states and so this process can only proceed through mass mixing interactions. Over all $m_X$ of interest this decay channel is negligible compared to other processes and is not included in the remainder of the analysis. In principle we should also consider decays to $\hat{h} \hat{h}$ pairs, but we will find below that the signature space of this model is limited to regions where $m_X \lesssim 2 m_{\hat{h}}$ and so we do not include this decay either (see~\cite{Craig:2016lyx} for a discussion of these processes).
The decay rates to the twin and visible sectors are then given by
\begin{align}
&\Gamma_{\text{tot}}(X \to {\text{SM}}) \approx \Gamma_{\text{mix}} (X \to \text{SM}) + \Gamma(X \to hh) \text{Br}(h \to \text{SM})^2 \\
&\Gamma_{\text{tot}}(X \to {\text{Twin}}) \approx \Gamma_{\text{mix}} (X \to \text{Twin}) + \Gamma(X \to hh) \text{Br}(h \to \text{Twin})^2,
\end{align}
and the rate for mixed decays (i.e. $X$ decays to two SM states and two twin states) is found to be
\beq
\label{eqn:mixed}
\Gamma_{\text{tot}}(X \to {\text{mixed}}) \approx 2 \Gamma(X \to hh) \text{Br}(h \to \text{SM}) \text{Br}(h \to \text{Twin}).
\eeq

The former two decay width estimates are illustrated in Figure \ref{fig:Xwidth} for $f/v = 3$ and $m_{\hat{h}} = 1500$ GeV, normalized by $(\lambda_X x)^{2}$ since this factor is common to all partial widths. We can see that across most of the range $10~\text{GeV} \leq m_X \leq 1000~\text{GeV}$, the branching ratio into the visible sector dominates over the mirror sector (with typically minor corrections owing to the mixed decays of Equation (\ref{eqn:mixed})). If $\lambda_X$ is chosen such that $X$ decays after the two sectors have thermally decoupled ($T\sim$ few GeV), then the visible sector will be preferentially reheated and $\Delta N_{\text{eff}}$ may be reduced to below experimental limits. The complete branching ratio into the twin sector is defined as 
\beq
\epsilon \approx \frac{\Gamma_{\text{tot}}(X \to \text{Twin}) + \frac{1}{2} \Gamma_{\text{tot}}(X \to \text{mixed})}{\Gamma_{\text{tot}}(X \to \text{SM})+\Gamma_{\text{tot}}(X \to \text{Twin}) + \Gamma_{\text{tot}}(X \to \text{mixed})}
\eeq
which can easily be smaller than $10^{-2}$, depending on $f/v$ and the mass of $X$.

\subsubsection*{\underline{Thermalization and cosmological bounds}}

At temperatures $T \gg m_X, m_h, m_{\hat{h}}$, the $X$ interactions with the bath scale as $\Gamma \sim T$ but the Hubble rate scales as $H \sim T^2$. Consequently, thermalization at high temperatures is not guaranteed.
$\lambda_X$ must be large enough for the $X$ to make good thermal contact with the SM/twin bath in the early universe, ensuring their number density reaches its equilibrium value, but small enough that the $X$ can freeze out with sufficient number density to eventually dominate the cosmology.\footnote{If $X$ is never in thermal contact until it decays, its initial abundance depends on the model of inflation, but to be conservative we do not consider this scenario here and instead require early thermalization.}

For a given $m_X$, there is a minimum coupling such that the interaction rate can surpass the Hubble rate at some early temperature. Larger couplings increase the time interval during which the $X$ are thermalized and lower the temperature at which the $X$ freezes out. If the coupling is too large, the $X$ particles stay in thermal equilibrium long enough to become Boltzmann-suppressed, reducing the amount of asymmetric reheating. 

\newpage

\begin{figure}[!h]
\vspace*{-10mm}
\begin{center}
\begin{tabular}{cc}
\includegraphics[width=6cm]{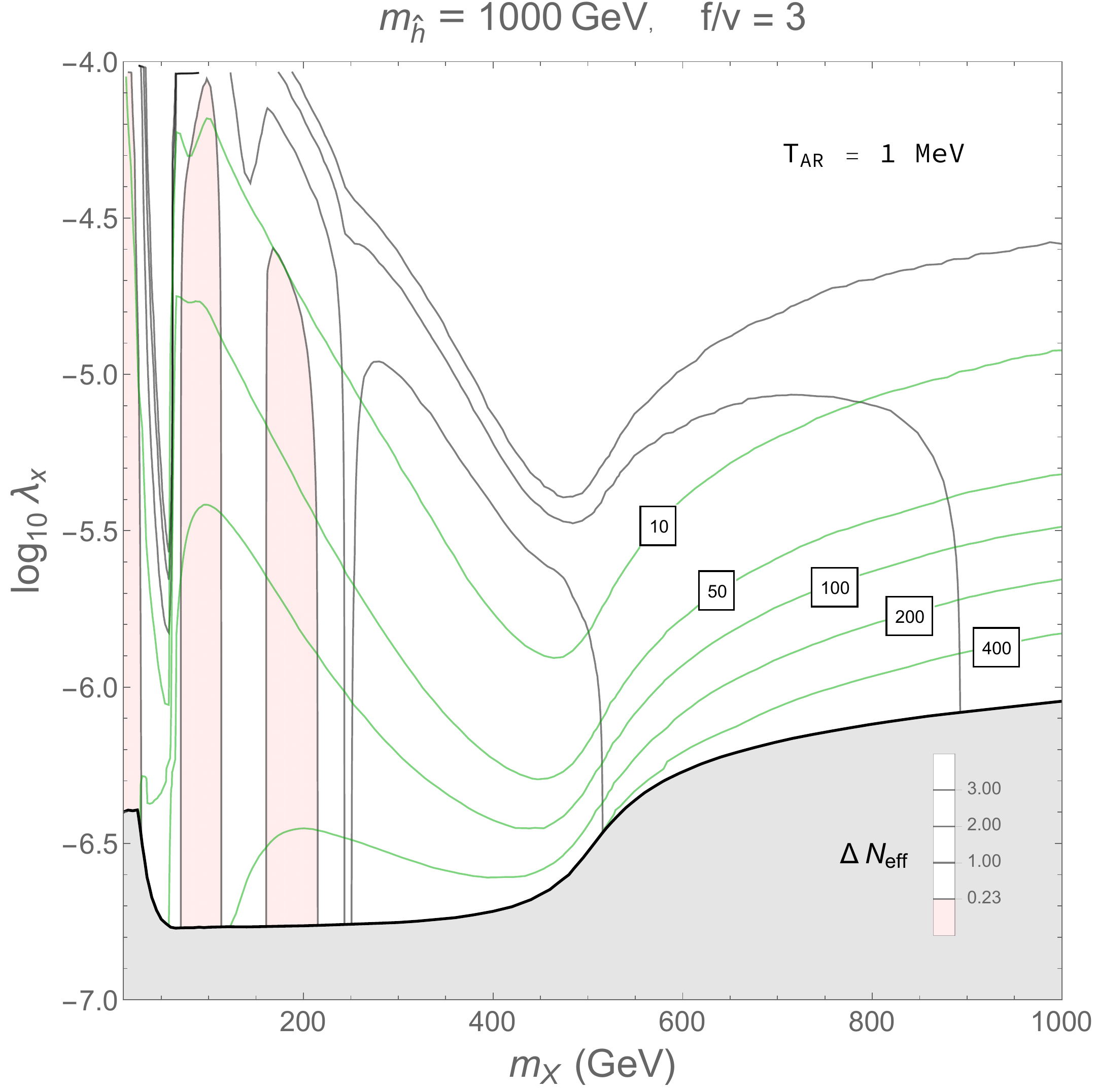} & \includegraphics[width=6cm]{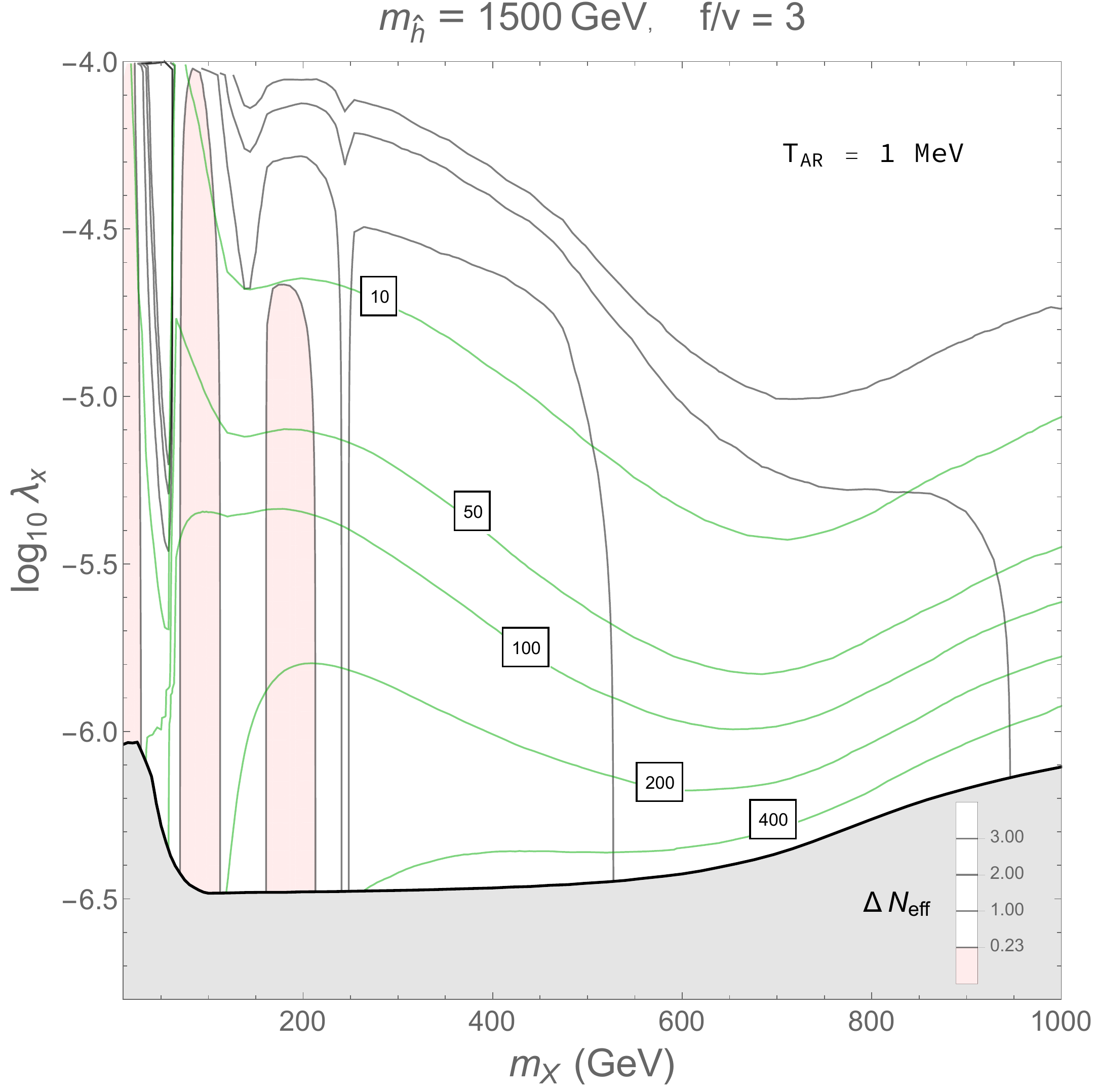}  \\
\includegraphics[width=6.15cm]{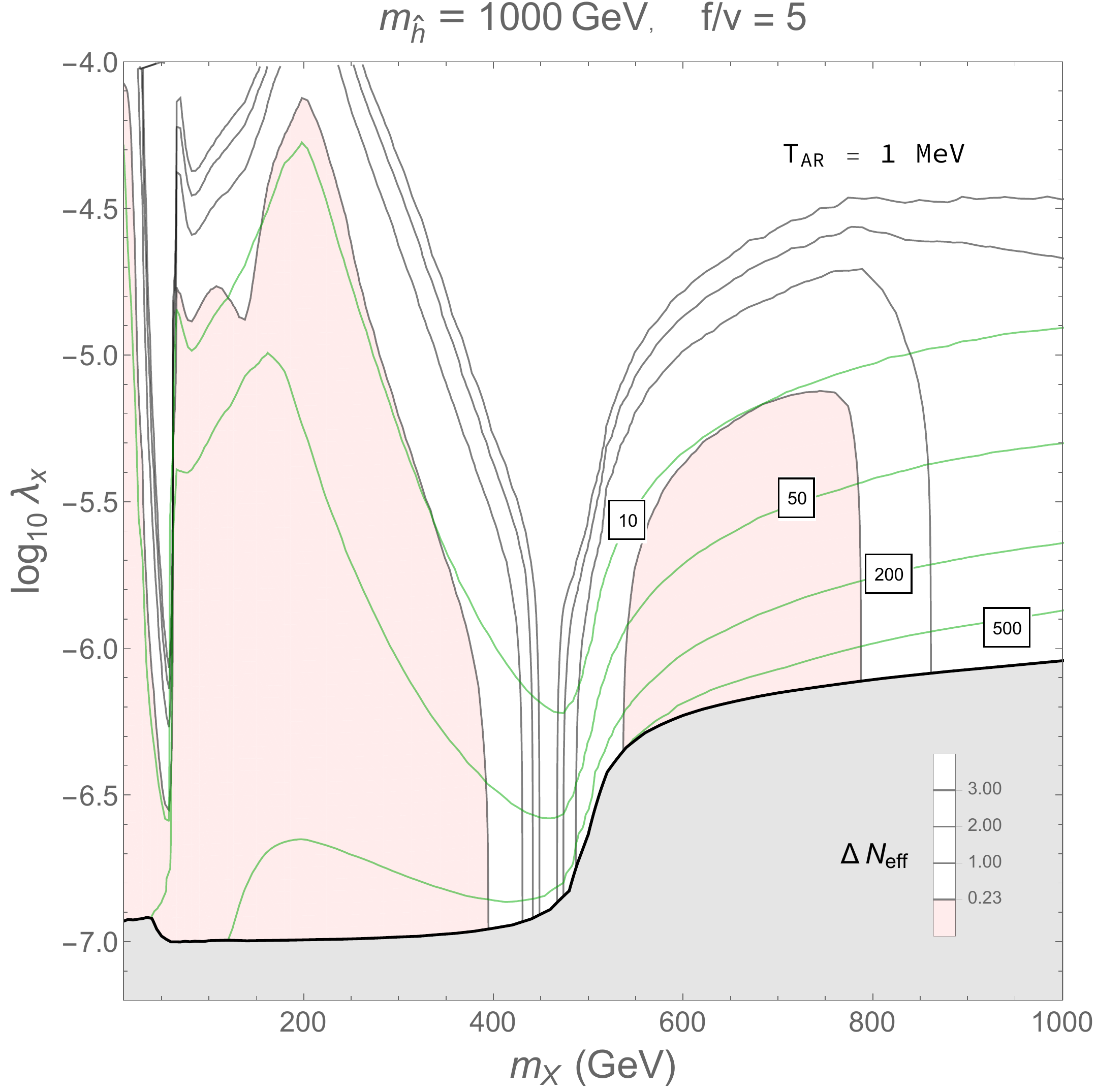} & \includegraphics[width=6cm]{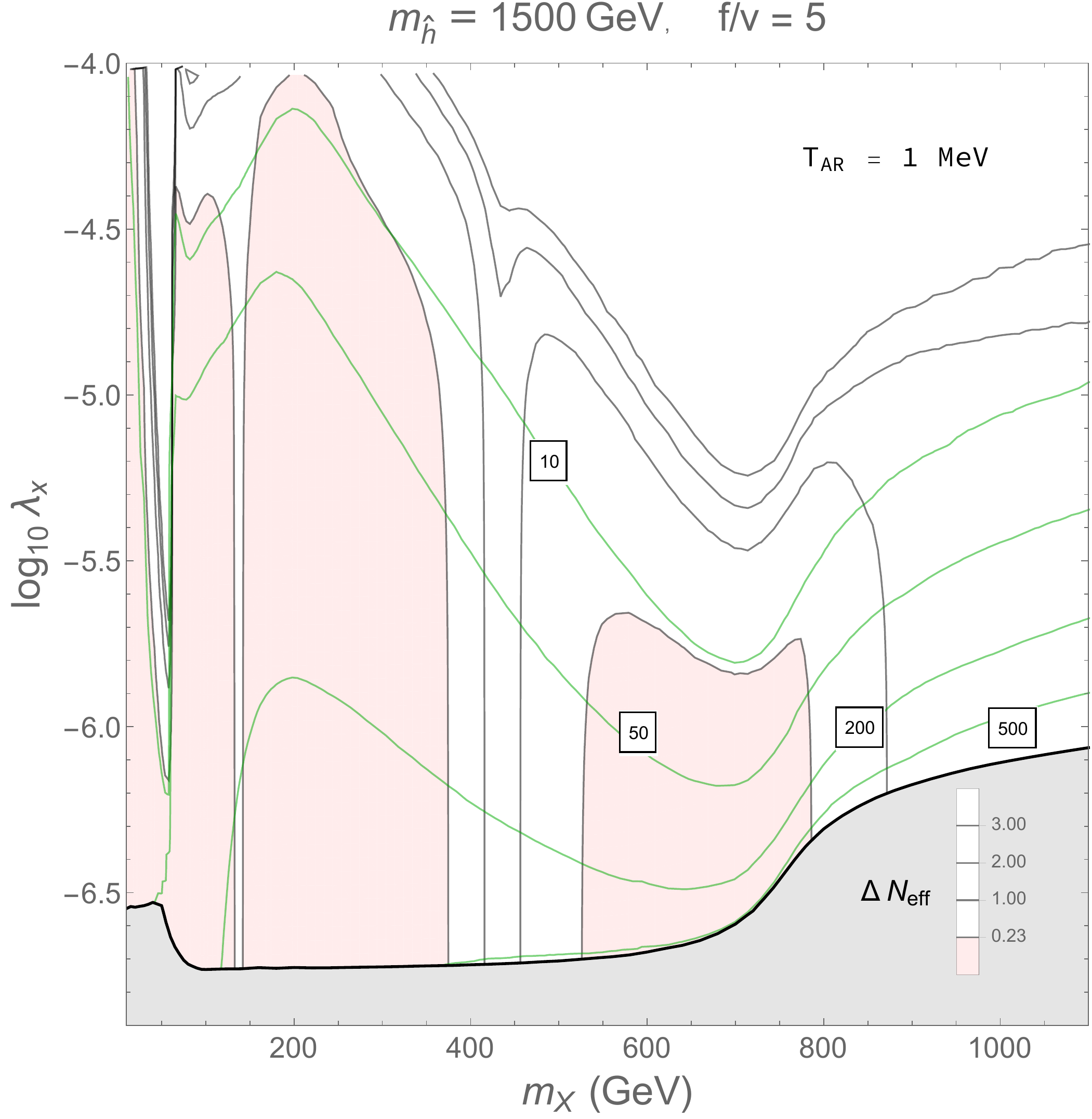}  \\
\includegraphics[width=6cm]{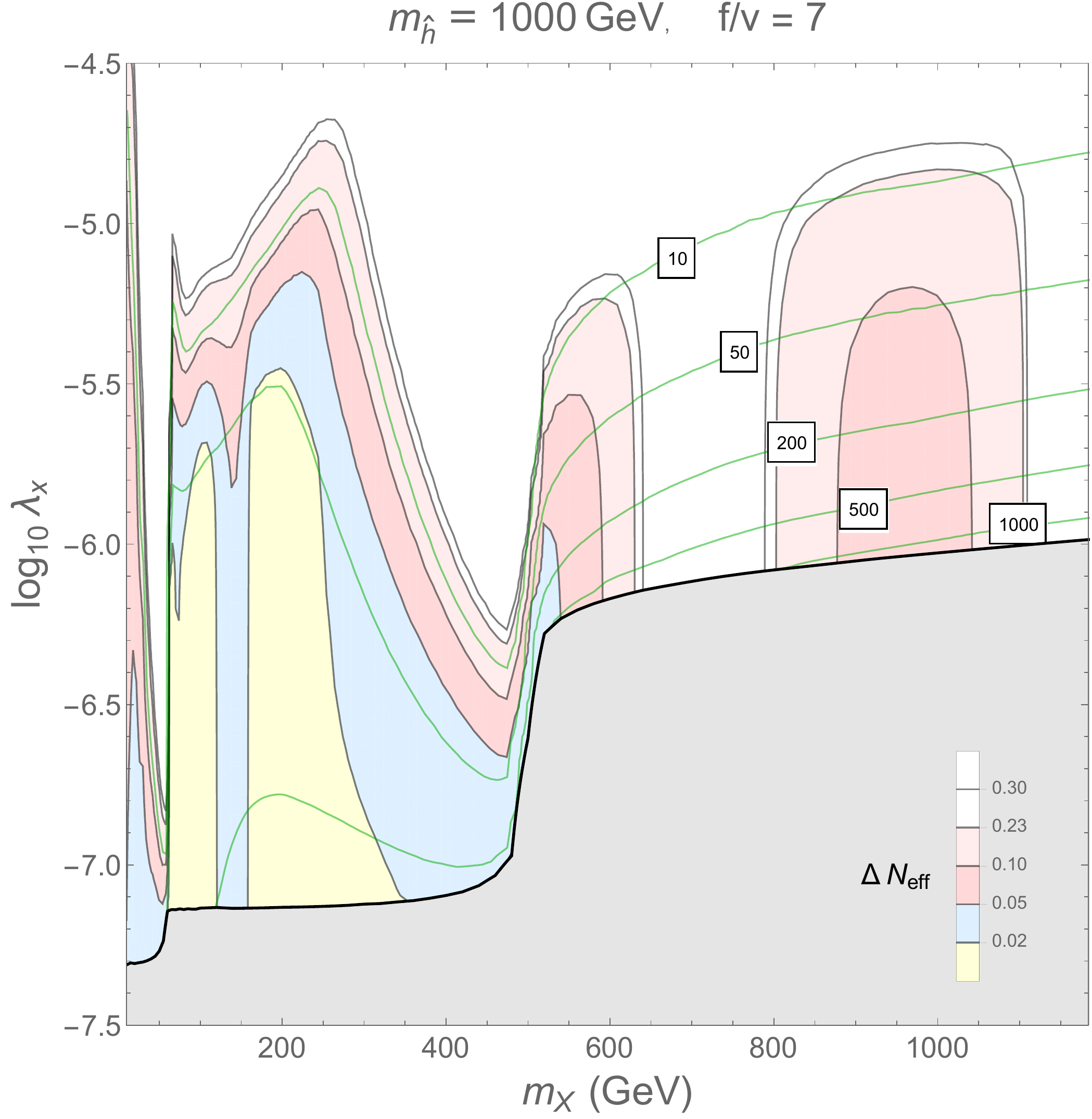} & \includegraphics[width=6.2cm]{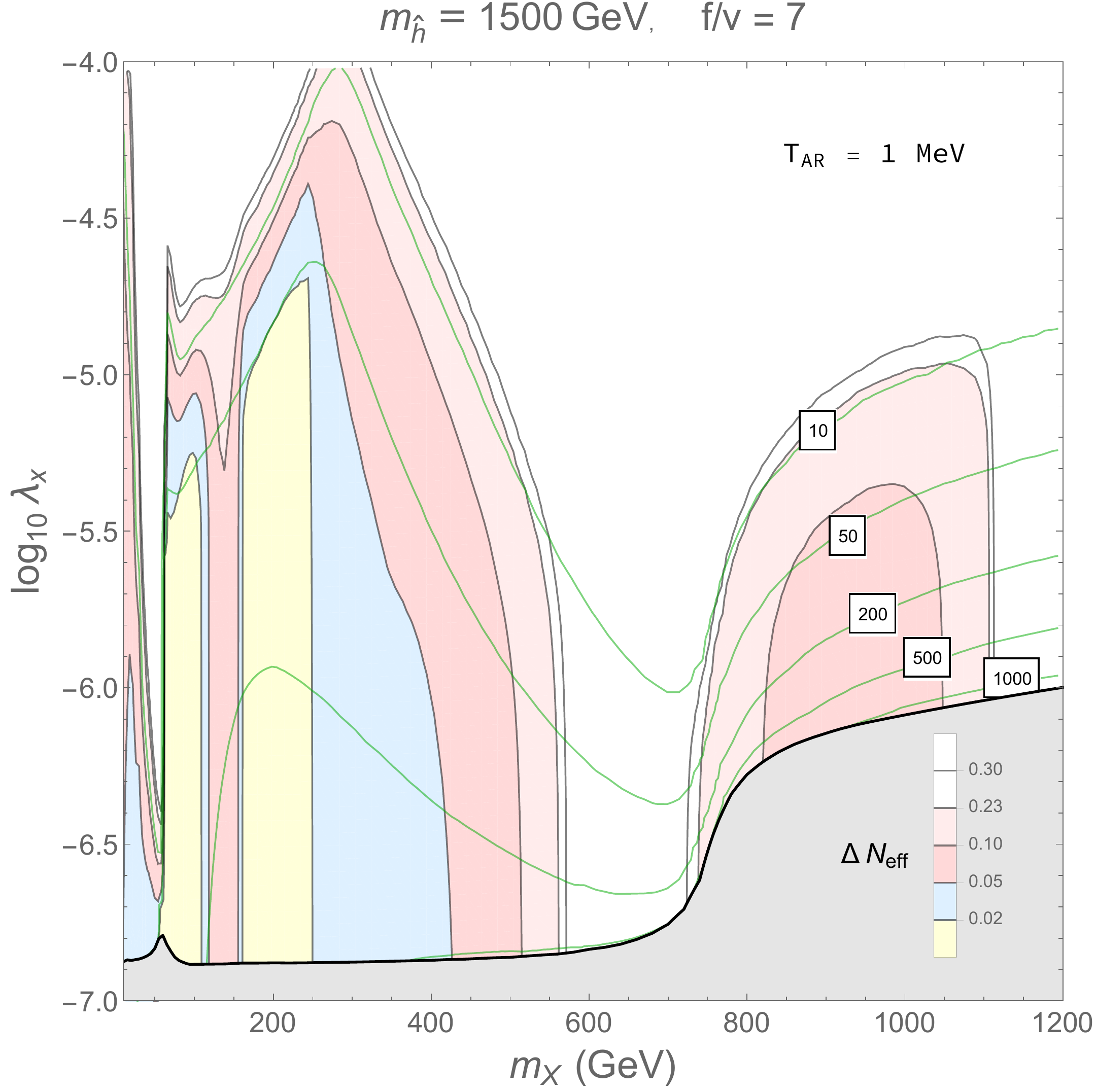}  \\
\end{tabular}
\caption{
Contours of $\Delta N_{eff}$ (coloured regions) and DM dilution factor $D$ (green contours) in the $X$MTH, where $x$ is set such that $X$ decays with a reheat temperature of $T_{A,R} = 1 \mev$, for different $f/v$ and $m_{\hat h}$. 
$X$ never thermalize in the gray region, and the blue, pink, and yellow shaded regions correspond to areas of parameter space where $\Delta N_{\text{eff}}$ is within current experimental limits.
All shaded regions except yellow are expected to be probed by CMB-S4 \cite{Abazajian:2016yjj}.}
\label{fig:XMTH_plots}
\end{center}
\end{figure}

\newpage
\begin{figure}[!h]
\vspace*{-10mm}
\begin{center}
\begin{tabular}{cc}
\includegraphics[width=6cm]{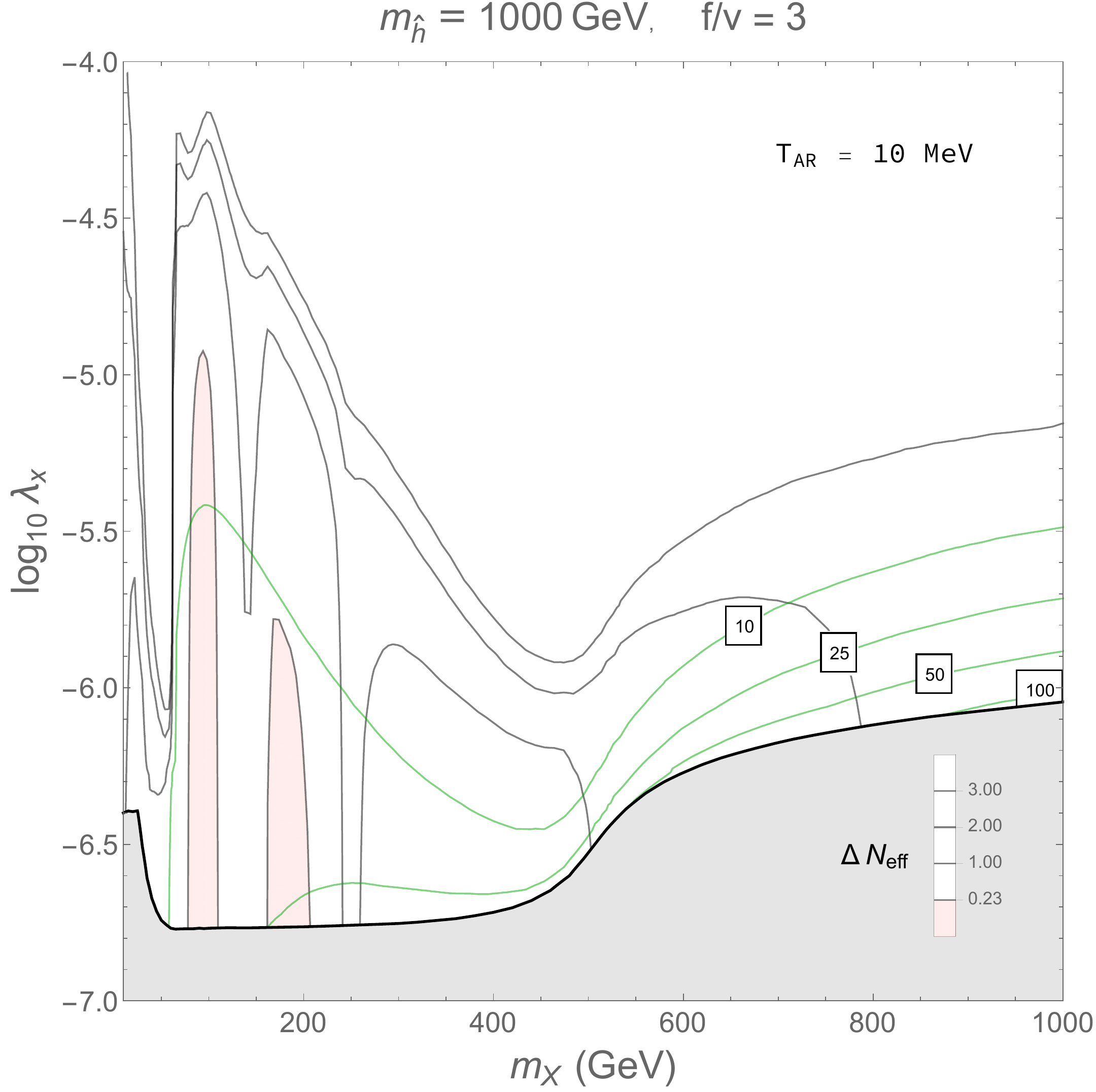} & \includegraphics[width=6cm]{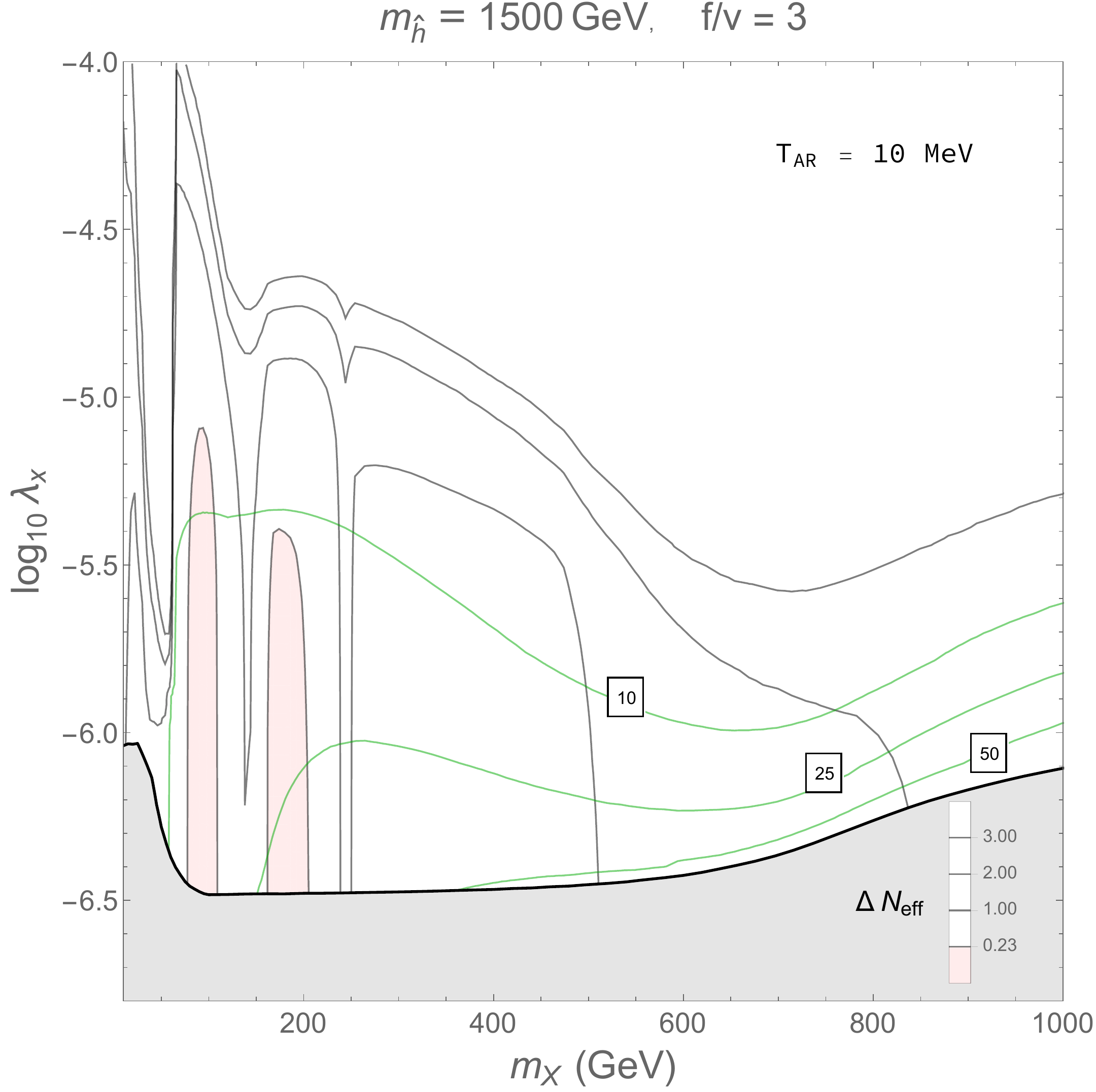}  \\
\includegraphics[width=6.15cm]{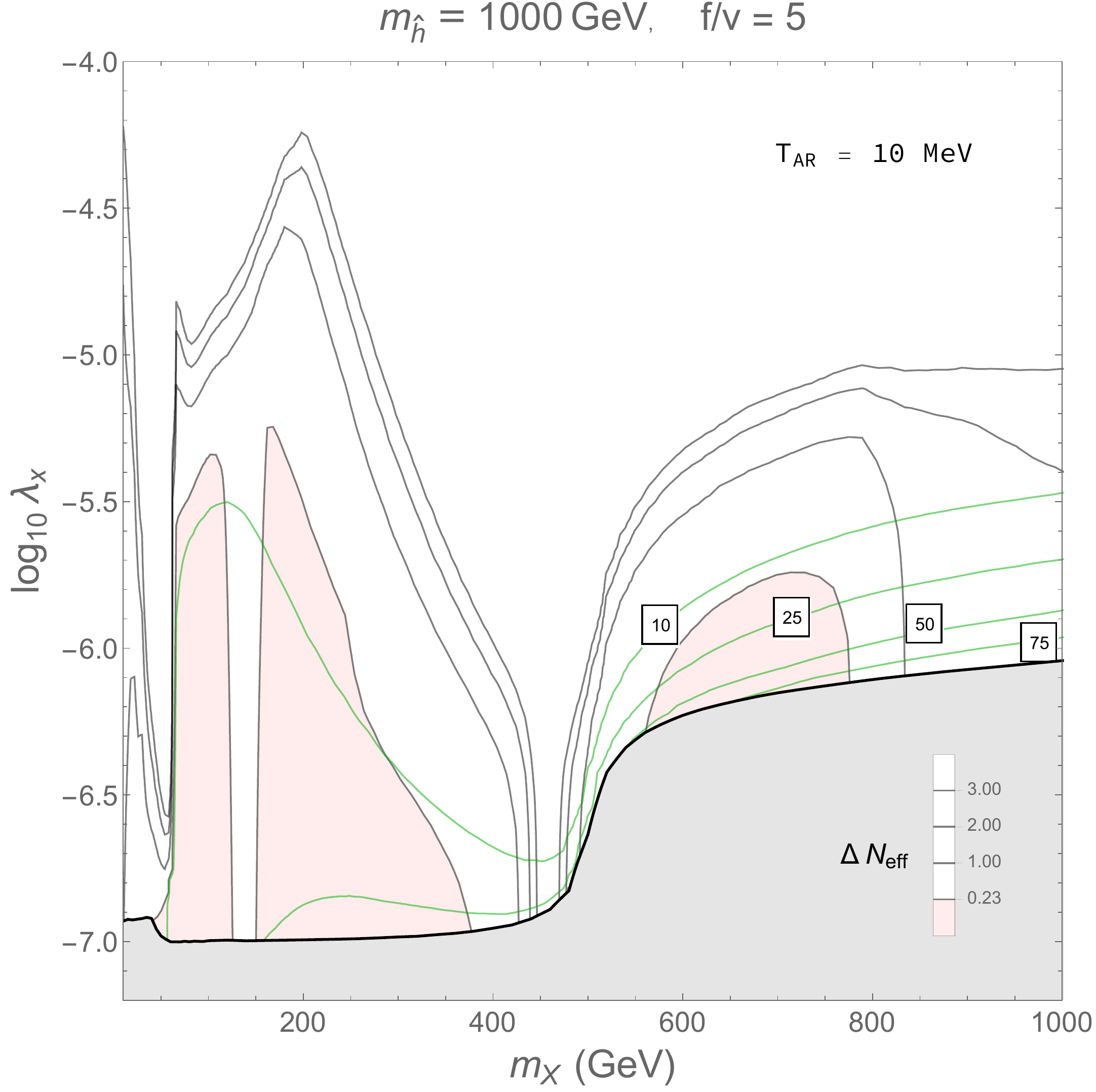} & \includegraphics[width=6cm]{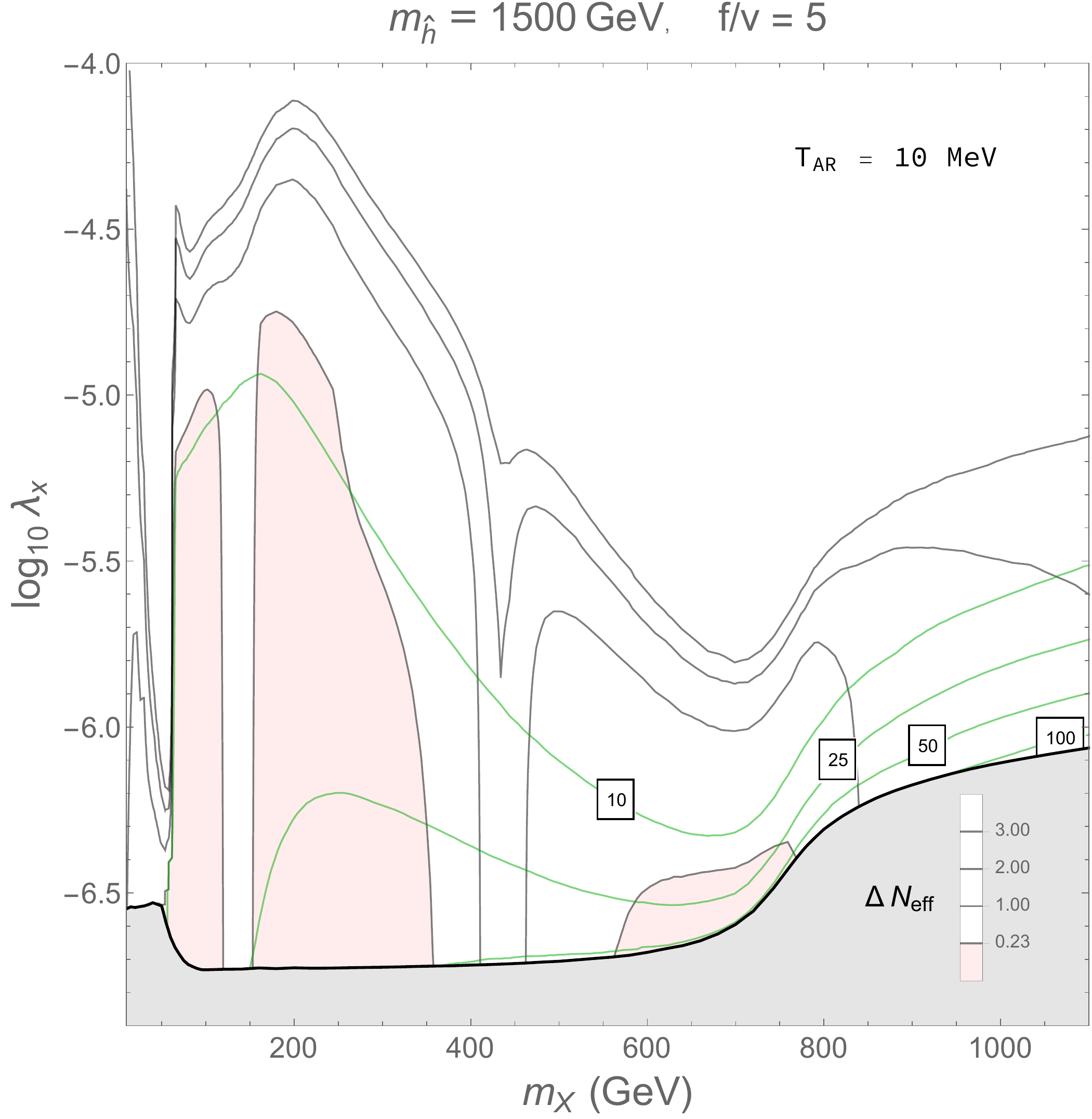}  \\
\includegraphics[width=6cm]{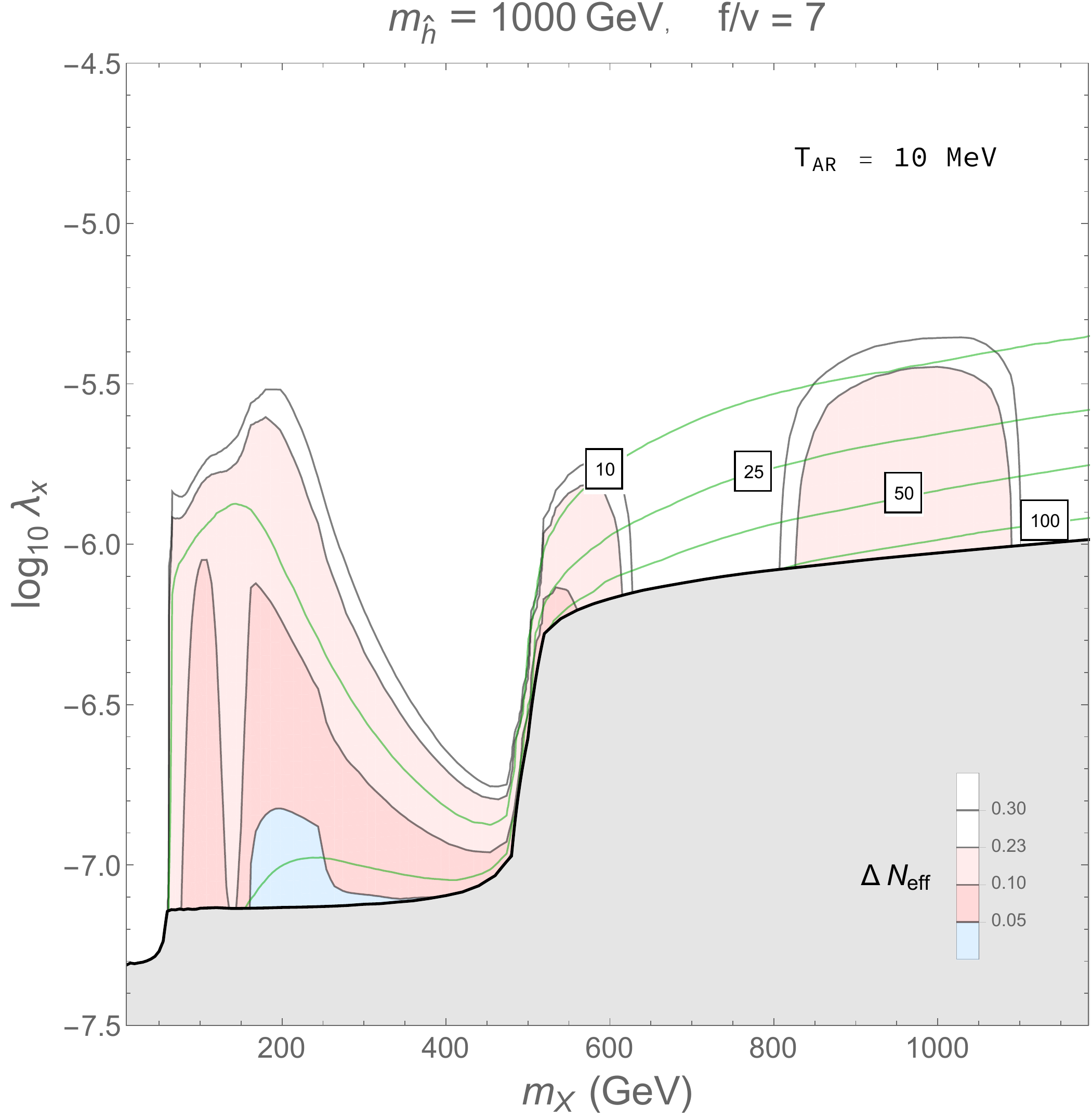} & \includegraphics[width=6.2cm]{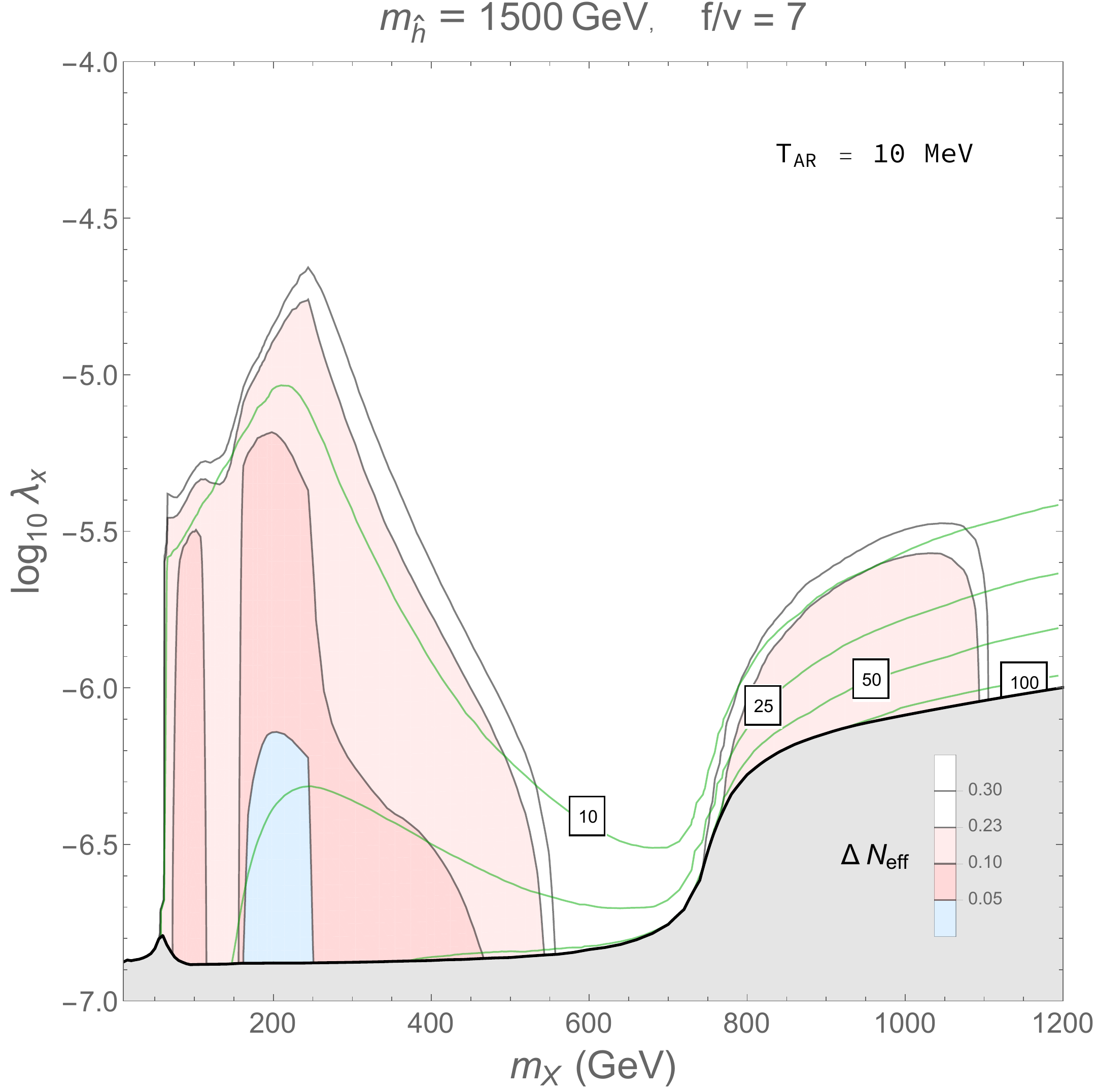}  \\
\end{tabular}
\caption{
Contours of $\Delta N_{eff}$ (coloured regions) and DM dilution factor $D$ (green contours) in the $X$MTH, where $x$ is set such that $X$ decays with a reheat temperature of $T_{A,R} = 10 \mev$, for different $f/v$ and $m_{\hat h}$. 
$X$ never thermalize in the gray region, and  the blue and pink shaded regions correspond to areas of parameter space where $\Delta N_{\text{eff}}$ is within current experimental limits.
All shaded regions are expected to be probed by CMB-S4 \cite{Abazajian:2016yjj}.}
\label{fig:XMTH_plots_tenmev}
\end{center}
\end{figure}

Figures \ref{fig:XMTH_plots} and \ref{fig:XMTH_plots_tenmev} show the values of $\Delta N_{\text{eff}}$ and DM dilution factor $D$ across the parameter range of interest, where $x$ was chosen such that the decay width is fixed at the lower bounds of $\Gamma_X \approx 4.4 \times 10^{-25}$ GeV and $\Gamma_X \approx 4.5 \times 10^{-23}$ GeV, corresponding to reheat temperatures in the visible sector of $T_{A,R} = 1$ MeV and 10 MeV respectively. These plots therefore show the largest amount of dilution that is possible in this model.
The black line depicts the limiting case where $X$ thermalizes and promptly decouples. In the gray regions, $X$ never thermalizes in the early universe. 
 As expected, the largest dilution and smallest $\Delta N_{\text{eff}}$ are found near the lower bound of $\lambda_X$, since a smaller coupling allows the $X$ to freeze out earlier and carry away more energy from the thermal bath. There are large regions of parameter space that satisfy bounds on $\Delta N_{\text{eff}}$, with dilution factors as large as $D \sim 1000$, which will significantly suppress direct detection signatures of THPDM.

\subsection{THPDM Direct Detection with Asymmetric Reheating}
\label{sec:results}

\begin{figure}
\begin{center}
\begin{tabular}{cc}
\includegraphics[height=5cm]{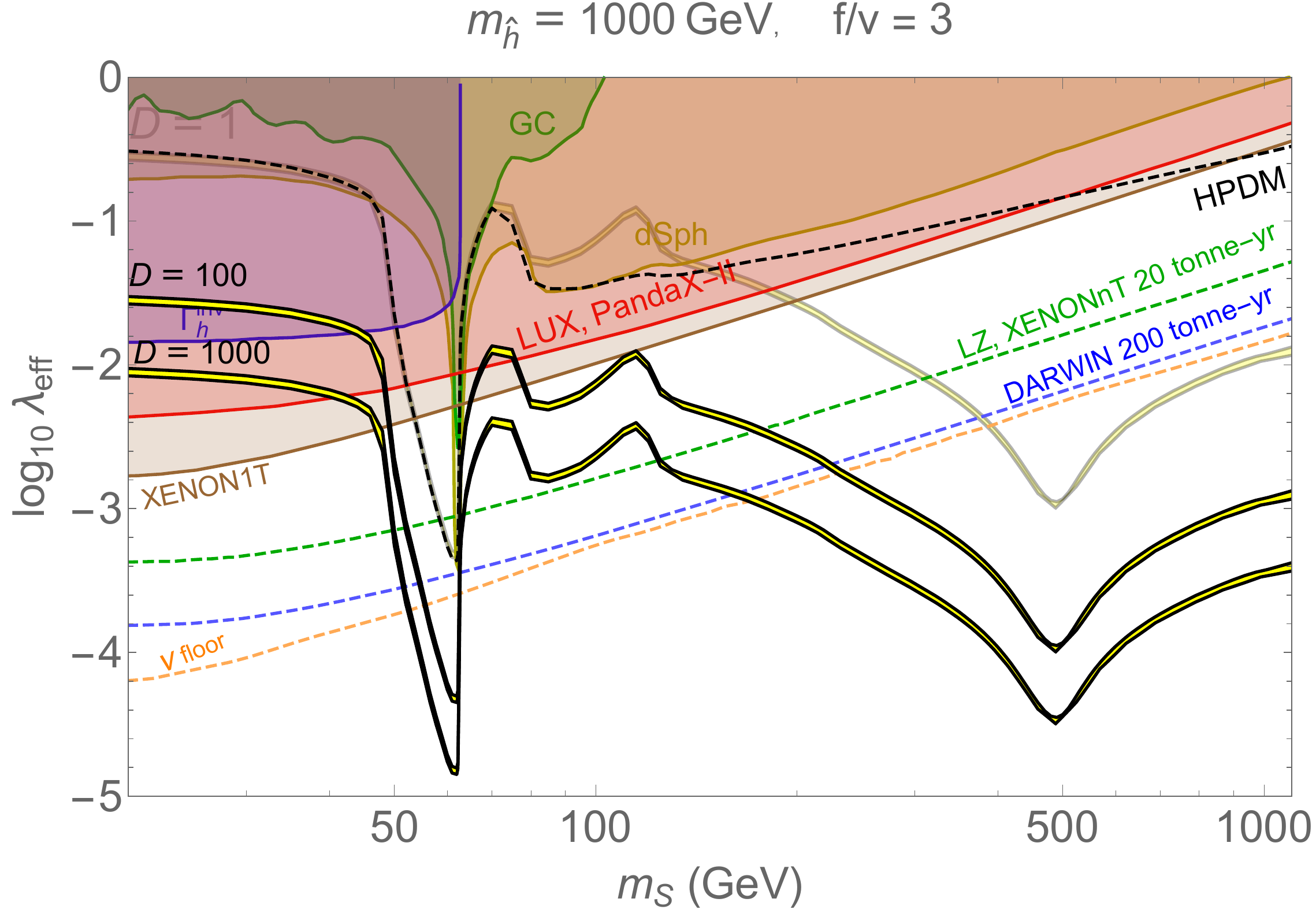} & \includegraphics[height=5cm]{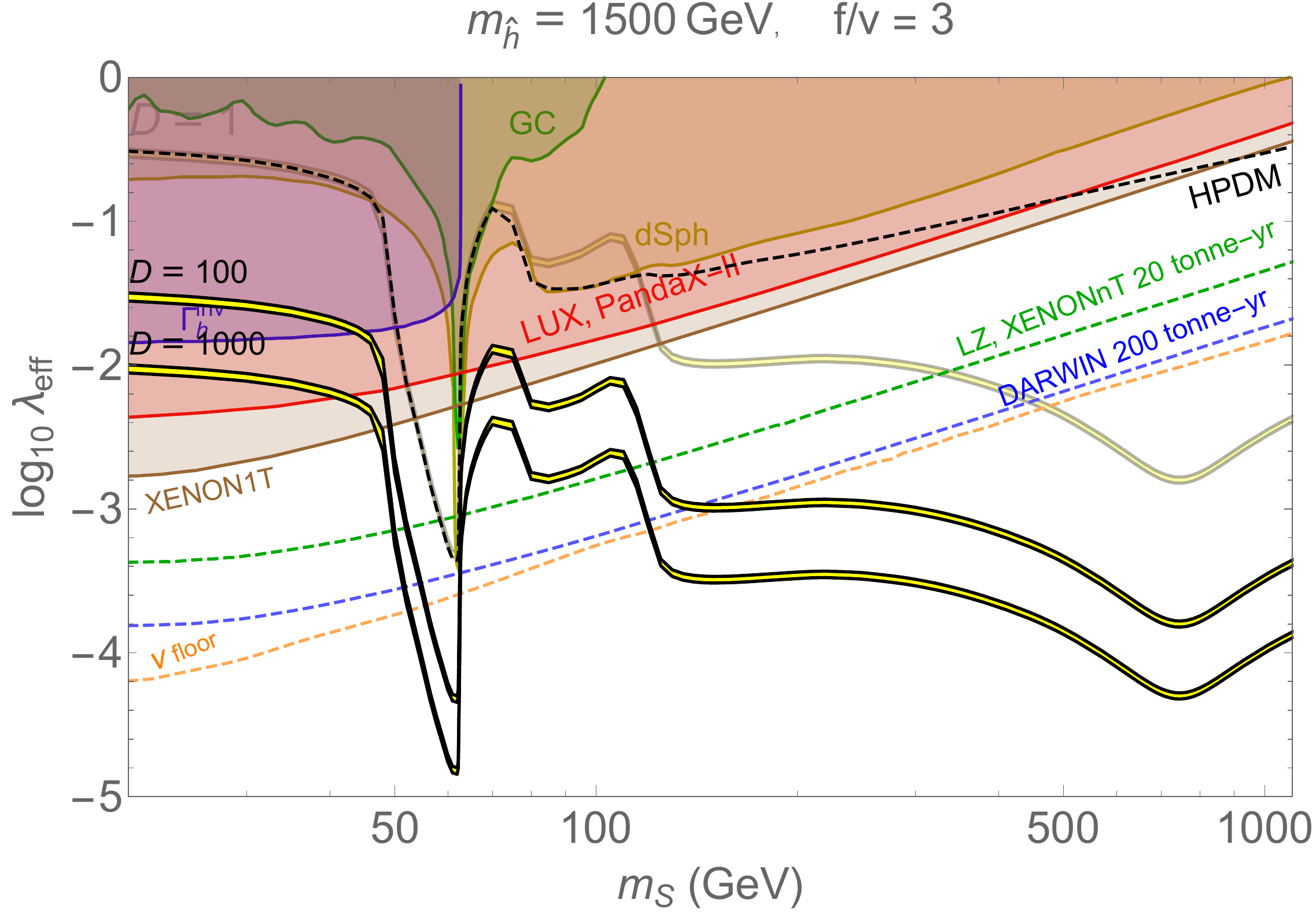} \\
\includegraphics[height=5cm]{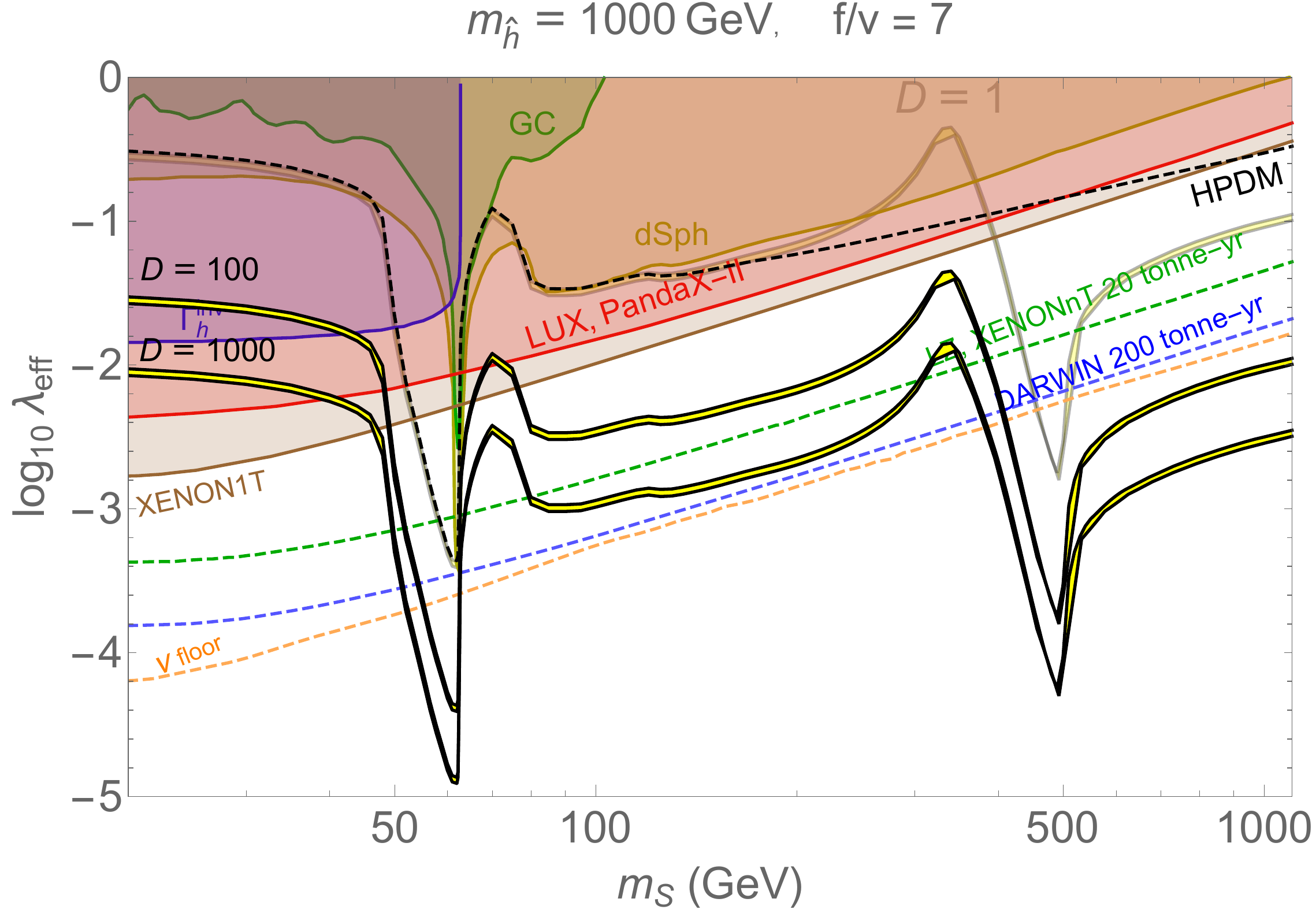} & \includegraphics[height=5cm]{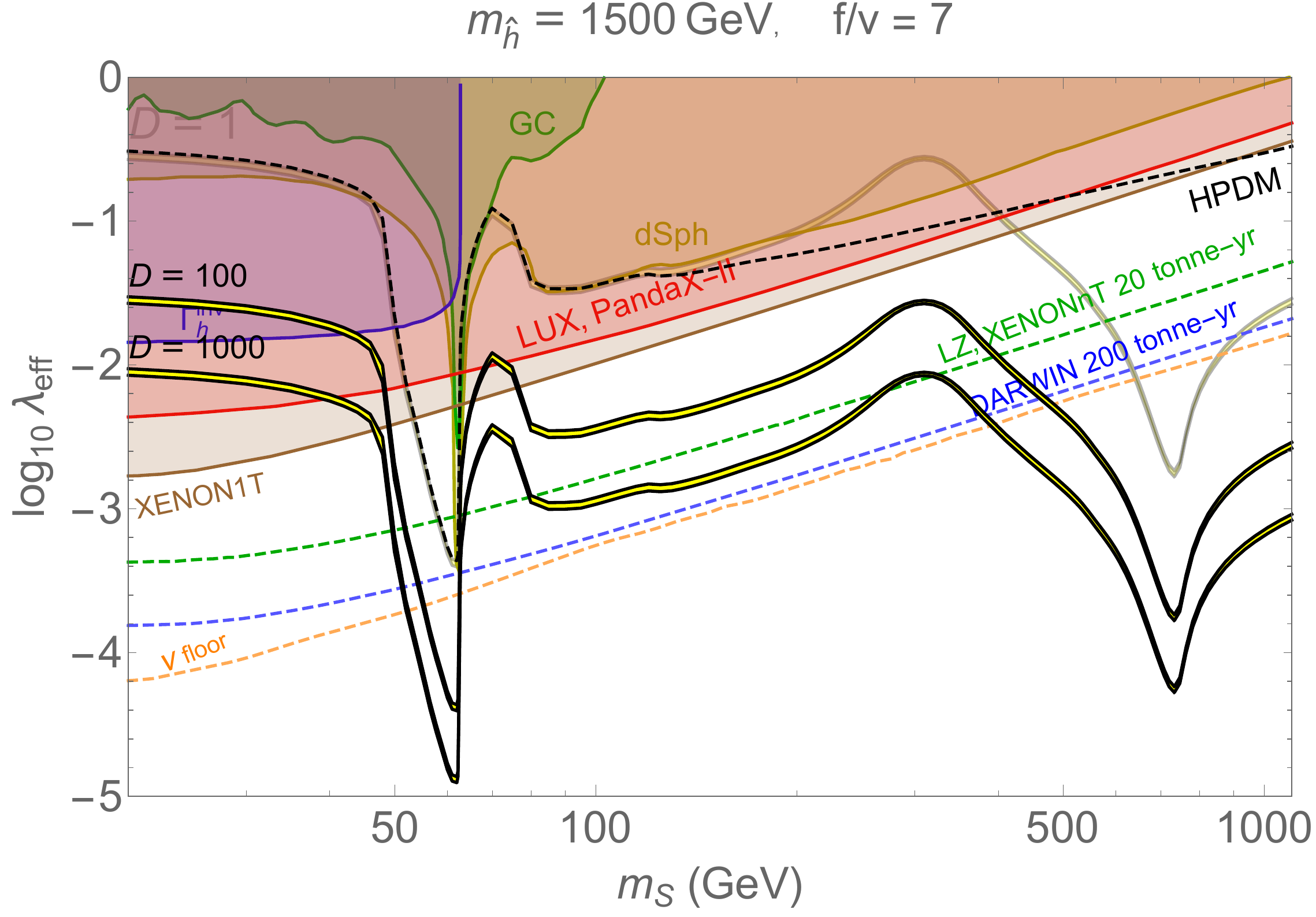} \\
\end{tabular}
\caption{
Direct detection parameter space of THPDM with asymmetric reheating. 
The opaque yellow band shows the effective THPDM coupling without dilution spanning FTH- and MTH-like mirror spectra, same as Figure~\ref{fig:THP_nodil}. 
The lower two yellow bands correspond to dilution factors from asymmetric reheating of $D = 100$ and $1000$, representing high dilution factors that can be produced in the $\nu$MTH and $X$MTH models.
}
\label{fig:results}
\end{center}
\end{figure}

In Section \ref{sec:nuMTH}, we saw that the $\nu$MTH can generate dilution factors in excess of $D \gtrsim 100$ that are consistent with current bounds on $\Delta N_{\text{eff}}$. For the $X$MTH, Figure \ref{fig:XMTH_plots} shows that dilution factors of $D \gtrsim 1000$ can be consistent with $\Delta N_{\text{eff}}$ bounds for $m_X \gtrsim 800$ GeV, and generally $D \gtrsim 300$ is possible. We therefore adopt dilution factors of $D = 100, 1000$ as reasonable benchmarks within the asymmetrically reheated MTH framework.

Figure \ref{fig:results} 
shows the predicted effective coupling for direct detection in THPDM with asymmetric reheating generating dilution factors of $D = 100, 1000$. 
Without asymmetric reheating, the DM mass is constrained to lie above $m_S \gtrsim 150, 400$ GeV for $f/v = 3,7$, with a small unexplored region in the Higgs funnel that is present in the predictions of the HPDM as well. 
When asymmetric reheating is included, the DM mass is unconstrained above $m_X \gtrsim 50 \gev$. 

As discussed in Section~\ref{sec:AR_signatures}, the dilution factor is in principle a separate observable, since in the event of a discovery (and possibly corroboration of the Twin Higgs model in collider measurements), the observed DM direct detection rate can be compared to the prediction from standard cosmology. 
$D$ can then be correlated with other cosmological measurements that may be sensitive to asymmetric reheating. For example, CMB-S4 measurements are expected to constrain $\Delta N_{\text{eff}} \lesssim 0.02$ \cite{Abazajian:2016yjj}. 
In the event of a nonzero $\Delta N_{\text{eff}}$ measurement, any given asymmetrically reheated THPDM scenario will yield a range of expected dilution factors. Consistency between these two observations would constitute significant evidence in support of these models. 

\section{Conclusion}
\label{sec:conclusion}

The Twin Higgs framework offers a compelling explanation for why new, coloured states have yet to appear in collider experiments. Its success as a theory of neutral naturalness is due to the pNGB nature of the 125 GeV Higgs, whose mass is protected from radiative corrections due to an accidental symmetry at one loop order. 

In this work we define the Twin Higgs Portal DM (THPDM) framework by coupling a stable scalar particle $S$ to the Higgs sector of a general Twin Higgs theory. The same accidental symmetry then suppresses interactions between the 125 GeV Higgs and the scalar DM at low energies. 
The scalar is a thermal WIMP, and the fact that freeze-out proceeds through the full Twin Higgs portal coupling, while direct detection is suppressed, leads to predictions for direct detection cross sections that can be several orders of magnitude smaller than is the case in regular scalar Higgs Portal DM (HPDM). 
As can be seen from Figure~\ref{fig:THP_nodil}, HPDM is almost completely excluded, while THPDM is consistent with current constraints for $m_S \gtrsim \mathcal{O}(100 \gev)$. Indeed, this is the favoured parameter space of the model, based on naturalness considerations of the DM mass within the IR effective Twin Higgs theory. 
THPDM naturally predicts null results at current direct detection experiments but positive detections in next-generation experiments -- a very compelling scenario. 
Furthermore, Twin Higgs models that predict dark radiation in excess of current experimental limits on $\Delta N_{\text{eff}}$ are consistent with CMB bounds if they include a source of asymmetric reheating, diluting the contribution of the twin sector. 
This dilution also affects any DM candidate, which further suppresses direct detection signatures for freeze-out DM since a smaller Twin Higgs portal coupling is needed to obtain the correct relic abundance after dilution.
 
We study two explicit models of asymmetric reheating, the $\nu$MTH~\cite{Chacko:2016hvu} and $X$MTH~\cite{Craig:2016lyx}, and find that in addition to alleviating the tension with cosmology, both lead to predicted DM dilution factors of $\mathcal{O}(100 - 1000)$.
This further opens up the direct detection parameter space of the model, with THPDM masses as low as 50 GeV allowed by current constraints (see Figure~\ref{fig:results}). 
The dilution factor is also in principle observable  by comparing the expected direct detection event rate for a given DM mass to observation, if the rest of the model is corroborated in collider experiments. The measured dilution therefore provides another probe of the cosmology of the asymmetrically reheated Twin Higgs. 

THPDM is a simple and general extension of the Twin Higgs that naturally predicts DM heavier than $\sim \mathcal{O}(100 \gev)$ and evades current direct detection experiments, but predicts a signal at next generation detectors such as  LZ, XENONnT, and DARWIN in large regions of parameter space. 
Twin Higgs models will be additionally constrained by measurements of $\text{Br}(h \to \text{inv})$ and precision electroweak physics, searches for LLPs \cite{Beacham:2019nyx,Lubatti:2019vkf,Curtin:2018mvb,Gligorov:2017nwh,Cheng:2015buv}, and searches for additional scalars \cite{Katz:2016wtw,Chacko:2017xpd,Ahmed:2017psb,Buttazzo:2015bka,Aad:2019zwb,Kilic:2018sew,Alipour-fard:2018mre,Batell:2019ptb}. New results from CMB-S4 will put stringent constraints on twin radiation and the dilution factor $D$. 
If naturalness and dark matter are explained by the Twin Higgs, it is therefore possible that a deluge of discoveries are on the horizon, which will not only provide compelling evidence of new physics but, taken together, tell a detailed story of how the Twin Higgs 
explains several fundamental mysteries of our universe.

 \begin{acknowledgments}

DC would like to especially thank Timothy Cohen and Matthew McCullough for important conversations early in this project.
We are also grateful to Nathaniel Craig and Seth Koren for useful comments on a draft version of this paper, and to Jack Setford for particularly helpful consultations.
We additionally thank Marco Drewes, Zackaria Chacko, and Patrick Fox for useful discussions and correspondence. 
The research of DC and SG
was supported in part by a Discovery Grant from the Natural Sciences and Engineering
Research Council of Canada, and by the Canada Research Chair program.
The research of SG was supported in part by a Postgraduate Scholarship -- Doctoral (PGS D) from the National Sciences and Engineering Research Council of Canada.
 \end{acknowledgments}


\bibliography{bibliography}
\bibliographystyle{JHEP}


\end{document}